\documentclass[journal=apchd5,manuscript=review,layout=twocolumn]{achemso}

\usepackage{chemformula} 
\usepackage[T1]{fontenc}
\usepackage{textcomp} 

\usepackage{amsmath,amssymb,amsfonts}%
\usepackage{mathtools}
\usepackage{braket}
\usepackage{siunitx}
\usepackage{bm} 

\usepackage{graphicx}%
\usepackage{xcolor}%
\usepackage{array}
\usepackage{svg} 

\usepackage{booktabs, makecell}
\usepackage{threeparttable}
\usepackage[symbol]{footmisc}

\usepackage[colorlinks=true,citecolor=cyan]{hyperref}

\newcommand{\eps}{\epsilon}

\newcolumntype{C}[1]{>{\centering\arraybackslash}m{#1}} 
\newcommand\T{\rule{0pt}{2.6ex}}       
\newcommand\B{\rule[-1.2ex]{0pt}{0pt}} 

\DeclarePairedDelimiter\abs{\lvert}{\rvert}%
\DeclarePairedDelimiter\norm{\lVert}{\rVert}%

\makeatletter
\let\oldabs\abs
\def\abs{\@ifstar{\oldabs}{\oldabs*}}
\let\oldnorm\norm
\def\norm{\@ifstar{\oldnorm}{\oldnorm*}}
\makeatother

\newcommand{\jac}{\mathbf{J}}
\newcommand{\statevec}{\mathbf{q}}

\author{Jadon Y. Lin}
\affiliation[The University of Sydney]{%
 School of Physics, The University of Sydney, Sydney, 2006, NSW, Australia
}%
\alsoaffiliation[The University of Sydney]{%
Institute of Photonics and Optical Science, The University of Sydney, Sydney, 2006, NSW, Australia
}%

\author{C. Martijn de Sterke}%
\affiliation[The University of Sydney]{%
 School of Physics, The University of Sydney, Sydney, 2006, NSW, Australia
}%
\alsoaffiliation[The University of Sydney]{%
Institute of Photonics and Optical Science, The University of Sydney, Sydney, 2006, NSW, Australia
}%

\author{Ognjen Ilic}
\affiliation[University of Minnesota]{Department of Mechanical Engineering, University of Minnesota, Minneapolis, MN 55455, USA}

\author{Boris T. Kuhlmey}
\email{boris.kuhlmey@sydney.edu.au}
\affiliation[The University of Sydney]{%
 School of Physics, The University of Sydney, Sydney, 2006, NSW, Australia
}%
\alsoaffiliation[The University of Sydney]{%
Institute of Photonics and Optical Science, The University of Sydney, Sydney, 2006, NSW, Australia
}%
\alsoaffiliation[The University of Sydney]{The University of Sydney Nano Institute, The University of Sydney, Sydney, 2006, NSW, Australia}

\title{Photonic lightsails: Fast and Stable Propulsion for Interstellar Travel}


\begin{document}








\begin{abstract}
Lightsails are a highly promising spacecraft concept that has attracted interest in recent years due to its potential to travel at near-relativistic speeds. Such speeds, which current conventional crafts cannot reach, offer tantalizing  opportunities to probe nearby  stellar systems within a human lifetime. Recent advancements in photonics and metamaterials have created novel solutions to challenges in propulsion and stability facing lightsail missions. This review introduces the physical principles underpinning  lightsail spacecrafts and discusses how photonics coupled with inverse design substantially enhance  lightsail performance compared to plain reflectors. These developments pave the way through a previously inaccessible frontier of space exploration.
\end{abstract}

\section{\label{sec:intro}Introduction\protect}
Interstellar travel, a concept that has been fantasized in science fiction for decades, is entirely impractical with conventional, chemically propelled spacecraft. 
To travel even a modest interstellar distance within a human lifetime, such as the distance to our nearest-neighbor star Proxima Centauri at 4.2 light years, requires a spacecraft moving at more than 10\% of lightspeed ($0.1c$). According to the Tsiolkovsky rocket equation~\cite{tsiolkovsky1903rocket}, accelerating even a single proton to those speeds with chemical propulsion would require a propellant mass that is larger than the mass of the observable Universe. Therefore, to reach relativistic velocities, the source of spacecraft momentum (and energy) must be external to the craft. Recent work has shown that the use of radiation pressure and gigantic but realistic lasers could allow a reflective lightsail spacecraft to be accelerated to $0.2c$, thus reaching Proxima Centauri within 21 years~\cite{Lubin:2016aa,Lubin:2024aa}.

Maxwell himself predicted that light carries momentum~\cite{Maxwell:1865_dynamical}, and thus that a force should be exerted on matter upon absorption or reflection of light. This radiation pressure, however, is of such a small magnitude that it took over 30 years to be experimentally verified~\cite{Lebedew:1901,Nichols:1903_2}. It is only with the advent of lasers, which have high intensities, that radiation pressure became useful: Ashkin~\cite{ashkin:1970} demonstrated that micron-sized particles could be accelerated and trapped using radiation pressure of tightly focused laser beams. Such optical tweezers have since found wide applications across scientific disciplines. The trapping technique was extended to trap and cool atoms~\cite{hansch:1975,chu:1986}. 
More recently, radiation pressure has been harnessed in cavity optomechanics to measure minute mirror displacements~\cite{Aspelmeyer:2014aa,Kippenberg:2008aa}. 

However, radiation pressure is not limited to small length and mass scales. In principle, if light can exert observable forces on microscopic particles, then sufficiently high intensities of light can produce forces to levitate, rotate or propel meter-scale objects. For instance, a spacecraft composed of a highly reflective thin film orbiting the Sun can be rotated in select ways by using radiation pressure in order to change the radius of orbit. Such spacecraft, ``solar sails'', are precursors to the lightsail and have had numerous successful experimental tests, the most prominent of which being the 2010 IKAROS launch~\cite{Mori:2010aa}.  

The success of solar sails demonstrates the exciting prospect of radiation-pressure-based spacecraft. However, solar sails are limited mostly to orbital transfers  within the solar system, or exploration of the near interstellar medium. Indeed, the limited power density of solar radiation, even upon close approach to the sun, limits solar sails to velocities at most of order a fraction of a percent of the speed of light~\cite{Vulpetti:1996,bailer-jones:2021,Karlapp:2024}. To reach higher velocities as required for interstellar travel, dramatically larger radiation pressure is needed, which can be achieved using a high-power laser on Earth (or in Earth's orbit)~\cite{Marx:1966aa,Forward:1984aa,Redding:1967aa}. Combined with ultra-thin, low mass sails, a high-power laser could enable substantial spacecraft acceleration, overcoming the fundamentally weak nature of radiation-pressure forces. However, such an endeavor faces numerous other design and system challenges, which were recently formalized in the Breakthrough Foundation's Starshot Initiative~\cite{Starshot:2024}. The initiative aimed to send a lightsail (often shortened to \textit{sail}) to our nearest-stellar neighbor, Proxima Centauri, at speed $0.2c$. The spacecraft contains a small payload for imaging the exoplanets surrounding Proxima Centauri, determining their habitability by means that are impossible when observing from Earth. Some of the most critical challenges to the mission success are associated with the sail-membrane, which must be: lightweight and highly reflective to reach high speeds; radiatively cooled to survive the laser heating and; dynamically and structurally stable within the propelling laser beam to assure its trajectory towards Proxima Centauri. Satisfying these objectives simultaneously is only possible with innovative and optimized photonic designs.

In this review, we explore the novel photonic designs and leading analysis techniques that address the lightsail propulsion and stability challenges. We begin in Section~\ref{sec:mission} by discussing the requirements for a successful lightsail mission, including topics on the geometry of the sail, launch phase, order of magnitude of mission parameters and key design challenges. In Section~\ref{sec:radiation_pressure}, we explain the physics of radiation pressure in the canonical models of light, before introducing effects from special relativity that influence the radiation pressure. With this understanding, we discuss in Section~\ref{sec:Materials} the material requirements for the lightsail membrane and outline some promising candidates and their optical, thermal and mechanical properties. The discussed materials feature heavily in Section~\ref{sec:propulsion_designs}, where we explain the need for potential sails to have efficient propulsion, highlighting methods and examples from the literature, and in particular the transition towards inverse photonic design using nanostructuring. Section~\ref{sec:thermal} follows with a literature review of lightsail thermal modeling and techniques using photonics that enhance passive sail cooling. Following in Section~\ref{sec:stability} is a deep investigation into the lightsail stability problem, containing the fundamental physics, key analysis methods and clever photonic designs. We round out the lightsail topic by exploring crucial experimental work in Section~\ref{sec:experiments}, before concluding the review in Section~\ref{sec:conclusion}.

\section{Mission parameters\label{sec:mission}}

\subsection{Geometry}
\begin{figure*}[htb]
    \centering
    \includegraphics[width=0.95\textwidth]{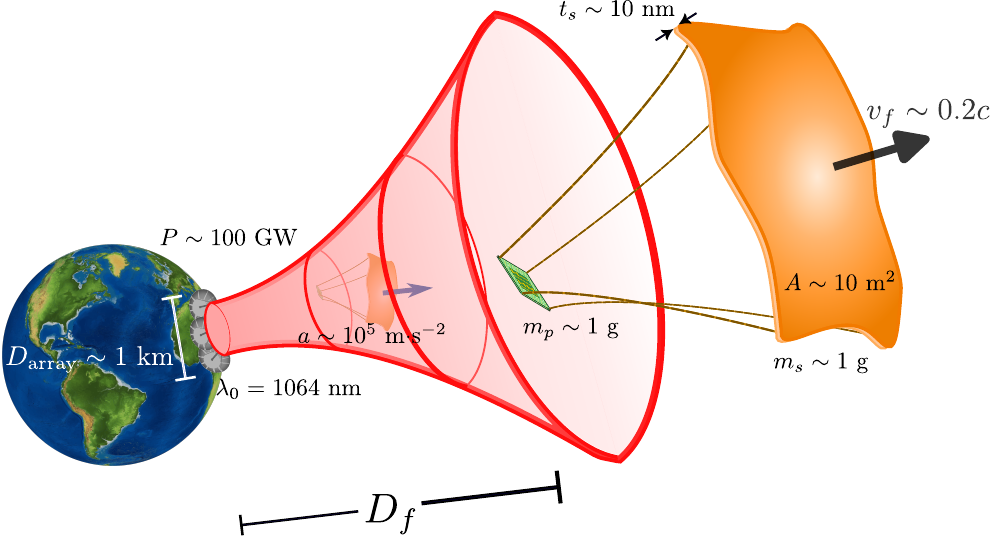}
    \caption{Lightsail concept art and order of magnitude of parameters.}
    \label{fig:parameters}
\end{figure*}

Figure~\ref{fig:parameters} shows concept art of a lightsail in the accelerating laser beam. The craft consists of two main elements: the payload (mass $m_p$), containing instruments for data collection and communication to Earth, and the sail membrane (mass $m_s$, area $A$) that is responsible for propulsion, stabilization and heat dissipation. The payload could be positioned at the center of the membrane, distributed across the membrane or even dragged behind the sail through nano-filament tethers akin to a parachute -- the latter being difficult given the stresses involved, but bringing advantages in terms of mechanical stability (discussed in Section~\ref{sec:stability}). The sail membrane can have curvature by design or because of the balances of forces, although it is often approximated as a flat membrane for first estimates. Detailed examples of geometry will be discussed in Sections~\ref{sec:propulsion_designs} and~\ref{sec:stability}.

This review focuses almost entirely on the photonic design of the sail membrane, however, the payload will factor into the analysis of both the propulsion, via the payload mass, and the stability, by providing a means of offsetting the spacecraft center-of-mass.  

\subsection{Launch}
Here, we consider an Earth-based laser array with a lightsail in orbit above the laser source. The laser array is aligned towards a chosen destination ({\em e.g.} Proxima Centauri) and switched on. The acceleration phase of the mission is the period during which the laser remains activated, and ends when the sail reaches its target velocity. 
The laser is designed such that the beam can be focused on the sail up to the distance required for acceleration to the target velocity. Beyond that distance, the laser diffraction rapidly reduces the fraction of power hitting the sail. Therefore, once the target velocity is reached, the laser is turned off to mitigate further operational cost, and the sail cruises on its own momentum towards its destination. 

\subsection{Orders of magnitude}
The total force on the sail is proportional to the sail-intercepted laser power $P_\text{intercept}=IA$, with $I$ the laser intensity at the sail (see Figure~\ref{fig:parameters}). The sail acceleration is then the radiation pressure force divided by the craft's mass $m=m_s+m_p$, where $m_s$ is the sail mass and $m_p$ is the payload mass. The minimization of the payload mass is restricted by the ability to miniaturize the on-board imaging system, power source and communication laser, with common estimates of order $m_p\sim 1\,$g.

The propelling laser is predicted to dominate the project cost~\cite{Parkin:2018aa}. To maximize the radiation pressure force, as much of the laser beam must be intercepted as possible. The laser must thus be focused with a beam waist of similar size as the sail diameter. For a fixed laser diameter on Earth, diffraction limits the distance over which such focusing can be effective, determining the  \textit{acceleration distance} $D_f$ over which the lightsail travels to reach its final speed.
Minimizing the array cost is a matter of minimizing its size~\cite{Parkin:2024aa}, which is achieved by minimizing the acceleration distance (maximizing the acceleration), ultimately limited by the g-forces the craft can endure and the total power that the laser can deliver. Therefore, there is a compromise between the laser-array size and the lightsail size: the smaller the lightsail, the bigger (and more expensive) the laser, but larger sail sizes add mass, reducing acceleration and increasing the required acceleration distance. 

The membrane must be as lightweight and reflective as possible to achieve optimal thrust. Simultaneously, it must also be minimally absorptive at the laser wavelength to avoid overheating under the extreme intensity of the laser. 
Keeping the mass low requires decreasing the sail thickness as much as possible, which is limited by material strength and fabrication considerations. The sail material must thus be chosen wisely, and needs to posses a combination of low mass density, high tensile strength, low absorption at the laser wavelength, and ideally large permittivity to enable strong reflection (see Section~\ref{sec:Materials}).

Given the interplay of sail and laser design, the system must be optimized as a whole, and detailed parameters will depend on future progress in laser and material sciences. Based on the cost of current or foreseeable technology, 
a reasonable set of order-of-magnitude parameter estimates~\cite{Lubin:2016aa, Parkin:2018aa} are shown in Figure~\ref{fig:parameters}. 
In short, the target velocity $0.2c$ and maximum g-forces (order $10^4$ g's) determine acceleration distance. The spacecraft mass (gram-scale) and minimum thickness (tens of nanometers) determine the membrane surface area. Sail mass and acceleration determine the required laser power, and acceleration and sail area determine the laser-array diameter.

For the laser, near-infrared (NIR) wavelengths are favorable due to mature high-power laser technology and low atmospheric absorption, a key consideration for ground-based laser arrays. However, other schemes such as an orbiting or Moon-based laser (and lightsail) are possible alternatives that do not suffer from atmospheric absorption, albeit at a tremendously higher cost~\cite{Lubin:2024aa}. For the laser-array size, if we assume the laser is a diffraction-limited Gaussian beam, then the array diameter is governed by 
\begin{equation} \label{eq:w_array}
    D_\text{array} = \frac{\lambda_0}{\pi \vartheta} \,, 
\end{equation}
where $\lambda_0$ is the laser-operating wavelength and $\vartheta$ is the beam's angular divergence. Assuming a laser with NIR wavelength incident on a 10~m$^2$ sail over an estimated  acceleration distance $D_f=\SI{10}{\giga\meter}$ (about 7\% of an astronomical unit), the sail can only intercept the full beam power for the entire $\SI{10}{\giga\meter}$ of travel if the array diameter is of order 1~km. 

Although we highlight here the orders of magnitude for a light-weight interstellar probe mission, there are many other viable missions. Test flights for Solar System exploration with modest-mass probes and reasonable laser-array power~\cite{Tung:2022aa} would be the first steps to demonstrate the feasibility of interstellar missions. Once the laser is constructed (taking the majority of the flight-system budget), many sails can be deployed, potentially including human-carrying crafts for interplanetary travel~\cite{Lubin:2016aa,Lubin:2024aa}.

\subsection{Challenges} 
The strict constraints necessary for high-speed interstellar travel and sheer orders of magnitude involved create several hurdles for any lightsail mission. Here, we briefly introduce some of the most important challenges, before discussing in subsequent sections photonic solutions to several of those challenges specific to the sail membrane. 

\begin{figure}[htb]
    \centering
    \includegraphics[width=0.95\linewidth]{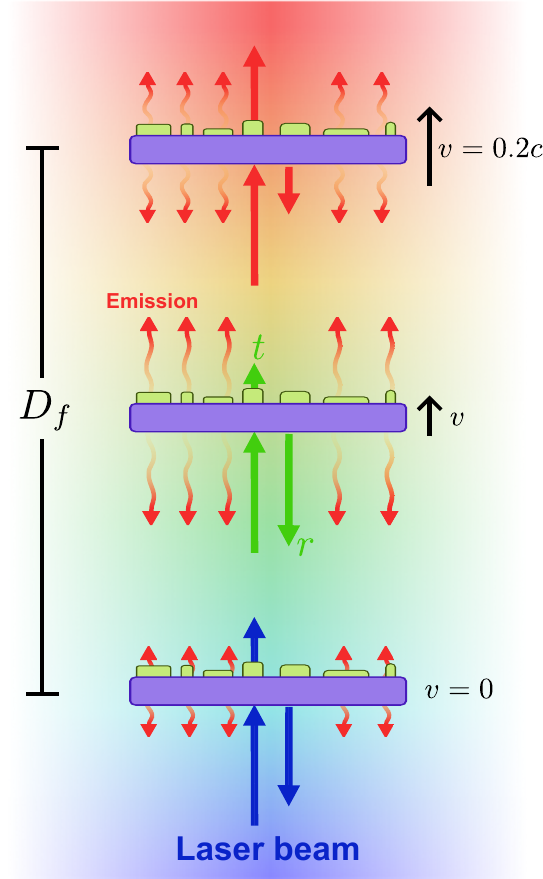}
    \caption{Lightsail reflection, transmission and absorption depend on the wavelength of a continually Doppler-shifting laser beam.}
    \label{fig:propulsion}
\end{figure}

One aspect that must be accounted for in all the lightsail mission challenges is relativity. In particular, due to the Doppler effect, the laser wavelength, when measured in a reference frame attached to the lightsail, increases with the sail's velocity relative to the laser source (depicted in Figure~\ref{fig:propulsion}). Hence, all sail candidates must be designed to address the mission challenges over the entire Doppler-broadened spectral band starting at the laser wavelength $\lambda_0$ and ending at~\cite{Kulkarni:2018aa} $\lambda_f = \lambda_0 \sqrt{(1+\beta_f)/(1-\beta_f)}$ for $\beta_f$ the normalized final sail velocity ($\beta_f=v_f/c$). A significant consequence of the finite laser bandwidth is that the solutions to the lightsail challenges cannot rely on narrow-band, resonant effects. 

\textbf{Laser array.} The laser is anticipated to be the most expensive and difficult facet of the mission~\cite{Parkin:2018aa, Worden:2021aa, Worden:2024aa, Parkin:2024aa}. Given the size of the beam required to concentrate the full power onto the relatively small sail membrane for a prolonged distance, the laser will need to be an aperture-filling, extremely large array of coherent lasers. A potential configuration using today's technology~\cite{Bandutunga:2021aa} involves a hierarchical array of $10^8$ lasers that are kept in phase via digitally enhanced heterodyne interferometry~\cite{Sibley:2020aa} from an orbiting-beacon laser~\cite{Peretz:2022aa}, making the system robust against atmospheric and mechanical destabilization. The method is derived from techniques involving guide stars used for adaptive optics in astronomy. A further requirement for many potential missions is that the laser beam be continually rotated during the sail flight to compensate for the Earth's rotation and keep the beam aligned with the mission target. Therefore, the array must have adjustable phase and directionality that accounts for the path difference between lasers as they rotate. Finally, the laser wavelength should be set within a range corresponding to high atmospheric transparency, for instance, somewhere within 1064~nm--1550~nm. Some lightsail features, such as stabilization discussed in Section~\ref{sec:stability}, can be substantially enhanced by resonant narrow-band effects: it would thus be ideal if the laser wavelength could be tuned down during acceleration to compensate for the Doppler shift, therefore allowing the sail to operate at fixed wavelength. However, this may be unrealistic with current and emerging laser technology. As a result, the sail must carry the burden of operating in the full Doppler band to address the mission challenges. 

With current technology, the cost of the propelling laser would be close to one trillion US dollars~\cite{Lubin:2024aa}, with a substantial fraction of the cost attributed to local-energy-storage requirements and precision optics. The cost becomes more palatable taking into account the cost scaling of fiber-coupled diode lasers, whose price per coherent watt has halved every four years based on observed trends~\cite{Worden:2021aa}. At this rate, the cost would be in the billion-dollar range within 30 years, comparable to existing large-scale scientific experiments or space missions. 
Another promising avenue are the advances in large, coherent, surface-emitting semiconductor lasers~\cite{yoshida:PCSEL2023}: if semiconductor lasers can be made coherent at a wafer scale, deploying kilometer-scale lasers becomes as realistic as today's large-scale photovoltaic power plants. Directionality and phase control could potentially be achieved on the same planar semiconductor platform using tunable metamaterials.

\textbf{Propulsion.} Propulsion performance is characterized by the ability for the sail to reach its final speed in the minimum possible acceleration distance. Maximal propulsion is achieved with a high-power laser, maximally reflective sail material and minimal spacecraft mass. Nanostructured photonic membranes are strong candidates for sails which simultaneously have small mass (compared to uniform membranes) and broadband, near-unity reflectance, as discussed in Section~\ref{sec:propulsion_designs}.

\textbf{Sail temperature.} A large laser intensity, while beneficial for spacecraft acceleration, results in the sail absorbing large laser power. Therefore, the lightsail must have sufficiently small absorption in the entire Doppler-shifted-laser-wavelength band (residing within $\SI{1}{\micro\meter}$--$\SI{2}{\micro\meter}$) to survive the laser irradiation. Furthermore, the sail must be cooled to within material-limited temperatures (typically around $\SI{1000}{\kelvin}$--$\SI{1500}{\kelvin}$) via passive radiative processes. As a consequence of Wien's displacement law, sails operating from $\SI{50}{\kelvin}$ to these upper temperature limits require a large emissivity (hence absorption) in the mid-infrared (MIR) wavelength region, roughly corresponding to the range $\SI{2}{\micro\meter}$--$\SI{10}{\micro\meter}$. Unlike the laser absorption, the emission in the MIR wavelengths is not influenced by the Doppler shift. Balancing these thermal considerations against the need for maximized propulsion can be achieved by photonic sail designs with simultaneously optimized reflectivity and wide-spectrum absorption/emissivity, or even by incorporating active photonic cooling (see Section~\ref{sec:thermal}).

\textbf{Structural integrity.} A spacecraft mass of order $\SI{1}{\gram}$ and a sail membrane with area around 10~m$^2$ constrains the membrane thickness to the order of 10s of nm for typical dielectrics. Such a thin sail will be flexible, and must survive the intense stresses and strains induced by the laser irradiation over several minutes. Additionally, the sail must retain its intended shape; perturbations in the sail membrane could be disastrous, especially for photonic sails with tailored reflections that are disrupted by shape deformations. Addressing this requires a combination of strong materials, judicious design of radiative-pressure distributions, and exploring new regimes of optomechanical coupling, all discussed in Section~\ref{sec:flexible_sails}.

\textbf{Stability.} One of the most crucial challenges for the lightsail mission is keeping the sail within the laser beam and maximizing its optical cross section. The sail must stay centered in the laser beam for maximum propulsion, but also to stay on target. This is not trivial because the laser beam will be continually rotated during the sail flight to compensate for the Earth's rotation. For a sail constructed as a planar mirror (Figure~\ref{fig:rp_force}a), any misalignment between the beam center and the sail would lead to forces and torques ejecting the sail from the beam that must be countered.
Since there is no mass budget for any on-board, active stabilizer ({\em e.g.} thrusters), the sails should be stabilized passively, which is most efficiently achieved through clever light-scattering design, generating photonic restoring forces and torques. This can be accomplished using specific sail shapes, or with metasurfaces and gratings, discussed in Section~\ref{sec:stability}. 

In order to overcome the problem of sail stability, the sources of destabilizing perturbations should be considered. We anticipate that laser-beam intensity fluctuations, imperfect laser-beam tracking with the Earth's rotation, atmospheric beam distortions (for Earth-based lasers), fabrication imperfections (causing diffuse light scattering) and optomechanic coupling in flexible sails to be some sources of instability. However, the exact nature of perturbations and their relative influence on the sail motion has not yet been investigated in detail due to the many variable mission parameters (such as the precise laser array configuration, beam shape, sail material, etc.).

\textbf{Communication.} Even after a successful launch, the spacecraft must be capable of transferring data in the vicinity of Proxima Centauri, over vast distances, and in a narrow beam to reach detectors on Earth. For transmission from the spacecraft itself, the 10~m$^2$ sail, in addition to providing propulsion and stability, can form the aperture for downlink communication to Earth. This is especially true of metasurface sails, which can be optimized to satisfy multiple objectives for propulsion, stability and communication simultaneously~\cite{Taghavi:2022ab,Taghavi:2024aa}. Substantial work must also be carried out to determine how best to receive the data on Earth, for example through an array of ground receivers or a low-mass-nanophotonic collecting aperture in space~\cite{Mauskopf:2023aa}. 

Of the challenges just outlined, all but the laser design and communication will be discussed in this review. For the remaining considerations, namely propulsion, thermal, structural and stability, the core underpinning concept is radiation pressure, which we explain in the following section.

\section{Radiation pressure physics\label{sec:radiation_pressure}}

\subsection{Calculation}
\begin{figure}[htb]
    \centering
    \includegraphics[width=0.85\linewidth]{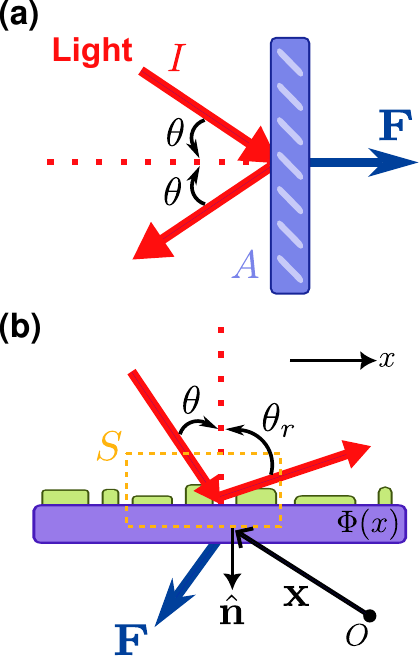}
    \caption{Radiation pressure forces. (a) Light rays/photons incident on a mirror are specularly reflected, and impart a radiation-pressure force normal to the mirror. (b) The radiation-pressure forces and torques on metasurfaces can be calculated by integrating the Maxwell Stress Tensor over the surface. Metasurfaces can be constructed with phase gradient $d\Phi/dx$ such that incident light is reflected at an angle that differs from the incident angle.}
    \label{fig:rp_force}
\end{figure}

Light incident on matter exerts a force called radiation pressure, a phenomenon known since the early days of electromagnetic theory~\cite{Maxwell:1865_dynamical,Maxwell:1873aa}. 
Radiation pressure can be explained and characterized in the three canonical pictures of light: ray-optics, wave-optics and photon-optics. In all pictures, Einstein's special relativity~\cite{Einstein:1905aa} predicts that light carries energy $E$ and thus momentum $p$ through the energy-momentum relation $E=pc$. In the quantum picture, each photon carries momentum $\hbar \mathbf{\omega}/c$. In the ray picture each ray carries a momentum flux density $I/c$ with $I$ the intensity of the ray.
Upon reflection or absorption, light obeys the same momentum-conservation law as matter; a light ray or stream of photons incident on a perfectly reflecting mirror at angle $\theta$ reflects at the same angle (Figure~\ref{fig:rp_force}a). Light transmission through matter without changing angle does not influence the radiation-pressure force and is thus not considered. For a perfectly reflecting mirror, the radiation-pressure force is proportional to the change in momentum of the photons.
The force due to light incident at angle $\theta$ on a mirror of area $A$ is then
\begin{equation} \label{eq:rp_flat_reflection}
    \mathbf{F}  
    = \frac{2IA}{c}\cos\theta \,\hat{\mathbf{n}} \,,
\end{equation}
which is normal to the mirror surface.
The factor 2 appears because the photons' momentum is reversed upon reflection, so the mirror takes away an equal and opposite momentum component normal to its surface.

The ray picture is adequate when light can locally be modeled as a plane wave with a single wavevector, and furthermore when reflected, transmitted and scattered waves can be decomposed as a discrete number of plane waves with well-defined momenta. This is the case for structures that vary slowly in space relative to the light wavelength and relative to the spatial structure of the light source, such that optical properties can be treated as locally invariant (in translation) or periodic. For structures of size comparable to the wavelength, structures containing aperiodic features, or for finite light beams that contain a continuum of wavevector directions, the ray picture is inadequate. Instead, the more complete wave-optics picture must be used, involving integrating the momentum transfer due to the electromagnetic fields.
%
The forces on an object due to the electromagnetic energy density can be quantified through the Maxwell stress tensor (MST)~\cite{Jackson:1962aa}
\begin{equation} \label{eq:MST}
    \overleftrightarrow{\mathbf{T}} = \epsilon_0 \mathbf{E} \otimes \mathbf{E} + \mu_0 \mathbf{H} \otimes \mathbf{H} - \frac{1}{2}(\epsilon_0 \mathbf{E}^2 + \mu_0 \mathbf{H}^2) \mathbf{I}_3 \,,
\end{equation}
where $\mathbf{E}$ and $\mathbf{H}$ are the electric and magnetic fields, respectively. The factor $\epsilon_0$ is the vacuum electric permittivity, $\mu_0$ the free-space magnetic permeability and $\mathbf{I}_3$ the $3\times3$ identity tensor. The symbol $\otimes$ denotes the dyadic product. 
The radiation-pressure forces and torques induced on a surface $S$ (Figure~\ref{fig:rp_force}b) are derived by projecting the stress tensor onto the surface normal $\hat{\mathbf{n}}$ and integrating over the surface positions $\mathbf{x}$ as follows~\cite{Jackson:1962aa}:
\begin{align}
    \mathbf{F} &= \int_S \overleftrightarrow{\mathbf{T}} \cdot \hat{\mathbf{n}} \,dA \label{eq:MST_force} \,, \\
    \bm{\tau} &= \int_S \mathbf{x} \times (\overleftrightarrow{\mathbf{T}} \cdot \hat{\mathbf{n}}) \,dA \label{eq:MST_torque} \,.
\end{align}
The integration surface is chosen to encompass the part of the structure where forces and torques are required, for example, over a few building blocks in a metasurface depicted in Figure~\ref{fig:rp_force}b. For extended bodies that are much larger than the wavelength of the incident plane wave, all three pictures predict equivalent forces and torques, with the ray and photon models being easiest to calculate. However, integrating the MST is necessary for calculating radiation pressure when the electromagnetic fields are more complicated ({\em e.g.} non-plane-wave light sources) or when the structure is designed with non-periodic-wavelength-scale elements ({\em e.g.} metasurfaces or structures with gradients in their periodicity or effective properties).

For meter-scale lightsails with aperiodic nanostructures, calculating the local fields everywhere to enable integration in Eqs~\eqref{eq:MST_force} and~\eqref{eq:MST_torque} can be computationally prohibitive. Instead, a common approach is to decompose the sail into a mosaic of locally periodic structures, where local forces and torques are computed (using ray optics or MST) assuming a single incident plane wave and fully periodic structure. The local forces and torques are then integrated across the mosaic to obtain the total force and torque on the lightsail. This approach ignores scattering effects at the junctions between mosaic elements and also ignores diffraction effects at the edge of the sails, which is reasonable when variations of effective properties are slow compared to the wavelength. In conjunction with these assumptions, the light beam is often approximated as having a single wavevector (all photons with momentum along the laser beam), but with a finite spatial intensity distribution ({\em e.g.} Gaussian). In most cases, this is a very good approximation for lightsails because the beam waist and sail are very large relative to the wavelength. However, for the study of more subtle relativistic effects accumulating over the entire acceleration phase, such an approximation might not be sufficient.

\subsection{Special relativity}
The relativistic nature of the lightsail mission further complicates radiation-pressure modeling. Due to relativity, the energy/momentum of observed photons depends on the reference frame of the observer; the laser source and lightsail must be treated as separate reference frames to correctly calculate the radiation-pressure forces. The change in photon energy between reference frames is the well-known relativistic Doppler shift, but is just one of two phenomena arising from the Lorentz transformation between two reference frames. The second phenomenon is the lesser-known relativistic aberration~\cite{Einstein:1905aa}, whereby photon propagation directions change with an object's motion relative to a light source. Altogether, the Lorentz transformation between laser source and sail reference frames transforms both the temporal (Doppler effect) and spatial (relativistic aberration) components of photon four-momenta, and both should be included in lightsail modeling. These relativistic effects apply to all light sources, requiring complicated beam transformations~\cite{Yessenov:2023aa} between reference frames for all sources except plane waves. 

Of these relativistic effects, the Doppler shift is the most prominent in the literature, being incorporated as a global wavelength shift that varies with the sail center-of-mass velocity. So far, other effects have rarely been considered since they are relatively small: relativistic aberration only emerges when accounting for sail-velocity components transverse to the laser-beam direction, leading to sub-microradian incident-angle shifts even for transverse speeds as large as $\SI{e3}{\meter\per\second}$. However, such transverse velocities are expected to be common in the mission, as they are easily caused by perturbations such as laser misalignment or as a natural byproduct of sail-restoring forces (see Section~\ref{sec:stability}). 
Therefore, although such angles may appear negligibly small, their effect can accumulate over the acceleration phase, which can be harnessed to increase stability (see Section~\ref{sec:geometric-stabilization}). Similarly, as soon as the sail has rotational motion, the Doppler shift becomes non-uniform across the sail surface: portions of the sail rotating towards the laser receive a shorter wavelength than portions of the sail rotating away from the laser.

\subsection{Other propulsion methods}
Before returning to lightsails, it is worth mentioning that other forms of laser-based propulsion have been proposed that do not involve radiation directly imparting momentum onto matter. For instance, with laser ablation propulsion, the spacecraft is composed of a material that, when irradiated by an Earth-based laser, evaporates and expands at high pressure to generate thrust~\cite{Rezunkov:2021aa}. The reactive mass provides a much higher energy-to-momentum conversion than using purely the reflection of photons, and could in principle be used to lift heavier craft than lightsails discussed here (even through the atmosphere). However, the reactive mass must be carried with the craft, so the final velocity is still severely limited according to Tsiolkovsky's rocket equation~\cite{tsiolkovsky1903rocket}. For interstellar missions, laser ablation propulsion is thus only marginally better than conventional chemical rockets. In contrast, lightsails do not need to carry their source of momentum, allowing them to reach near-relativistic speeds.

With an understanding of the radiation pressure central to the lightsail mission, we begin the review with the first lightsail design consideration: materials.

\section{Materials\label{sec:Materials}}
The stringent requirements on material optical, thermal and mechanical properties presents an imposing challenge to finding appropriate sail materials. No unique set of materials has yet been identified as having the perfect combination, but studies have converged on several candidates~\cite{Atwater:2018aa}. Here we discuss the key required material properties and most promising candidates, presented in Table~\ref{tab:materials}. The columns of Table~\ref{tab:materials} show relevant properties of the materials, which were obtained from experimental measurements. However, note that some properties depend strongly on details of material preparation and fabrication processes ({\em e.g.} purity for absorption, defect-free large surfaces for tensile strength, etc.). In Table~\ref{tab:materials}, material properties are presented at $\SI{300}{\kelvin}$, however, because of inevitable sail heating during acceleration, optical and mechanical properties must be characterized at elevated temperatures to properly gauge the thermal influence on propulsion. A detailed study of the material's thermal limit (likely well below the melting point) in vacuum conditions is needed to ascertain if the materials can stay intact in the entire lightsail thermal operating regime. 

\begin{table*}[!ht] 
\caption{Properties of studied lightsail materials at $\SI{300}{\kelvin}$.}
\label{tab:materials}
\centering
\begin{threeparttable}
    \resizebox{\textwidth}{!}{%
    \begin{tabular}{C{0.12\linewidth} C{0.15\linewidth} C{0.2\linewidth} C{0.2\linewidth} C{0.2\linewidth} C{0.1\linewidth} C{0.2\linewidth} C{0.2\linewidth}}
        \toprule\midrule
        \T\B \textbf{Material}\tnote{*} & \textbf{Mass density}\tnote{\textdagger}, $\rho_V$ ($\SI{}{\gram\per\centi\meter^3}$) & \textbf{NIR refractive index}\tnote{\textdaggerdbl}, $\braket{n_r}$ & \textbf{NIR} $\braket{\alpha_\text{abs}}$\tnote{\textdaggerdbl} ($\SI{}{\per\centi\meter}$) & \textbf{MIR} $\braket{\alpha_\text{abs}}$\tnote{\S} ($\SI{}{\per\centi\meter}$) & \textbf{MP}\tnote{\textdagger,\P}\;\; (K) & \textbf{Young's modulus}\tnote{\#} (GPa) & \textbf{Tensile strength}\tnote{\#} (GPa) \\ \hline
        \T\B Si (crystal) (\cite{Atwater:2018aa,Ilic:2018aa,Jin:2020aa,Kudyshev:2022aa,Santi:2022aa,Chang:2024aa}) & 2.33 & 3.53 (\cite{Green:1995aa}) & 1.53 (\cite{Green:1995aa}) & $2.32\times10^{-1}$ (\cite{Palik:1998aa}) & 1687 & 169 & 2.1 \\ \hline
        \T\B \ch{SiO2} (glass) (\cite{Ilic:2018aa,Santi:2022aa}) & 2.65 & 1.55 (\cite{Kashan:2001aa})  & $4.93\times10^{-7}$ (\cite{Kashan:2001aa}) & $4.70\times10^{3}$ (\cite{Franta:2016aa,Polyanskiy:2024aa}) & 1986 & 77 & 
        Up to 1.0 \\ \hline
        \T\B \ch{Si3N4} (\cite{Jin:2020aa,Brewer:2022aa,Lien:2022aa,Chang:2024aa}) & 3.17 & 1.98 (\cite{Philipp:1973aa,Polyanskiy:2024aa}) & $1.3\times10^{-2}$ (\cite{Kumar:2023aa})\tnote{**} & $4.66\times10^{3}$ (~\cite{Luke:2015aa})\tnote{**} & 2173 & 270 & 6.4 \\ \hline
        \T\B \ch{MoS2} (bulk) (\cite{Atwater:2018aa,Brewer:2022aa}) & 5.06 & 4.05 (\cite{Song:2019aa,Polyanskiy:2024aa}) & $1.58\times10^4$ (\cite{Song:2019aa,Polyanskiy:2024aa}) & $2.85\times10^{2}$ (\cite{Ermolaev:2020aa})\tnote{**} & 2023 & 238 (\cite{Feldman:1976aa}) & 21\tnote{\textdagger\textdagger} \\ \hline
        \bottomrule
    \end{tabular}
    }
\begin{tablenotes}\footnotesize
\item[*] References indicate the publications studying the material for lightsail applications.
\item[\textdagger] All entries for this column obtained from Reference~\cite{Hammond:2016aa}.
\item[\textdaggerdbl] Averaged over 1064~nm--1303~nm.
\item[\S] Averaged over $\SI{2}{\micro\meter}$--$\SI{10}{\micro\meter}$.
\item[\P] Melting point.
\item[\#] All entries in this column are from data collated in Reference~\cite{Gao:2024aa}.
\item[**] Data could not be found over the entire wavelength range.
\item[\textdagger\textdagger] This value is for multilayer \ch{MoS2} tensile strength.
\end{tablenotes}
\end{threeparttable}
\end{table*}

The reflectivity of the sail is one of the most significant properties in the lightsail mission. The most immediate choice for highly reflective sail materials are metallic reflectors ({\em e.g.} aluminum or gold). Indeed, all solar sails launched so far have been made of aluminum-coated polymers that reflect more than 90\% of incident sunlight~\cite{Davoyan:2021aa}. However, metallic reflectors have non-negligible absorption at NIR wavelengths, which is of no concern when operating under sunlight, but would lead to almost instant vaporization when irradiated by a $\SI{100}{\giga\watt}$ propelling laser. Therefore, all parts of the sail exposed to the laser must have vanishingly small absorption over the NIR Doppler-shifted wavelength range. In addition, the sail must be effective at radiating away any residual absorbed heat, which requires high emissivity in the MIR (discussed in more detail in Section~\ref{sec:thermal}). Thus, combinations of dielectrics and semiconductors with gap energy well above the propelling laser's photon energy are the most promising contenders, having potential for small absorption and substantial reflectivity in the NIR Doppler band. Higher refractive-index dielectrics are desirable as they can have larger reflectivity and resonant reflectivity at thinner sail thicknesses (hence masses) than lower-index materials.

Another key consideration for lightsail candidates is the material's Young's modulus and tensile strength. Solid materials with a larger Young's modulus likely have larger bending stiffness, which helps the sail resist shape deformations~\cite{Savu:2022aa}. The material's tensile strength should be large to ensure forces transverse to the beam-propagation direction do not tear the sail apart. The transverse structural stability is especially important because centrifugal forces (from, for example, radiation pressure or sail spinning) are likely required to counteract the tendency for the sail to collapse under irradiation by the nonuniform beam. Sail spinning in particular may be a significant stabilizing mechanism as will be discussed in Section~\ref{sec:stability}. For a more detailed discussion of material structural properties, along with other materials not discussed here, see Reference~\cite{Gao:2024aa} and the associated Supplementary Information.

The final aspect we discuss here, and one likely to be decisive in the final lightsail design, is the ease of fabrication and patterning of the lightsail material. For all materials, the membrane must be sufficiently thin to constitute a low-mass spacecraft. Therefore, the chosen lightsail material must have an associated method of fabrication that enables nanometer-thin, large-area and minimal-defect patterning. Any defect is likely to decrease the sail's tensile strength, and could add unwanted perturbation in the radiation-pressure distribution during flight. Techniques such as nanoimprint lithography~\cite{Guo:2007aa} are, in principle, capable of patterning on the square-meter scale, but may be difficult to refine to the low level of defect density required. Nanolithography is capable of making near-defect-free diffractive optics at wafer-scale~\cite{Chang:2024aa}, but cannot currently be applied to meter-scale membranes. We discuss recent examples of fabricated lightsail structures in Section~\ref{sec:experiments}, but it is clear that substantial advances in high-resolution patterning over large, square-meter areas with a minimal number of defects are required for successful lightsail missions. 

\begin{figure*}[htb]
    \centering
    \includegraphics[width=0.95\textwidth]{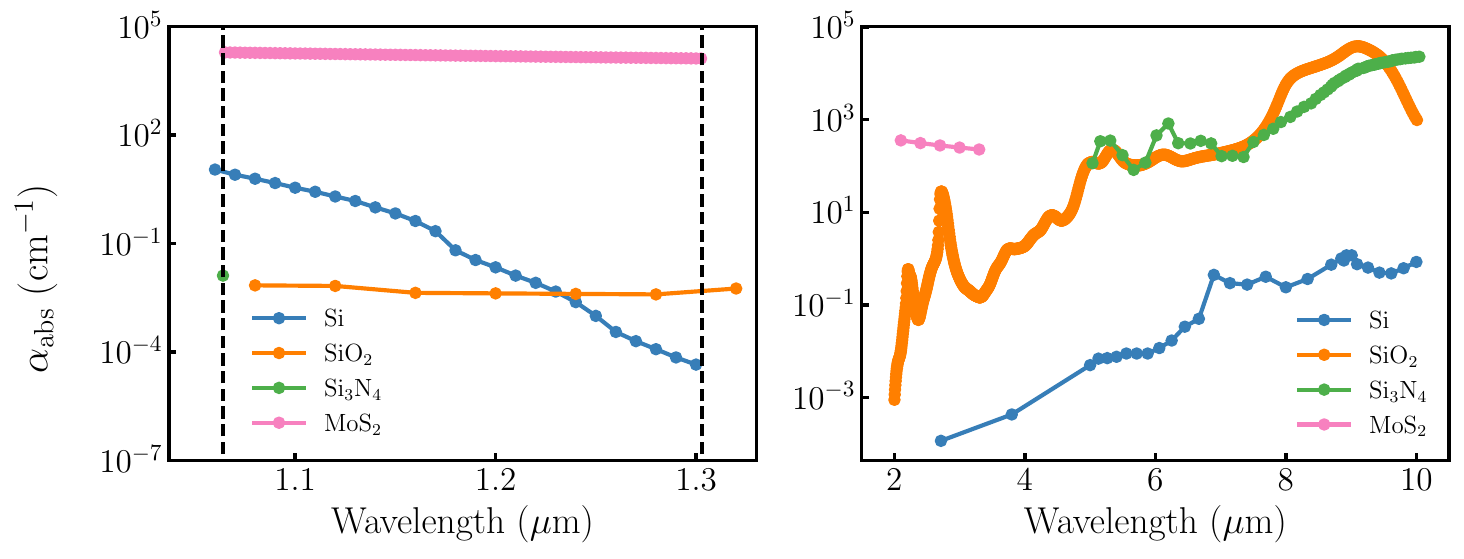}
    \caption{Absorption spectra at $\SI{300}{\kelvin}$ of materials in Table~\ref{tab:materials} in the (a) NIR~\cite{Green:1995aa,Kashan:2001aa,Kumar:2023aa,Song:2019aa,Polyanskiy:2024aa} and (b) MIR~\cite{Palik:1998aa,Franta:2016aa,Luke:2015aa,Ermolaev:2020aa,Polyanskiy:2024aa} spectral regions. In (a), the region between the dotted black lines is the Doppler-shifted wavelength band starting at 1064~nm. In (b), the \ch{Si3N4} data was not obtainable and hence extracted using an image data picker~\cite{PlotDigitizer} on Figure~1 of Reference~\cite{Luke:2015aa}.}
    \label{fig:absorption_spectra}
\end{figure*}

One of the most studied lightsail materials is silicon (Si), which has a relatively high refractive index and hence great potential for substantial reflectivity. Additionally, the refractive index of Si is reported to be relatively constant over large temperature changes, only increasing by approximately 0.08 at $\SI{300}{\kelvin}$ above room temperature~\cite{Li:1980aa}. The issue with Si is its thermal performance: Si has quite large average absorption in the NIR Doppler band as shown in Figure~\ref{fig:absorption_spectra}a, which is exacerbated at higher temperatures. Furthermore, Si has poor emissivity in the MIR spectral region (Figure~\ref{fig:absorption_spectra}b) compared to other lightsail candidate materials.

Silica (\ch{SiO2}) is another promising sail material, which has the opposite strengths and weaknesses of Si. Indeed, \ch{SiO2} has orders of magnitude smaller absorption in the NIR and large emissivity in the MIR, as shown in Figure~\ref{fig:absorption_spectra}. However, \ch{SiO2} also has a much lower refractive index than Si. 

Other promising material contenders include silicon nitride (\ch{Si3N4}) and molybdenum disulfide (\ch{MoS2}), which have similar characteristics to \ch{SiO2} and Si, respectively, with regards to NIR reflectivity and MIR emissivity. However, to gauge the thermal performance of \ch{Si3N4} and \ch{MoS2}, experimental data for absorption coefficients over a much broader wavelength range is required. In terms of the Young's modulus and tensile strength, \ch{MoS2} is a strong contender compared to the other materials listed in Table~\ref{tab:materials}. \ch{MoS2} is a transition metal dichalcogenide, which, along with materials such as graphene, have recently gained traction as two-dimensional (2D) materials~\cite{Li:2015aa}. \ch{MoS2} has a similar Young's modulus to graphene, but unlike graphene, \ch{MoS2} has a 1.8~eV bandgap and hence less absorption in the NIR spectral region. The \ch{MoS2} structure can exist as a single-atom layer or a stack of layers connected by van der Waals forces, where each layer is roughly 0.65~nm thick. 

With an understanding of the candidate materials, we turn to discussing the lightsail structures that enable the sail to be efficiently propelled. The designs to be discussed make use of the material properties, but more crucially, they enhance the desirable sail characteristics by harnessing nanostructuring through optimization.

\section{Propulsion designs\label{sec:propulsion_designs}}

\subsection{Acceleration distance} 
For a given laser, the acceleration distance to reach a target velocity depends predominantly on spacecraft mass and reflectivity over the Doppler-shifted laser-wavelength range. 
Minimizing the acceleration distance is critical because it determines the laser-aperture size and cost. The acceleration distance is thus a good figure of merit to quantify propulsion-focused sail designs. 

\begin{figure}[htb]
    \centering
    \includegraphics[width=0.99\linewidth]{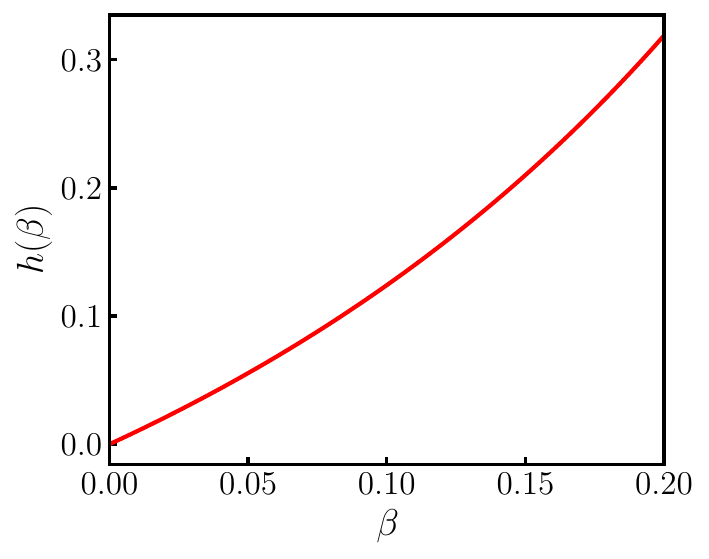}
    \caption{Functional form for $h(\beta)$ (eq~\eqref{eq:D_FOM}).}
    \label{fig:h_beta}
\end{figure}

The acceleration distance is derived using special relativity by summing the laser-photon momenta imparted on a lightsail, then integrating the resultant equation of motion with respect to velocity from the (normalized) sail velocity $\beta=v/c=0$  to the target velocity $\beta_f$~\cite{Kulkarni:2018aa}. In the derivation, it is assumed that the sail remains perfectly aligned at the center of the beam, and that the laser-spot size of the beam remains smaller than the sail size throughout the acceleration phase (so that all of the power is intercepted by the sail). The resultant acceleration distance figure of merit is~\cite{Atwater:2018aa}
\begin{align} 
\begin{split} \label{eq:D_FOM}
    D_f 
    &= \frac{mc^3}{P} \int_0^{\beta_f} \frac{\gamma\beta d\beta}{2r(\lambda')(1-\beta)^2} \\
    &= \frac{mc^3}{P} \int_0^{\beta_f} \frac{h(\beta) d\beta}{2r(\lambda')} \,,
\end{split}
\end{align}
where $m = m_p + m_s$ is the spacecraft mass, $P$ is the laser output power and $\gamma = 1/\sqrt{1-\beta^2}$ is the Lorentz factor. The reflectivity $r(\lambda'$) is taken at the Doppler-shifted wavelength, related to the laser wavelength ($\lambda_0$) in the laser frame by  $\lambda' = \lambda_0 \sqrt{(1+\beta)/(1-\beta)}$. In the second line of eq~\eqref{eq:D_FOM}, we define $h(\beta) = \gamma\beta/(1-\beta)^2$ for convenience, encapsulating all sail-independent velocity factors into a single term. The functional form for $h(\beta)$ is shown in Figure~\ref{fig:h_beta}. This function operates as a weight when integrating reflectivity, showing that reflectivity reductions at the tail end of the acceleration phase ({\em i.e} at longer Doppler-shifted wavelengths) harm the acceleration distance disproportionally.   

Photon momentum transfer due to absorption can be included by substituting  $2r(\lambda')$ with $2r(\lambda') + a(\lambda')$ in the denominator of eq~\eqref{eq:D_FOM}, where $a(\lambda')$ is the wavelength-dependent material absorption. However, for the lightsail mission, laser absorption should be minimized as much as possible, so the $a(\lambda')$ contribution to the propulsion is negligible. 

Equation~\eqref{eq:D_FOM} allows comparing acceleration distances for a given final velocity for different sail geometries and sail materials simply from their mass and reflectivity spectrum $r(\lambda')$. A first glance suggests that minimizing the sail mass $m_s$ to zero will minimize the acceleration distance. However, the sail must have finite thickness, so decreasing sail mass means decreasing sail area, which reduces the power the sail can intercept from the laser beam. By balancing an increasing sail area against increasing sail mass, it was shown that for a fixed sail thickness,  the minimum acceleration distance is achieved when  sail and payload mass are equal, {\em i.e.}  $m_s=m_p$~\cite{Lubin:2016aa,Kulkarni:2018aa}. Since the onboard imaging and communication components are currently unknown, the payload mass is usually set to a reasonable value of order $\SI{0.1}{\gram}$--$\SI{10}{\gram}$. 

For a target velocity $\beta_f=0.2$, the lower limit on $D_f$ is reached for a sail with unit reflection across the spectrum: in that case, assuming the smallest payload mass used in the literature~\cite{Jin:2020aa} of $m_p=\SI{0.1}{\gram}$ and the ideal-mass condition of $m_s=m_p$, eq~\eqref{eq:D_FOM} yields an acceleration distance $D_f = \SI{0.73}{\giga\meter}$. This value corresponds to just under $1/150$ of an astronomical unit, and can be considered as a reference for the minimum achievable value of $D_f$ (Table~\ref{tab:acceleration_distance}, first row).

Different designs~\cite{Atwater:2018aa,Ilic:2018aa,Jin:2020aa,Brewer:2022aa,Jin:2022aa,Kudyshev:2022aa,Lien:2022aa,Santi:2022aa} minimize the figure of merit by balancing a decreasing sail thickness and/or material filling fraction (hence mass) against an increasing sail reflectivity. In Table~\ref{tab:acceleration_distance}, we present the acceleration distance for standard laser parameters for a variety of sail structure/material combinations that have been proposed to address propulsion or thermal management. The designs range from simple monolayer slabs to judiciously crafted layers of photonic structures, which can resonantly enhance NIR reflectivity and MIR emissivity. Each design assumes either a fixed sail area (often 10~m$^2$) or a fixed sail mass ($m_s=\SI{1}{\gram}$) as a constraint for optimization. The payload mass is usually fixed to some constant, or fixed equal to the sail mass to satisfy the optimal $D_f$ condition. The key column in Table~\ref{tab:acceleration_distance} is the acceleration distance $D_f$ for a target velocity $\beta_f=0.2$, which has been either recorded from the literature~\cite{Atwater:2018aa,Jin:2020aa,Brewer:2022aa,Kudyshev:2022aa,Lien:2022aa} or calculated from alternative figures of merit~\cite{Ilic:2018aa,Santi:2022aa} that are related to eq~\eqref{eq:D_FOM} by  scaling factors such as mass and laser power. Due to the large variety of sail mass and sail area assumptions, we also show the laser-array diameter $D_\text{array}$ (eq~\eqref{eq:w_array}) that would be required to accelerate each design to their associated acceleration distance. Results range from plausible laser aperture sizes of 460~m to a staggering 175~km. The most promising results rely on disconnected dielectric structures suspended in vacuum (without substrates) or multilayers separated by vacuum, making them unrealistic in their current configuration~\cite{Ilic:2018aa,Jin:2020aa,Kudyshev:2022aa}, but possibly achievable with aerogel-type materials in the future. The dramatic penalty to the acceleration distance due to added mass from thicker layers is apparent in, for example, the last row~\cite{Lien:2022aa} of Table~\ref{tab:acceleration_distance} (although, it is one of the few lightsail structures thus far that has been characterized experimentally, see also Section~\ref{sec:experiments}).

In the following subsections, we highlight the key photonic designs in Table~\ref{tab:acceleration_distance} and techniques used to optimize sail membranes for effective propulsion, and discuss the physics behind them. 

\begin{table*}[!ht] 
\caption{Acceleration distances of several sail designs from the literature.}
\label{tab:acceleration_distance}
\centering
\begin{threeparttable}
    \resizebox{\textwidth}{!}{%
    \begin{tabular}{C{0.02\linewidth} C{0.2\linewidth} C{0.15\linewidth} C{0.04\linewidth} C{0.2\linewidth} C{0.06\linewidth} C{0.04\linewidth} C{0.05\linewidth} C{0.1\linewidth} C{0.08\linewidth} C{0.05\linewidth}}
        \toprule\midrule
        \T\B \textbf{Ref} & \textbf{Material} & \textbf{Pattern} & $A$ (m$^2$) & \textbf{Thickness} (nm) & $m_s$ (g) & $m_p$ (g) & $\lambda_0$ (nm) & $\braket{r}$\tnote{*} & $D_f$\tnote{\textdagger} (Gm) & $D_\text{array}$\tnote{\textdaggerdbl} (km) \\ \hline
        \T\B -- & Ideal reflector & Slab & 10 & -- & 0.1 & 0.1 & -- & 1.0 & 0.73 & 0.17 \\ \hline
        \T \cite{Ilic:2018aa} & \ch{SiO2} & Slab & 2.2 & 206 & 1.0 & 1.0 & 1200 & 0.12 & 60.6 & 31.15 \\ 
        ~ & $4\times$(\ch{SiO2}/Vac)/\ch{SiO2} & Slab multilayer & 0.7 & 130/484/140/473/143 /473/140/484/130 & 1.0 & 1.0 & 1200 & 0.76 & 9.4 & 8.87 \\ 
        ~ & Si/\ch{SiO2} & Slab multilayer & 6.5 & 61/5 & 1.0 & 1.0 & 1200 & 0.65 & 11.5 & 3.44 \\ 
        \B ~ & Si/Vac/Si/\ch{SiO2} & Slab multilayer & 6 & 34/506/33/5 & 1.0 & 1.0 & 1200 & 0.82 & 9.0 & 2.81 \\ \hline
        \T\B \cite{Santi:2022aa} & Si/\ch{SiO2}/Si & Slab multilayer & 1.5 & 88/96/89 & 1.0 & 1.0 & 1064 & 0.9 & 8.3 & 4.61 \\ \hline
        \T \cite{Jin:2020aa} & Si & 1D grating\tnote{\S} & 10 & 107 & 0.3 & 0.1 & 1200 & 0.75\tnote{\P} & 1.9 & 0.46 \\ 
        \B ~ & \ch{Si3N4} & 1D grating\tnote{\S} & 10 & 243 & 1.9 & 0.1 & 1200 & 0.57\tnote{\P} & 13.0 & 3.14 \\ \hline
        \T\B \cite{Kudyshev:2022aa} & Si & 1D grating\tnote{\S} & 10 & 110 & $\approx0.4$\tnote{\#} & 0.1 & 1200 & 0.81 & 1.9 & 0.46 \\ \hline
        \T\B \cite{Atwater:2018aa} & Si & PhC\tnote{**} circular holes & 10 & 59 & 0.8 & 0.1 & 1200 & 0.7 & 3.2 & 0.77 \\ \hline
        \T \cite{Brewer:2022aa} & \ch{Si3N4}/\ch{MoS2}/\ch{Si3N4} & PhC circular holes & 10 & 5/90/5 & 1.8 & 1.0 & 1200 & 0.65 & 10.6 & 2.56 \\ 
        \B ~ & \ch{Si3N4}/\ch{MoS2}/\ch{Si3N4} & PhC circular holes + Mie islands\tnote{\textdagger\textdagger} & 10 & 30/90/30 & $\approx2.3$\tnote{\textdaggerdbl\textdaggerdbl} & 1.0 & 1200 & NR\tnote{\S\S} & 16.7 & 4.03 \\ \hline
        \T \cite{Lien:2022aa,Lien:2023aa} & \ch{Si3N4} & PhC circular holes & 10 & 690 & 19.3 & 0.1 & 1064 & 0.35\tnote{\P\P} & 177.5 & 38.02 \\ 
        ~ & \ch{Si3N4}/\ch{TiO2} & PhC circular holes/coating & 10 & 690/10 & 20.3 & 0.1 & 1064 & 0.35\tnote{\P\P} & 180.6 & 38.68 \\ 
        ~ & \ch{Si3N4}/PDMS & PhC circular holes/coating & 10 & 690/8000 & 96.5 & 0.1 & 1064 & 0.35\tnote{\P\P} & 817.2 & 175.05 \\
        \bottomrule
    \end{tabular}
    }
\begin{tablenotes}\footnotesize
\item[*] Reflectivity averaged over the Doppler-broadened laser band.
\item[\textdagger] Calculated using eq~\eqref{eq:D_FOM} assuming a fixed $\SI{100}{\giga\watt}$ power intercepted by all sails.
\item[\textdaggerdbl] Calculated using eq~\eqref{eq:w_array}.
\item[\S] No supporting substrate attached.
\item[\P] $h(\beta)$-weighted reflectivity average (see eq~\eqref{eq:D_FOM} and Figure~\ref{fig:h_beta}).
\item[\#] Estimated value from optimized material distribution (Figure 3e bottom in Reference~\cite{Kudyshev:2022aa}).
\item[**] Photonic crystal.
\item[\textdagger\textdagger] See Figure~\ref{fig:propulsion_lit}c.
\item[\textdaggerdbl\textdaggerdbl] Estimated value ignoring the change in mass from the Mie island design.
\item[\S\S] Not recorded.
\item[\P\P] Estimated from measured transmission spectra (Figures~2, 3 and 5 in Reference~\cite{Lien:2022aa}).
\end{tablenotes}
\end{threeparttable}
\end{table*}

\subsection{Monolayers and multilayers}
The simplest candidate lightsail is the monolayer uniform slab. It is characterized by just two parameters: the slab material and thickness. Compared to more advanced photonic structures, the uniform monolayer should be easily understood physically, have an easier-to-explore parameter space, and be easier to fabricate. On the other hand, the restricted parameter space only allows for limited optimization. 
The slab material and thickness determine the sail's surface-mass density, which, using the optimal mass condition $m_s=m_p$, also determines the sail area for a given payload mass. 
The optical response of a monolayer slab is that of a Fabry-Perot (FP) etalon.
Generally, increasing the sail thickness narrows each FP resonance but increases the number of resonances. In the limit of small thickness, before the first FP resonance, the reflectivity increases rapidly with thickness. For large thicknesses, multiple resonances over the Doppler-broadened spectrum can lead to increased averaged reflectivity. In any case, increasing the sail thickness also increases the sail mass, so optimizing eq~\eqref{eq:D_FOM} is a battle between maximizing the averaged sail reflectivity and minimizing sail mass.

A promising method to achieving large reflection is to construct a lightsail with multiple thin layers at roughly quarter-wavelength scale. Such multilayers consist of stacks of monolayer slabs, which can benefit from resonant-Bragg effects to enhance reflection, potentially over a broad wavelength band. However, adding layers increases the mass of the sail, so candidate designs must still carefully balance the added reflectivity against increasing mass, through optimization of eq~\eqref{eq:D_FOM}. Furthermore, stacks with different materials and thicknesses can satisfy multiple requirements simultaneously, such as resonant-Bragg reflection over the Doppler-shifted-laser-wavelength range {\em and} enhanced emissivity in the MIR for radiative cooling. Finally, multilayers are simpler to fabricate than patterned photonic crystals, especially for very-thin films, however, photonic crystals offer substantial advantages as discussed in subsequent sections.

\begin{figure*}[htb]
    \centering
    \includegraphics[width=0.9\textwidth]{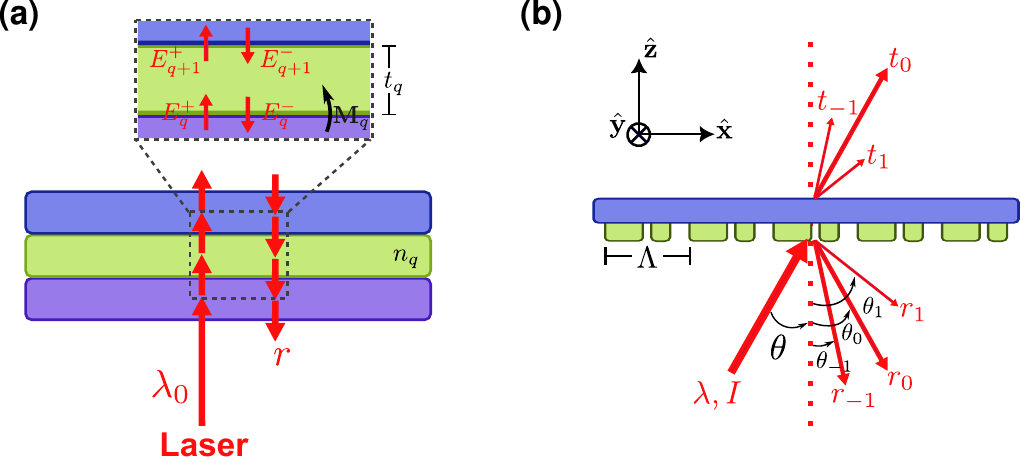} 
    \caption{Propulsion-focused lightsail design concepts. (a) Reflection from a monolayer or multilayer is determined using the transfer-matrix method, which, at each layer boundary $q$, superimposes all waves into a single forward- and reverse-traveling wave. (b) Diffraction grating, depicted with up to $\pm1$ orders.}
    \label{fig:propulsion_designs}
\end{figure*}

\begin{figure*}[htb]
    \centering
    \includegraphics[width=0.99\linewidth]{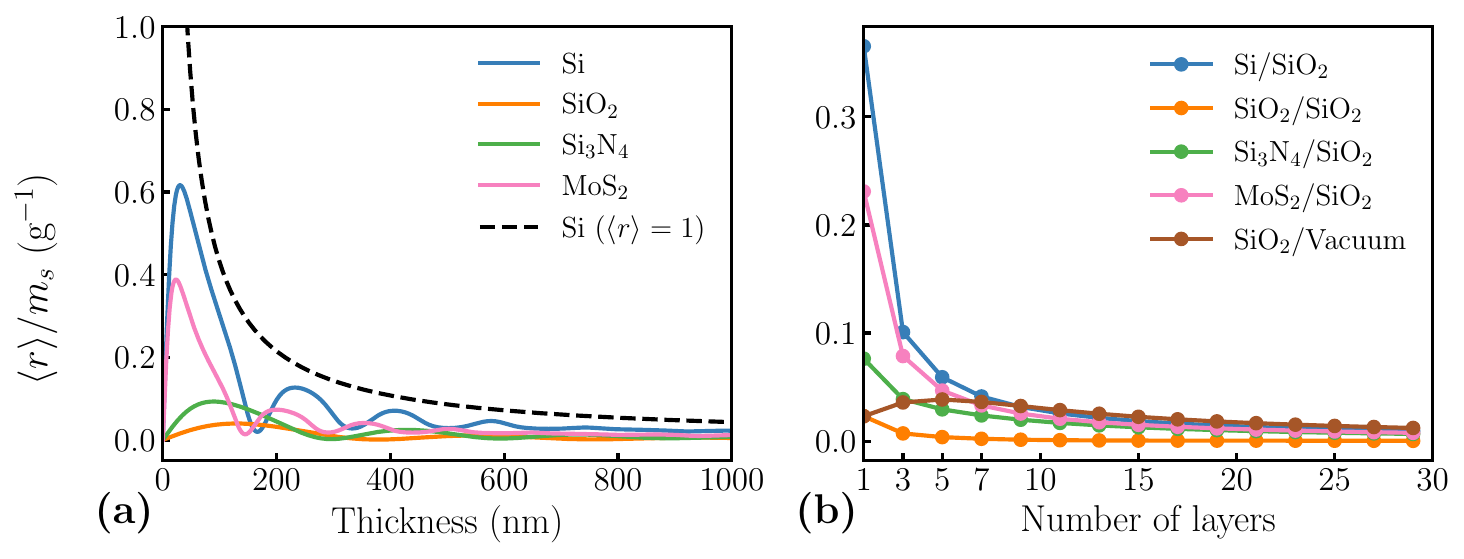}
    \caption{Ratio of reflectivity averaged over the Doppler spectrum (1064~nm--1303~nm) to sail mass for (a) monolayer and (b) multilayer structures composed of materials from Table~\ref{tab:materials}. The sail area is assumed to be 10~m$^2$. In (a), the dotted black line represents the reflectivity-to-mass ratio for an idealized reflector with the same mass density as Si.}
    \label{fig:multilayer_r_to_m_ratio}
\end{figure*}

\textbf{Mass-reflectivity trade-off.}
Evaluating the reflection of plane waves from a multilayer structure is easily done using the transfer-matrix method (TMM, depicted in Figure~\ref{fig:propulsion_designs}a)~\cite{Macleod:2018aa,yeh:optical_waves_in_layered_media}. Subsequently, the trade-off between reflectivity and mass for monolayers and multilayers can be evaluated.
For various monolayer slab materials, Figure~\ref{fig:multilayer_r_to_m_ratio}a shows the ratio of reflectivity to mass ($\braket{r}/m_s$) averaged over the Doppler-broadened wavelength range (1064~nm--1303~nm), and as a function of thickness. In such plots~\cite{Lubin:2016aa}, a larger reflectivity-to-mass ratio corresponds to a smaller acceleration distance, advantageous to the mission. For each material, we see several peaks corresponding to FP-enhanced reflections averaged over the Doppler band. The largest $\braket{r}/m_s$ peak comes from the smallest-thickness FP reflection resonance, sitting at thicknesses less than 200~nm for all materials shown. Here, a pure Si layer of thickness $t=30$~nm has the best performance. However, note that the peak is achieved for a thickness that is smaller than the thickness $t=\lambda/(4n)\approx 84$~nm associated with peak FP reflectivity, because the loss in reflectivity is compensated by mass reduction.

In the large-thickness limit, mass is added to the sail but not compensated by any increase in reflectivity because the reflectivity is bounded by 1. In the limit of zero thickness, the TMM predicts reflectivity vanishes faster than mass, as seen in Figure~\ref{fig:multilayer_r_to_m_ratio}a by the curves tending to zero. In reality, the smallest thickness corresponds to a discrete, one-atom-thick layer, which has non-zero reflection but the minimal-possible mass.

For a fixed layer thickness, relying on FP-resonant reflectivity is unwise because the effect is generally weak and narrow band compared to the breadth of the Doppler-broadened spectrum. 
Indeed, all candidate monolayers studied thus far~\cite{Ilic:2018aa, Lien:2022aa, Santi:2022aa},
have 10--100 times larger acceleration distance than an ideal reflector (Table~\ref{tab:acceleration_distance}). 

\begin{figure*}[htb]
    \centering
    \includegraphics[width=0.7\linewidth]{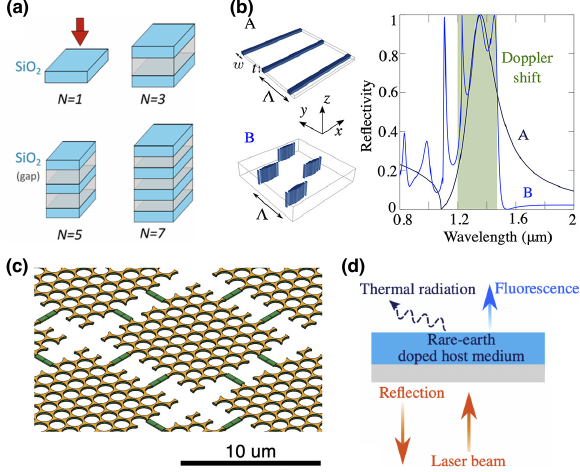}
    \caption{Candidate lightsail designs for propulsion and thermal management. (a) \ch{SiO2} monolayer or multilayer slabs. Reproduced from Reference~\cite{Ilic:2018aa}. Copyright 2018 American Chemical Society. (b) Left: design A represents the optimized 1D grating. Right: the optimized grating reflectivity spectrum. Reproduced from Reference~\cite{Jin:2020aa}. Copyright 2020 American Chemical Society. (c) Circular-hole photonic crystal with a MIR-scale-Mie-island design, depicted with green structural-support scaffolds connecting each island. Reproduced from Reference~\cite{Brewer:2022aa}. Copyright 2022 American Chemical Society. (d) Diagram showcasing laser cooling. Reproduced from Reference~\cite{Jin:2022aa}. Copyright 2022 American Chemical Society.}
    \label{fig:propulsion_lit}
\end{figure*}

Broader spectrum resonant reflectivity can, in principle, be achieved using coherent back scattering (Bragg reflection) with multilayer structures such as those shown in Figure~\ref{fig:propulsion_lit}a. Each additional layer adds mass, but the increased reflectivity can outweigh the mass penalty for a small number of thin layers. As an example, Figure~\ref{fig:multilayer_r_to_m_ratio}b shows the reflection-to-mass ratio as a function of layer number for stacks of different materials interspersed with \ch{SiO2}. The thickness of each layer is fixed equal to $\lambda/(4n)$, where $n$ is the refractive index of the layer and $\lambda$ is the wavelength midway through the Doppler shift associated with a final velocity $v_f=0.2c$ (1184~nm). In this plot, a single layer corresponds to a single slab composed purely of the first material in the pair named in the legend of Figure~\ref{fig:multilayer_r_to_m_ratio}b. Subsequent odd-numbered layers add two layers with the materials shown in the named pairs. Here, the benefit of Bragg reflection is only evident for \ch{SiO2} spaced by vacuum layers, with optimum reflection-to-mass ratio at 5 layers. With increasing layer number, all other material combinations are penalized more by the increasing mass than they gain from coherent back scattering. 

Authors of Reference~\cite{Ilic:2018aa} showed that \ch{SiO2} layers interspersed with vacuum layers (Figure~\ref{fig:propulsion_lit}a) can indeed achieve smaller acceleration distances than monolayer \ch{SiO2} due to resonantly enhanced reflections over the Doppler band. This is consistent with Figure~\ref{fig:multilayer_r_to_m_ratio}b up to 5 layers, but note that the work in Reference~\cite{Ilic:2018aa} numerically optimized the layer thicknesses for broadband reflection, resulting in layer thicknesses deviating substantially from the guideline $\lambda/(4n)$ assumed here. The acceleration distance was further decreased by an order of magnitude by swapping out \ch{SiO2} layers for Si layers to leverage the higher refractive index (see Table~\ref{tab:acceleration_distance}). In a more recent study, a wider variety of materials were compared in two-, three- and four-stack multilayer structures~\cite{Santi:2022aa}. Materials such as Si, SiC, \ch{TiO2}, \ch{Al2O3}, \ch{SiO2} and \ch{MgF2} were chosen for their readily retrievable optical and physical properties, and combined into mix-and-match multilayer sails to determine optimal acceleration distances. 
It was found that multilayer structures with more layers decreased the acceleration distance relative to a monolayer, consistent with earlier results~\cite{Ilic:2018aa}. However, the improvement saturated at four layers for the chosen materials because the average reflectance tended to unity, and more layers would increase the mass and hence acceleration distance, as suggested by Figure~\ref{fig:multilayer_r_to_m_ratio}b.

\subsection{Gratings, photonic crystals and metasurfaces}
For the lightsail, nanopatterning can radically modify the optical response. Patterning can be periodic along one dimension (gratings), periodic along two dimensions (photonic crystals, PhC) or aperiodic (metasurfaces~\cite{QiuKivshar:2021_meta_review,Zhou:2024_meta_review}, typically constructed with nanoresonators or ``meta-atoms'' on subwavelength scales). Nanopatterned surfaces, while harder to fabricate than unpatterned membranes at the scale of lightsails, do offer huge design advantages. Primarily, the vast photonics parameter space enables multi-objective optimizations over all of the lightsail mission requirements, including resonant membrane reflectivity at the laser wavelength, low membrane mass, enhanced sail emissivity in the MIR for thermal management, and restoring and damping forces for stability within the laser beam. 
Here, we follow the literature on propulsion-focused membrane designs, which only feature periodic structures, leaving metasurfaces for Section~\ref{sec:stability}.

\textbf{Forces and acceleration distance.} 
In gratings and PhC, the simplified picture of photons imparting momentum solely in the direction of acceleration is no longer valid. Instead, forces parallel to the surface of the sail (``in-plane'' forces) can appear, requiring modification of eq~\eqref{eq:D_FOM}. 

Consider the simplest patterned surface, shown in Figure~\ref{fig:propulsion_designs}b: a diffraction grating, which is periodic in only one direction ($x$) (being continuous in the orthogonal $y$ direction). Gratings scatter light into approximately discrete directions (``orders'') governed by the grating equation~\cite{Petit:1980aa,Loewen:2017aa,Joannopoulos:2008aa}
\begin{equation} \label{eq:grating_equation}
    \sin\theta_q = \sin\theta + \frac{q\lambda}{\Lambda} \,,
\end{equation}
where $q = 0,\pm1, \pm2, \ldots$ labels the diffraction order, $\lambda$ is the wavelength of light incident on the grating, $\Lambda$ is the grating period, $\theta$ is the angle of light incidence and $\theta_q$ is the angle of light scattered into the $q$-th order (see Figure~\ref{fig:propulsion_designs}b).

If the periodicity is greater than the wavelength, there is at least one non-specular order for a range of incident angles. By deflecting photons at angles different from that of the incident light, non-specular scattering orders redistribute some of their momentum into the direction orthogonal to the desired $z$ acceleration, leading to in-plane force components. In-plane scattering can be very useful for redirecting the sail towards the laser beam's center~\cite{Ilic:2019aa}, but by the same token, in-plane scattering reduces the propulsion force produced by reflected orders. Thus, when considering propulsion as the only goal, grating sails should be purely reflective and operate in the specular-diffraction regime (where only $q=0$ is diffracted), occurring when $\Lambda<\lambda$. Note that in the opposite regime ($\Lambda>\lambda$), non-specular transmitted orders ($q>0$) also contribute to the propulsion because some of the photons' $+z$-momentum components are transferred into in-plane momentum components. Since completely eliminating transmitted orders is difficult with thin dielectric gratings, there may be instances where redistributing power (and momentum) from the transmitted specular order to transmitted scattered orders benefits propulsion.

The power distributed into the orders is represented by the reflection efficiencies $r_q$ and transmission efficiencies $t_q$. The efficiencies are defined as the power flow normal to the grating in the $q$-th diffraction order normalized to the power incident normal to the grating. 
Photonic crystals can be seen as a generalization of gratings, with additional diffraction orders along the second dimension of periodicity. 

In the geometric-optics regime, the total radiation pressure on a grating can be calculated by subtracting the incoming laser momentum from the momentum scattered by all transmitted and reflected orders. For a grating sail whose surface (hence periodicity) is parallel to the $\hat{\mathbf{x}}$ direction, the result is~\cite{Ilic:2019aa}
\begin{align}   
\begin{split} \label{eq:grating_rp_force}
    \mathbf{p}=&\frac{I}{c}\mathbf{\hat{z}}\left(\cos\theta + \sum_q  \left(-t_q+r_q
    \right)\cos(\theta_q)\right) \\
     & - \frac{I}{c}\mathbf{\hat{x}} \sum_q\left(
    t_q+r_q
    \right)\sin(\theta_q)
   ,
   \end{split}
\end{align}
where $I$ is the intensity in the sail reference frame, and the axes directions follow the convention shown in Figure~\ref{fig:propulsion_designs}b. When non-specular orders are propagating, finding the acceleration distance then requires integrating eq~\eqref{eq:grating_rp_force} instead of using eq~\eqref{eq:D_FOM} directly.
 
\textbf{Evaluating scattering efficiencies.}
Evaluating the sail acceleration performance requires calculating the optical forces and thus the grating or PhC transmission and reflection efficiencies ($t_q$ and $r_q$). These can be obtained using a variety of commercial or open-source general electrodynamics solving tools including finite-element and finite-difference solvers. However, a numerical method particularly well suited for planar PhC and grating geometries is rigorous-coupled-wave analysis (RCWA), which efficiently solves the full-wave problem to a specified accuracy. Several easy-to-use open-source libraries for RCWA are available, including some aimed for integration in inverse design optimizations~\cite{Song:2018aa,Jin:2020aa,kim:2023torwa}. RCWA assumes the structure under illumination is periodic and infinite in the transverse plane and is composed of layers with finite height in the longitudinal direction, each layer being invariant in this direction. The RCWA method transforms both the periodic permittivity profile of each layer and the electromagnetic fields into Fourier space, and solves the associated algebraic Maxwell equations to obtain the field coefficients. From there, the efficiencies are obtained from the field amplitudes for the appropriate Fourier order.

\subsection{Optimization}
The immense parameter space of photonic structures for  lightsail missions are only explored efficiently using well-designed optimization methods~\cite{Venter:2010aa}. Inverse design~\cite{Molesky:2018aa},  the process of using a numerical optimizer to increment structures that increasingly improve upon some design criteria (figure of merit, {\em e.g.} eq~\eqref{eq:D_FOM}), typically requires a better strategy than naïvely applying local gradient optimizers.

Gradient-free optimization methods are argued by some to be optimal for photonic optimization, especially in the presence of many resonances~\cite{Bennet:2024optimization_tutorial}. An example is the  genetic algorithm, which  has been used to optimize the thicknesses of multilayer stack lightsails in order to locate resonant-Bragg effects for enhanced reflectivity~\cite{Santi:2022aa}.
Outside genetic algorithms, derivative-free optimization has also been used for lightails PhC: 
Suitably perforating a monolayer slab with periodic holes can lower its mass while simultaneously increasing its reflectance, even reaching unity reflectance over a narrow band via interference effects between guided and scattered modes~\cite{Fan:2002Fano}.
Specifically, it was shown that a PhC sail composed of a crystalline-Si slab patterned with circular holes can achieve significantly shorter acceleration distances than monolayer or multilayer sails (see Table~\ref{tab:acceleration_distance})~\cite{Atwater:2018aa}. The improvement came from a combination of minimized mass (circular holes reduce the Si filling fraction by 50\% compared to a Si slab) and optimized reflection via the unit cell.  

However, with adjoint methods and automatic gradient calculations recently making calculations of gradients over large number of dimensions very efficient,  gradient-based optimizers have become popular for optimization of large parameter  photonic designs. One commonly employed gradient-based optimizer for lightsail parameter spaces is the method-of-moving asymptotes (MMA)~\cite{Svanberg:1987aa}. MMA has been applied to find optimal layer thicknesses in multilayer \ch{Si}/\ch{SiO2} structures~\cite{Ilic:2018aa}, but the real power of the method lies in optimizing photonic crystal structures. Interestingly, optimizing a single-layer 2D photonic crystal (with refractive index constrained to be less than that of either Si or \ch{Si3N4}) for minimal acceleration distance resulted in optima with a one-dimensional (1D) profile~\cite{Jin:2020aa,Kudyshev:2022aa}. The best candidate found was the 1D Si grating depicted in Figure~\ref{fig:propulsion_lit}b, which had almost half the acceleration distance of the previously studied circular-hole-patterned photonic crystal~\cite{Atwater:2018aa} (see Table~\ref{tab:acceleration_distance}). 
Unrealistically, these 1D-gratings resulting from optimization are suspended in vacuum; a substrate is necessary for structural support, and would increase the acceleration distance compared to the reported $\SI{1.9}{\giga\meter}$. The optimization results suggested that reducing the unit-cell filling ratio (hence overall sail mass) had a larger effect on reducing the sail's acceleration distance than increasing the reflectivity. However, reflectivity still influences the acceleration distance in a more subtle way: by studying eq~\eqref{eq:D_FOM}, we see that $h(\beta)$ is monotonically increasing. Therefore, sails with greater reflectivity towards the larger-wavelength portion of the Doppler spectrum have lower acceleration distances than sails with larger reflectivity towards the smaller-wavelength portion, corroborated by optimization results~\cite{Jin:2020aa}. As expected, all gratings found by optimizing eq~\eqref{eq:D_FOM} have subwavelength periods, therefore operating in the specular diffraction regime (see eq~\eqref{eq:grating_equation}) that is advantageous for maximal sail acceleration. 

\section{Thermal management\label{sec:thermal}}
Throughout the acceleration, the sail temperature must remain below the sail's thermal limit. In the vacuum of space, heat transfer is limited to absorption of the laser light in the NIR spectral region, and radiative emission dominantly at MIR wavelengths. The temperature evolution is thus determined by the balance of these two mechanisms, which are both photonic in nature. Since the temperature is defined in the sail's rest frame, the laser-light absorption in the NIR {\em is} influenced by the Doppler shift, but the MIR emission is not.  

The absorbed power is determined by the laser power $P$ and wavelength-dependent, total-sail absorption. For a monochromatic laser, the wavelength dependence only appears because of the Doppler effect, so it is convenient to express the power dependence in terms of the (normalized) velocity instead~\cite{Ilic:2018aa}:
\begin{equation} \label{eq:P_abs}
    P_\text{abs}(\beta,T) = P' a(\lambda_0,\beta,T) \,,
\end{equation}
where $P' = P (1-\beta)/(1+\beta)$ is the laser power in the sail frame, $\lambda_0$ is the laser wavelength, and $T$ is the sail temperature. 
The sail's absorption $a(\lambda_0,\beta, T)$ is not simply a material property, but is also affected by the electric-field distribution inside the sail and thus can be somewhat influenced by appropriate design. However, given the enormous laser intensity, it is essential all spacecraft materials have very low intrinsic absorption across the Doppler-shifted laser spectrum, or are otherwise sufficiently shielded from the laser by low-absorption reflectors (especially important for the payload).

To obtain the radiated power, the sail is modeled as a thermal emitter with hemispherical spectral emissivity $\eps(\lambda, T)$ and blackbody spectral intensity $I_{b}(\lambda, T) = (2hc^2/\lambda^5) (e^{hc/(\lambda k_BT)}-1)^{-1}$, where $h$ is Planck's constant and $k_B$ is the Boltzmann constant. The hemispherical spectral emissivity $\eps(\lambda, T)$ is equal to the sum of emissivities from the laser-facing surface and non-laser-facing surface, averaged over the sail surface and over all angles of emission. The radiated power is then~\cite{Ilic:2018aa,Sparrow:radiation_heat_trnsfer_book}
\begin{equation} \label{eq:P_rad}
    P_\text{rad}(T) = \pi A \int_0^\infty \eps(\lambda, T) I_{b}(\lambda, T) \,d\lambda \,.
\end{equation}
Ignoring the emissivity, eq~\eqref{eq:P_rad} follows the Stefan-Boltzmann law with a $T^4$ temperature dependence. This is ideal for the mission because it strongly counteracts sail temperature increases, and potentially allows for a maximum limit on said temperature. Accounting for the emissivity, the key wavelength range for emissivity enhancement is between $\SI{2}{\micro\metre}$--$\SI{10}{\micro\metre}$, corresponding to a rough operating temperature range of $\SI{300}{\kelvin}$--$\SI{1500}{\kelvin}$ for the sail. 

Using photonic design, it is in principle much easier to increase the emissivity $\eps(\lambda,T)$ in the MIR (for example, using resonant effects enhancing material absorptivity) than it is to reduce $a(\lambda_0,\beta, T)$ in the near infrared~\cite{Cornelius:1999aa,baranov:2019,fan:2022review}. However, as seen in the context of propulsion optimization, mass constraints reduce the parameter space for photonic design. Instead, a simple but effective way to increase $\eps(\lambda,T)$ is to attach to the sail a thin layer with strong material absorption in the MIR, such as \ch{SiO2}~\cite{Ilic:2018aa}. 

Since there are only two mechanisms for heat transfer from the sail, the temperature evolution is governed by the net power intake~\cite{Santi:2022aa}
\begin{equation} \label{eq:temperature_evolution}
    \frac{\partial T}{\partial t} = \frac{P_\text{abs} - P_\text{rad}}{C_s} \,,
\end{equation}
where $T$ and $t$ are the sail temperature and time, respectively (in the sail reference frame), and $C_s$ is the heat capacity of the sail. Although eq~\eqref{eq:temperature_evolution} gives the full temperature dynamics, a common assumption in lightsail thermal modeling is that the sail reaches equilibrium at all points during acceleration (the ``lumped capacitance model'' assumption). Equilibrium is achieved at a certain velocity $\beta$ when the absorbed NIR power equals the radiated MIR power, {\em i.e} $P_\text{abs}(\beta,T_\text{eq}) = P_\text{rad}(T_\text{eq})$, which may be expressed as
\begin{equation} \label{eq:equilibrium}
    P' a(\lambda_0,\beta,T_\text{eq})
    = 
     \pi A \int_0^\infty \eps(\lambda,T_\text{eq}) I_{b}(\lambda, T_\text{eq}) \,d\lambda \,.
\end{equation}
This equation expresses the fact that there may be an equilibrium temperature $T_\text{eq}$, which changes with velocity $\beta$. The equilibrium-power-balance assumption is reasonable given the timescales of temperature changes. Indeed, using eq~\eqref{eq:temperature_evolution} as a rough guide, the timescale for temperature change of a 10~m$^2$, $\SI{1}{\gram}$ Si slab at $\SI{300}{\kelvin}$ is of order $\SI{e-6}{\second\per\kelvin}$ (assuming the sail NIR absorptivity and MIR emissivity are those of thin-film Si~\cite{langevin:2024_pymoosh} with absorption coefficients given by the averages in Table~\ref{tab:materials}). At $\beta = 0$ ($\lambda_0 = 1064$~nm), the timescale of wavelength change can be estimated using eq~\eqref{eq:rp_flat_reflection} and the Doppler-wavelength-shift equation to be of order $\SI{e-1}{\second\per\nano\meter}$. Therefore, the sail temperature generally evolves to equilibrium considerably faster than the timescale for significant variations in the spectral response of the sail under study. 

From here, there are several approaches to modeling the sail's thermal performance. The first method involves specifying a maximum-allowed sail temperature $T_s$ based on the sail's thermal limit, and enforcing a constraint to ensure this temperature is not exceeded. One way to achieve this is to, while optimizing eq~\eqref{eq:D_FOM}, require that $P_\text{abs}(\beta,T) \leq P_\text{rad}(T_s)$, guaranteeing that the sail radiates more power than it absorbs should it reach the temperature limit~\cite{Ilic:2018aa}. An intended consequence of this constraint is that the equilibrium temperature at all sail velocities does not surpass $T_s$. Assuming equilibrium was instantaneously achieved at all velocities during acceleration, it was shown that adding a \ch{SiO2} layer, which has strong MIR emissivity, to a Si slab can maintain a sub-$\SI{1000}{\kelvin}$-constrained temperature with only a moderate penalty to the acceleration distance (see Table~\ref{tab:acceleration_distance})~\cite{Ilic:2018aa}. This combination of materials illustrates the advantage of multilayers over monolayers: the \ch{SiO2}/Si design combines the strong MIR emissivity of \ch{SiO2} with the high index of Si to achieve passive cooling and substantial reflection simultaneously. However, for Si and other semiconductors that have bandgap energies only slighter larger than laser-photon energies, two-photon absorption must be accounted for. For a Si/\ch{SiO2}-based metasurface, it was estimated that $\SI{500}{\kelvin}$ is the limit beyond which thermal runaway occurs~\cite{holdman:2022aa}. This limit was obtained at a laser intensity $\SI{1}{\giga\watt\per\square\meter}$, with the typical Starshot laser intensity ($\SI{10}{\giga\watt\per\square\meter}$) resulting in thermal runaway at any temperature. This result stresses the need for alternative materials to be considered, for which linear and non-linear absorption data over a broad spectrum and temperature range must be acquired.

Another method for passive sail cooling is clever photonic design, which enables multiple objectives to be accomplished simultaneously. Photonic crystal sails can be patterned to operate with high reflectivity in the NIR band to create maximal longitudinal thrust, but can be additionally structured at the slightly larger MIR-wavelength scale to create resonant emissivity. In simulations, a multilayer photonic crystal with period in the micron scale was cut into periodic ``islands'' at the $\SI{10}{\micro\meter}$ scale, where the islands acted as Mie resonators (see Figure~\ref{fig:propulsion_lit}c)~\cite{Brewer:2022aa}. The multilayer consisted of two emissive \ch{Si3N4} layers sandwiching a high-index, reflective \ch{MoS2} layer, with all layers having identical circular-hole patterning. The Mie-island design had nearly triple the hemispherical exitance (radiant flux per unit area integrated over wavelength) in the $\SI{2}{\micro\meter}$--$\SI{6}{\micro\meter}$ range at $\SI{1000}{\kelvin}$ compared to a non-islanded photonic crystal design with equivalent areal density. These results show the promise of photonic, Mie-structured designs for enhancing sail emissivity in the MIR. Compared to their acceleration-distance-focused design, the thermal-focused-Mie-island design only mildly increased the acceleration distance, as seen in Table~\ref{tab:acceleration_distance}. 

For even better thermal performance, extra cooling processes can be added to the sail. For example, it was demonstrated that adding rare-earth-ion-doped layers to the sail can enable laser cooling (shown in Figure~\ref{fig:propulsion_lit}d), in which case the thermal radiation constraint becomes $P_\text{abs}(\beta,T) \leq P_\text{rad}(T_s) + P_\text{cool}(\beta, T_s)$~\cite{Jin:2022aa}. The rare-earth-ion-doped layer harnesses the accelerating laser beam as a pump to excite ions from thermally excited states to higher energy levels that subsequently fluoresces into a lower state than the initial thermally populated states, carrying away net energy. 
Sails that support laser cooling, as opposed to sails that do not,
can withstand higher net laser intensities, thus higher accelerations, decreasing the acceleration distance. 
A decreasing acceleration distance with increasing laser intensity corresponds to a favorable cost tradeoff between decreasing the required laser aperture size and increasing the laser-output power. The laser cooling was simulated for two \ch{Si3N4} layers attached to both faces of a \ch{Yb3+}-doped \ch{LiYF4} layer. As rare-earth cooling relies on transitions narrow relative to the Doppler shift, the additional cooling provided is only substantial over a narrow range of velocities. To maximize cooling power, 
the laser intensity was varied dynamically during the acceleration period such that the sail temperature was maintained at $\SI{300}{\kelvin}$. The acceleration distance of the doped sail was decreased by a factor of 10 or more for target velocities less than $0.03c$, compared to an identical sail with the doped layer swapped for undoped \ch{SiO2}. Unfortunately, the operating bandwidth for laser cooling is narrow relative to the laser Doppler shift, so thrust performance diminishes at target velocities approaching $0.1c$ due to the doped layer's significant mass. 

Extending the temperature modeling beyond the assumption of thermal equilibrium can be done by solving eq~\eqref{eq:temperature_evolution} in discrete time steps, assuming a given starting temperature. As an example, the temperature evolution of two slabs of \ch{TiO2} sandwiching a \ch{SiO2} slab was simulated with a starting temperature of $\SI{50}{\kelvin}$ (corresponding to an out-of-atmosphere launch)~\cite{Santi:2022aa}. The sail reached a maximum temperature of just $\SI{650}{\kelvin}$ for a $\SI{10}{\giga\watt\per\square\meter}$ laser intensity and an assumed \ch{TiO2} absorption coefficient of $\alpha_\text{abs} = \SI{1.2e{-2}}{\per\centi\meter}$ across the Doppler band starting from $\SI{1064}{\nano\meter}$. In almost all simulation time periods shown in Reference~\cite{Santi:2022aa}, the near constancy of the temperature lends credence to the power-equilibrium assumption used in previous studies~\cite{Ilic:2018aa,Brewer:2022aa}.

For both thrust and thermal management, the vast capabilities offered by photonic crystals far outweigh the capabilities of monolayer or multilayer slabs, especially when multiple layers of photonic crystals and slabs are combined. Photonic crystals are capable of high NIR reflectivity, high MIR emissivity and low mass, and are therefore the strongest candidates in terms of acceleration and cooling. However, the mission requires more than propulsion along the beam: the sail must also remain stable within the beam to reach its final speed and ultimately, its destination. 

\section{Stability designs\label{sec:stability}}

\begin{figure}[htb]
    \centering
    \includegraphics[width=0.99\linewidth]{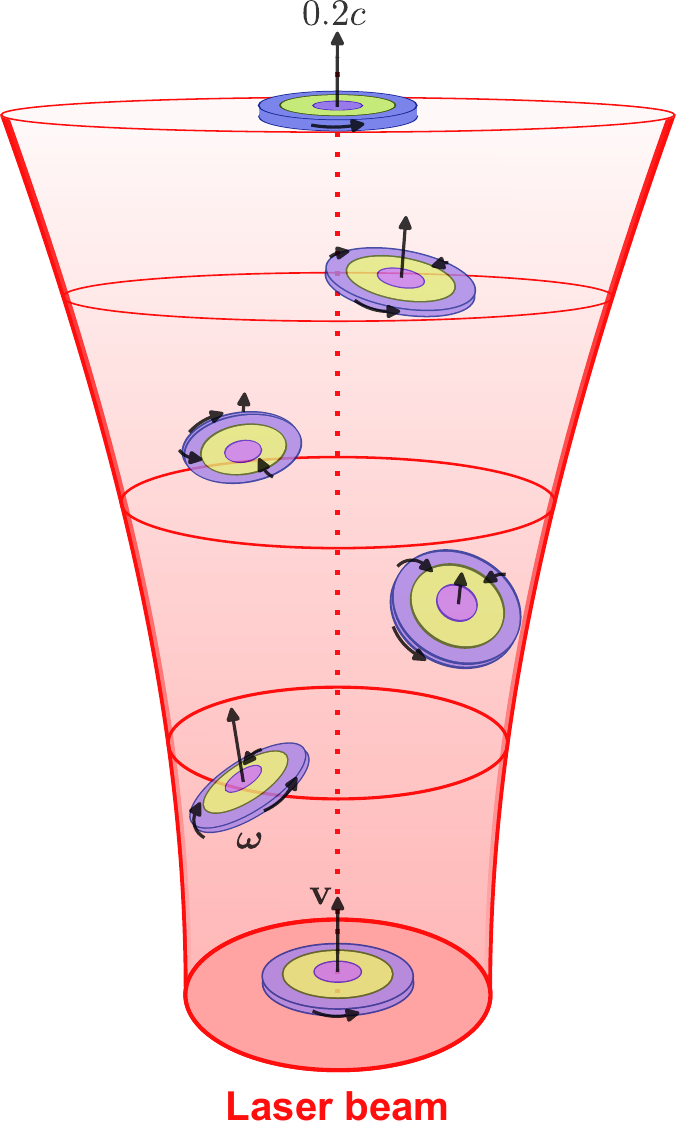}
    \caption{Self-stabilizing lightsail motion. The sail must be trapped around, and decay towards, the laser-beam axis (red dotted line) during the acceleration phase. The laser beam shown is a Gaussian with exaggerated beam width and diffraction.}
    \label{fig:stability}
\end{figure}

In addition to longitudinal thrust, the sail must remain stable within the beam. For a successful launch, there must be mechanisms in place to counteract any perturbations to the sail that may arise due to, for example, imperfect laser-beam alignment or residual atmospheric beam deformations. Stability can primarily be achieved by designing the sail to stay within the confines of the laser beam in the transverse/lateral plane (perpendicular to the laser-beam axis, the red dotted line in Figure~\ref{fig:stability}). However, a subsidiary requirement is that the sail retain its intended shape, which is challenging given the minuscule sail thickness and extreme acceleration: simple membranes tend to crumple up (like balancing a tissue on a pencil) rather than stay flat. To prevent this, the sail material and structure should have sufficiently large stiffness, tensile strength and spin frequency about the laser-beam axis. In Sections~\ref{sec:restoring_motion}--\ref{sec:stability_metasurfaces}, we will  examine stability within the beam assuming the sail is rigid, turning to flexible sails in Section~\ref{sec:flexible_sails}.

There are two mechanisms required for stability of a dynamical system: restoring and damping. Restoring forces and torques keep the sail trapped inside the laser beam, causing the sail's lateral and angular displacements to oscillate around an ``equilibrium''. Lightsails, which are constantly accelerating along the laser-beam axis, have an equilibrium in the sense of zero displacement in the plane perpendicular to the laser-beam axis (and zero rotation about all axes except possibly the laser-beam axis). For the sail to stay in this lateral equilibrium is the ultimate goal of stability-focused structures, and the focus of this section. Damping forces and torques dissipate the energy of sail oscillations, returning the sail to equilibrium. The right combination of restoring and damping leads to the condition of \textit{asymptotic stability}~\cite{Szidarovsky:2017aa}, under which perturbations to the sail do not eject the sail from the beam and furthermore, decay over time. Sail designs that only possess restoring mechanisms are, at best, \textit{marginally stable}, with oscillations that do not decrease in amplitude. Damping mechanisms are necessary because continued perturbations to marginally stable sails may add up,  increasing oscillation amplitude,  ejecting  sails out of the laser beam, and residual velocities at the end of the acceleration phase cause the sail to veer off course for the remainder of the mission.

Restoring and damping mechanisms must be implemented on-board the spacecraft, but cannot add mass beyond the mass budget, so active-feedback stabilizers such as thrusters are not viable. Other stabilization schemes have been considered, including on-board liquid crystals to modulate the optical phase~\cite{Chu:2021aa}, or viscous or magnetic induction~\cite{Shirin:2021aa}, but they too are likely to exceed the mass budget, and likely contribute to disastrous laser absorption. Instead, the best way to generate stabilizing forces and torques within the mass budget is to harness the laser momentum itself, depicted in Figure~\ref{fig:stability}. Sail-membrane architectures that utilize an appropriately chosen laser-intensity profile to passively correct their own motion are called ``beam-riding''~\cite{Singh:2000aa, Schamiloglu:2001aa, Popova:2016aa, Manchester:2017aa} or ``self-stabilizing''~\cite{Ilic:2019aa}. 

To accomplish passive stabilization, lightsail designs must trade prowess in longitudinal propulsion for prowess in transverse stability. This is achieved by partitioning some of the specular scattering into transverse/non-specular scattering to create lateral forces and torques; in other words, the lightsail design has an implicit tradeoff between propulsion and stability. Both aspects should be optimized simultaneously to determine ideal sail-payload-laser configurations that are self-stabilizing without completely sacrificing longitudinal thrust.

\subsection{Restoring motion\label{sec:restoring_motion}}
Fundamentally, lightsail restoring motion arises from a curated combination of the sail's scattering profile and the laser-beam intensity distribution. Here, we explain the principles behind those combinations that can lead to restoring forces and torques.

\begin{figure*}[htb]
    \centering
    \includegraphics[width=0.8\textwidth]{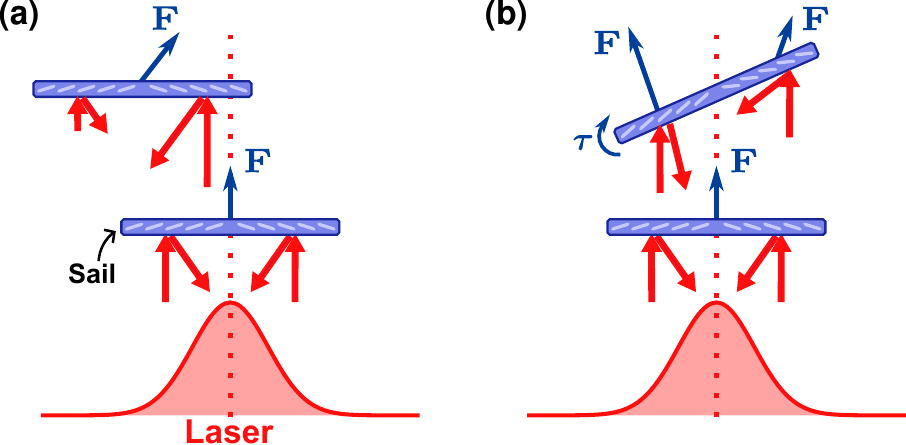}
    \caption{Using the laser beam and scattering profile to restore an arbitrary lightsail towards equilibrium. (a) Lateral displacements from the beam axis (red dotted line) are corrected by the spatially varying intensity distribution and resultant asymmetric scattering. (b) Angle-dependent scattering and thus torque fixes any rotational perturbation.}
    \label{fig:restoring_explained}
\end{figure*}

\textbf{Forces.} 
A restoring force is a force that depends on the displacement of a sail from an equilibrium, and acts to restore the sail to that equilibrium. In practice, the restoring force is, to lowest order, proportional to the sail's displacement from equilibrium, {\em i.e} $\mathbf{F}\propto -\mathbf{x}$. 
To illustrate the restoring force, consider Figure~\ref{fig:restoring_explained}a. The laser's spatial intensity distribution imparts a spatially varying radiation pressure across the sail's surface. Due to the presence of non-specular scattering necessary for restoring forces, certain conditions must be satisfied for the sail to operate in the ideal case, where the sail is not perturbed, and accelerates purely along the laser-beam axis. Firstly, the laser-beam profile should have some symmetry about the beam axis. Then, the sail's optical scattering response must be symmetric about the sail center of mass (COM). Finally, the COM must lie on the laser-beam axis. When these conditions are satisfied, the transverse forces on the sail cancel and the sail experiences only longitudinal thrust. However, in the inevitable case where the sail COM is displaced from the laser-beam axis, the symmetry is broken and net transverse scattering occurs. Therefore, with the right spatial-scattering profile for the sail, laser light can be redirected in the appropriate direction that counters the lateral displacement, via conservation of momentum. Such spatial scattering profiles can be achieved with shaped mirrors ({\em e.g.} cones), or more potently, with nanophotonic membrane designs, which can enhance the restoring effect.

\textbf{Torques.} Just as lateral displacements are countered by an appropriate spatially varying scattering profile, angular displacements can be corrected by an angularly varying scattering profile. Sail rotations change the angle between the incident laser light and the surface, which can be exploited as demonstrated in Figure~\ref{fig:restoring_explained}b: if an increase in incident angle leads to a decreased momentum transfer on the right-hand side of the sail, but an increased momentum transfer on the left-hand side, a torque is generated that counters the rotation of the sail. This angular dependence of momentum transfer can come from an angular dependence of scattering power or a change in scattering angles.

Restoring forces and torques can be understood separately like in Figure~\ref{fig:restoring_explained}, but the situation is more complicated due to coupling between translational and rotational degrees of freedom. For instance, a sail displacement such as that in Figure~\ref{fig:restoring_explained}a changes the intensity gradient of laser light across the sail surface, generating a torque independently of  the restoring torque displayed in Figure~\ref{fig:restoring_explained}b. That is, displacements in translation or rotation can further exacerbate one another. A consequence of the translation-rotation coupling is that sails with both restoring forces and torques still may not be marginally stable. Damping mechanisms can remedy this situation by damping the displacements in translation and rotation.

\subsection{Damping motion\label{sec:damping_motion}}
In order to have asymptotic stability, the sail must experience drag that reduces the velocity of translational and rotational oscillations. To lowest order, drag or damping is characterized by the forces and torques on the sail being proportional and opposite in sign to the velocities: $\mathbf{F} \propto - \mathbf{v}$ and $\bm{\tau} \propto - \bm{\omega}$, respectively. Unlike on Earth, no natural drag exists for damping the motion of objects in the vacuum of space. Just like the restoring mechanism, it would be ideal if the damping mechanism were implemented optically, using only the laser momentum and well-designed sail scattering. Here, we explain the requirements for an optical damping implementation in general terms, with specific examples discussed in later sections.

\textbf{Forces.} The sail most likely experiences perturbations that generate or add unwanted transverse velocities. For example, misalignment between the laser beam and sail initiates the restoring force into action, which subsequently creates potentially destabilizing transverse velocities. Therefore, if the sail has an unwanted velocity, then achieving damping through conservation of momentum requires the sail to scatter most of the laser power in the same direction as that velocity. Moreover, the power directed in such a manner should be proportional to the magnitude of the sail velocity, as required for a damping force. 

\textbf{Torques.} A damping torque is needed to slow the rotational motion of the sail. Relative to the sail COM, more laser momentum must strike the portion of the sail rotating towards the laser, and less laser momentum must strike the portion of the sail rotating away from the laser. In this manner, rotations are damped relative to the sail COM. As before, the characteristic of a damping torque is that the laser power imparted on the sail according to the above prescription be proportional to the sail angular velocity.

The physical arguments for the restoring and damping mechanisms apply equally well in three dimensions (3D), however, dynamics in 3D presents further challenges to the sail stability. The most significant effect of the increase in dimensionality is that there are more degrees of freedom, hence more coupling between rotations and translations that can lead to instability compared to the instability predicted by purely 2D interactions. For designs that are polarization sensitive and operate assuming a linearly polarized laser, a balancing-spin torque may be required to keep the sail aligned with the incident polarization (``spin locking''). 
However, some designs envision using spin as an aid to stabilization, both through  gyroscope effects and centrifugal forces keeping the sail stretched. Spinning of the sail can be achieved mechanically by an external launch vehicle prior to the laser acceleration phase. If the sail's optical response is designed for linearly polarized light, the polarization of the laser needs to rotate synchronously with the sail, which would be difficult to achieve without feedback, but could be facilitated by spin-locking torques.
Alternatively, the lightsail can be designed to have equivalent response for both linear polarizations, or a circular laser polarization could be used.

Having discussed the physical mechanisms for restoring and damping motion, we next explain how lightsail designs with these mechanisms can be characterized, which requires outlining the mathematics of lightsail-dynamics modeling.

\subsection{Lightsail dynamics\label{sec:lightsail_EOM}} 
\begin{figure}[htb]
    \centering
    \includegraphics[width=0.99\linewidth]{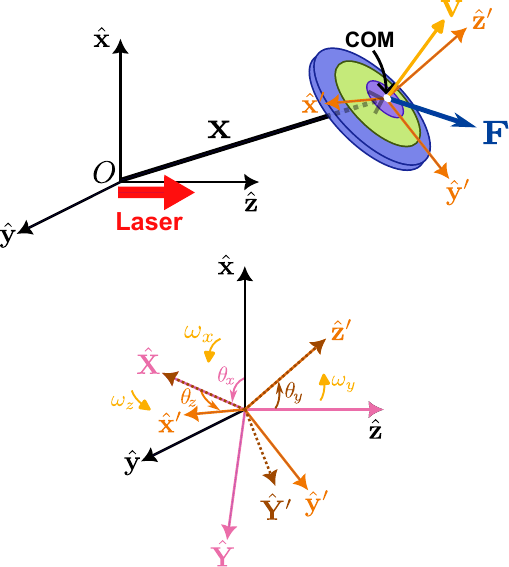}
    \caption{Three-dimensional sail reference frames. The laser frame axes are shown in black, whereas the sail frame axes are shown in orange (with Euler angles defined using the $z$-$X$-$z'$ convention).}
    \label{fig:sail_3d}
\end{figure}

Analyzing the full lightsail stability in 3D requires two frames of reference: the sail/body frame and the laser-source frame, which are shown in Figure~\ref{fig:sail_3d}. Stability is defined in the laser-source frame because the trajectory is determined by the laser direction for a particular mission target. However, the sail rotations are most easily characterized in the sail frame using Euler angles~\cite{Goldstein:2001aa}. For a rigid-body sail, the dynamics are then determined by solving the Newton-Euler equations, which combine the total forces $\textbf{F}$ and torques $\bm{\tau}$ on the sail into a single matrix equation:
\begin{equation} \label{eq:newton_euler}
    \begin{bmatrix}
        m\mathbf{I}_3 & 0 \\
        0 & \mathbf{I}_\text{COM} \\
    \end{bmatrix}
    \frac{d^2}{dt^2}
    \begin{bmatrix}
        \mathbf{x} \\ \bm{\theta}
    \end{bmatrix}
    +
    \begin{bmatrix}
        \mathbf{0} \\ \dot{\bm{\theta}} \times (\mathbf{I}_\text{COM} \dot{\bm{\theta}})
    \end{bmatrix}
    =
    \begin{bmatrix}
        \mathbf{F} \\ \bm{\tau}
    \end{bmatrix}
    \,.
\end{equation}
Here, $\mathbf{x} = [x,y,z]^T$ is the position vector and  $\bm{\theta}= [\theta_x,\theta_y,\theta_z]^T$ contains the Euler angles. The term $\mathbf{I}_3$ is the $3\times 3$ identity matrix, and $\mathbf{I}_\text{COM}$ is the principal moment of inertia about the sail COM. The forces and torques generally depend on the sail-position vector $[\mathbf{x},\bm{\theta}]^T$ and the sail-velocity vector $[\mathbf{v},\bm{\omega}]^T = d[\mathbf{x},\bm{\theta}]^T/dt$. However, in many cases, some dependencies can be removed, such as when there are no velocity-dependent forces ({\em i.e} no damping) or no torques about the axis normal to the sail body (the $z'$ axis in Figure~\ref{fig:sail_3d}). It is important to remember that eq~\eqref{eq:newton_euler} is a classical equation of motion, {\em i.e.} it does not incorporate special relativity. In special relativity, obtaining the equations of motion requires a rigorous Lorentz transformation and spatial-axes rotation between the laser-source and sail reference frames, which significantly complicates the problem. However, almost all of the literature thus far regarding stabilization~\cite{Popova:2016aa,Manchester:2017aa,Ilic:2019aa,Siegel:2019aa,Srivastava:2019aa,Srivastava:2020aa,Gieseler:2021aa,Kumar:2021aa,Shirin:2021aa,Gao:2022aa,Rafat:2022aa,Savu:2022aa,Gao:2024aa,Srivastava:2024aa} has ignored the influence of special relativity, or only included select relativistic effects~\cite{Salary:2020aa,Salary:2021aa,Taghavi:2022aa}. Therefore, these investigations operate in the non-relativistic-lightsail regime, appropriate for Solar System missions or travel just beyond the heliosphere, in which eq~\eqref{eq:newton_euler} is valid.

\textbf{Linear stability analysis.} Generally speaking, designing a sail for lightsail stability is more challenging than for simple lightsail propulsion. While comparing the effectiveness of designs for propulsion can be done using a single figure of merit (the acceleration distance, eq~\eqref{eq:D_FOM}), this is no longer the case for stability. The strength of restoring forces and torques are not a good measure by themselves because the rotational and translational degrees of freedom are coupled, and the full dynamics depend on initial misalignment and perturbations experienced during the acceleration phase. Instead, a good starting point for analyzing the stability of a lightsail (with a chosen set of sail and laser-beam parameters) is a linear stability analysis.

Linear stability analysis~\cite{Khalil:2002aa,Szidarovsky:2017aa} involves linearizing the equations of motion for a dynamical system about an equilibrium point. The solutions to the linearized equations are composed of analytic modes, which fall into three classes: bound (oscillating) modes, growing modes and decaying modes. Consider an arbitrary sail as in Figure~\ref{fig:sail_3d}. In this case, it can be assumed that the equilibrium point is at the origin $\mathbf{0}$ of the comoving-sail reference frame (marked ``COM'' in Figure~\ref{fig:sail_3d}) without loss of generality. Linear stability analysis is then conducted in the $x'y'$ plane, {\em i.e} ignoring the $z'$ components and rotations about the $z'$ axis. Under these assumptions, a general time-independent equation of motion has the form $\dot{\statevec} = \mathbf{f}(\statevec)$, where $\statevec = [x,\theta_x,y,\theta_y,v_x,\omega_x,v_y,\omega_y]^T$ is the state vector. In the limit where $\statevec$ is close to $\mathbf{0}$, $\mathbf{f}(\statevec) = [v_x, \omega_x, v_y, \omega_y,f_x,\tau_x,f_y,\tau_y]^T$, where $f$ and $\tau$ represent the forces and torques normalized by the sail mass and moment of inertia, respectively. The corresponding linearized equation of motion takes the form $\dot{\statevec} = \jac \statevec$, where $J_{ab} = \partial f_a/\partial q_b \vert_{\statevec = \mathbf{0}}$ is the $(a,b)$th component of the system-Jacobian matrix ($a$ and $b$ indexing the components of $\mathbf{f}$ and $\statevec$, respectively). The solution to the linearized first-order differential equation is 
\begin{equation} \label{eq:linear_solution}
    \statevec = e^{\jac t} \statevec_0\,,
\end{equation}
where $\statevec_0$ is the initial condition and $e^{\jac t} = \jac t + (\jac t)^2/2! + (\jac t)^3/3! + \ldots $ is a matrix exponential. Equation~\eqref{eq:linear_solution} shows that the linear solution about the equilibrium is a decomposition into eigenmodes $\exp(\xi_m t)$ that are characterized by $\jac$'s eigenvalues $\xi_m = \Re(\xi_m) + i\Im(\xi_m)$ (indexed by $m$), which are complex in general. The real parts $\Re(\xi_m)$ correspond to either growth or decay relative to the equilibrium, while the imaginary parts $\Im(\xi_m)$ correspond to oscillations about the equilibrium. A sufficient condition for asymptotic stability of the equilibrium in the full nonlinear system is that $\Re(\xi_m) < 0$ for all $m$ because the solutions decay to $\mathbf{0}$. Alternatively, if $\Re(\xi_m) > 0$ for any $m$, the system moves exponentially away from the origin and the equilibrium is unstable. If $\Re(\xi_m) \leq 0$ for all $m$ (with some $\xi_m$ having $\Re(\xi_m) = 0$), the system is deemed marginally stable. In this case, the solutions do not move exponentially away from the origin over time, but the true stability of the equilibrium can only be determined by solving the full nonlinear system for the entire integration period (the time taken for the sail to reach $v_f=0.2c$). In all but the simplest cases, integrating the nonlinear system requires numerical ordinary-differential-equation solvers such as fourth-order Runge-Kutta~\cite{Garcia:2000aa}. 

For the lightsail mission, we require solutions that decay towards the transverse equilibrium at the laser-beam center ($\Re(\xi_m) < 0$ for all $m$), which can be achieved by implementing some damping mechanism into the sail. In practice, the system is unlikely to be overdamped, so the sail should oscillate in the transverse plane upon implementing restoring mechanisms ($\Im(\xi_m) \ne 0$ for all $m$).

Most sail candidates studied in the literature experience only restoring forces and torques. Such sails are, at best, marginally stable, meaning the system Jacobian is independent of velocity and thus takes the form
\begin{equation} \label{eq:jacobian_restoring}
    \jac 
    = \frac{\partial \mathbf{f}}{\partial \statevec}\bigg\vert_{\mathbf{0}} 
    = 
    \begin{bmatrix}
        0 & 0 & 0 & 0 & 1 & 0 & 0 & 0 \\
        0 & 0 & 0 & 0 & 0 & 1 & 0 & 0 \\
        0 & 0 & 0 & 0 & 0 & 0 & 1 & 0 \\
        0 & 0 & 0 & 0 & 0 & 0 & 0 & 1 \\
        \frac{\partial f_x}{\partial x} & \frac{\partial f_x}{\partial \theta_x} & \frac{\partial f_x}{\partial y} & \frac{\partial f_x}{\partial \theta_y} & 0 & 0 & 0 & 0 \\
        \frac{\partial f_y}{\partial x} & \frac{\partial f_y}{\partial \theta_x} & \frac{\partial f_y}{\partial y} & \frac{\partial f_y}{\partial \theta_y} & 0 & 0 & 0 & 0 \\
        \frac{\partial \tau_x}{\partial x} & \frac{\partial \tau_x}{\partial \theta_x} & \frac{\partial \tau_x}{\partial y} & \frac{\partial \tau_x}{\partial \theta_y} & 0 & 0 & 0 & 0 \\
        \frac{\partial \tau_y}{\partial x} & \frac{\partial \tau_y}{\partial \theta_x} & \frac{\partial \tau_y}{\partial y} & \frac{\partial \tau_y}{\partial \theta_y} & 0 & 0 & 0 & 0 
    \end{bmatrix} \,.
\end{equation}
%
For a rotationally symmetric sail, eq~\eqref{eq:jacobian_restoring} simplifies further because the $x$ and $y$ components are interchangeable. In that case, we define $k_1 = -\partial f_x/\partial x$, $k_2 = \partial f_x/\partial \theta_y$, $k_3 = \partial \tau_y/\partial x$ and $k_4 = -\partial \tau_x/\partial \theta_x$ as the non-redundant stiffness coefficients, and it can then be shown~\cite{Kumar:2021aa} that the degenerate eigenvalues of $\jac$ are
\begin{align} \label{eq:jacobian_eigenvalues}
\begin{split} 
    \xi_{1\text{--}8}
    &=\pm \frac{1}{\sqrt{2}} 
    \Big[ -k_4-k_1
        \\ &\hspace{8mm} \pm \sqrt{(k_4+k_1)^2 + 4(-k_1k_4 + k_2k_3)}
        \Big]^{\frac{1}{2}} \,.
\end{split}
\end{align}
Under this convention, $k_1$ and $k_4$ correspond to the restoring force and restoring torque, respectively, while $k_2$ and $k_3$ represent the coupling between rotation and translation degrees of freedom. The eigenvalues in eq~\eqref{eq:jacobian_eigenvalues} have zero real part if the following conditions hold~\cite{Kumar:2021aa}: 
\begin{align}
    c_1 &\equiv k_4 + k_1 > 0 \label{eq:c1} \\
    c_2 &\equiv k_1k_4 - k_2k_3 > 0 \label{eq:c2} \\
    c_3 &\equiv (k_4+k_1)^2 + 4(-k_1k_4 + k_2k_3) > 0\,. \label{eq:c3}
\end{align}
Equations~\eqref{eq:c1}--\eqref{eq:c3} are the criteria for the marginal linear stability of a lightsail. In particular, eq~\eqref{eq:c2} intuitively indicates that marginal stability is acquired if the restoring forces and torques are stronger than the rotation-translation-coupling terms. Indeed, it has been suggested that sails with less coupling between rotation and translation for the entire Doppler band are more stable~\cite{Manchester:2017aa,Salary:2020aa,Salary:2021aa,Taghavi:2022aa}. Such sail designs can be achieved by minimizing $c_2$ over the Doppler band as a figure of merit. Alternatively, eqs~\eqref{eq:c1}--\eqref{eq:c3} can be set as constraints during optimization of other figures of merit, such as eq~\eqref{eq:D_FOM}.

However, it is important to remember that eqs~\eqref{eq:c1}--\eqref{eq:c3} are merely a guide for finding marginally stable structures in the optimization. The full dynamics are nonlinear and can deviate from linearity even at relatively small angles or displacements due to, for example, grating resonances. Large enough deviations in the initial sail position, angle or velocity could lead to oscillations with amplitudes outside the stability region. Exhaustive integration of the nonlinear dynamics over the full acceleration phase and for a range of realistic initial conditions is thus necessary to assess stability of specific sails designs. This highlights the difficulty with finding an appropriate figure of merit for lightsail stability that allows for both tractable numerical optimization and simultaneously, comparison between lightsail designs.

There are numerous ways to create transverse scattering and therefore transverse stability for the lightsail. The first and simplest way is to shape the sail geometrically (moving beyond flat-slab reflectors), which we discuss in the following subsection. Due to the complexity associated with 3D dynamics, we begin the discussion on 2D sail structures before transitioning to 3D structures.

\subsection{Geometric\label{sec:geometric-stabilization}}
\begin{figure*}[htb]
    \centering
    \includegraphics[width=0.8\linewidth]{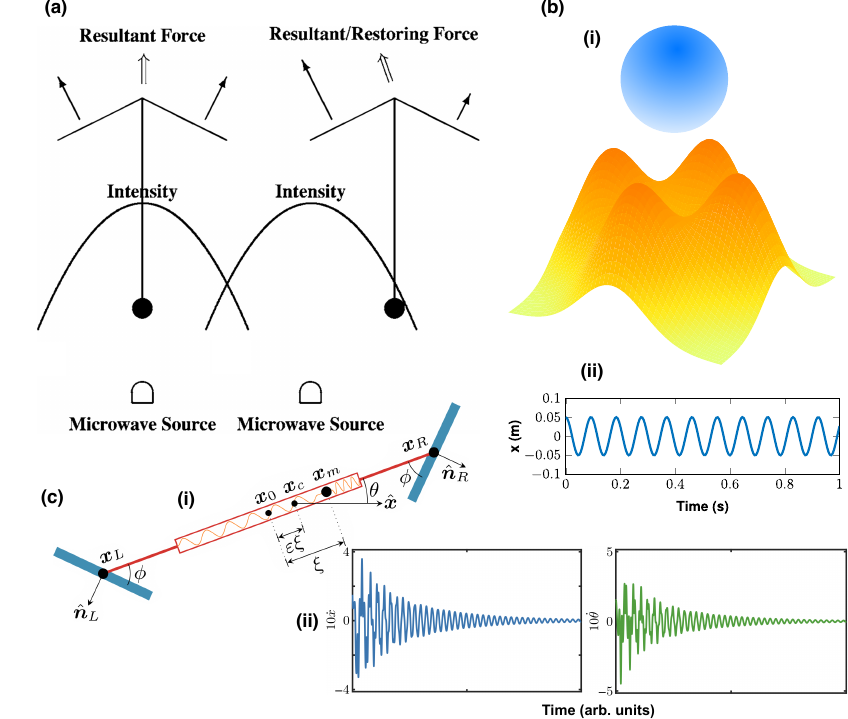}
    \caption{Geometric sail designs. (a) Conical sails attached to a boom and countermass ({\em e.g.} payload) demonstrating geometric-shape restoring forces. Reproduced from Reference~\cite{Abdallah:2003aa}, with the permission of AIP Publishing. Copyright 2002 AIP Publishing. (b) Self-restoring spherical shell sail irradiated by a multi-modal Gaussian beam (panel (i), sphere added) and associated position dynamics (panel (ii), horizontal axis label added). Adapted from Reference~\cite{Manchester:2017aa} with permission by the AAS. Copyright 2017 the AAS. (c) Self-restoring and self-damping V-shaped mirror with internal damped-mass-spring system (panel (i)). Panel (ii) shows the decaying translational and angular velocities over time (axis label added). Adapted from Reference~\cite{Rafat:2022aa}. Copyright 2022 American Physical Society.}
    \label{fig:geometric_stability_lit}
\end{figure*}

The concept of using geometrically shaped reflectors to generate self-restoring forces and torques was discovered not for NIR-laser propelled lightsails, but for microwave-irradiated hovering sails~\cite{Singh:2000aa, Schamiloglu:2001aa, Benford:2002aa, Abdallah:2003aa, Benford:2003aa}. The sails were chosen to be conical specular reflectors (apex pointing away from the light source), irradiated by a microwave beam whose intensity decreases with distance from a specified beam center (Figure~\ref{fig:geometric_stability_lit}a). The combination of a conical sail and light beam whose power decays from a central axis results in restoring forces, which can be understood as a specific implementation of the restoring motion illustrated in Figure~\ref{fig:restoring_explained}a. Sail rotation perturbations on the conical sail would ordinarily be amplified, but offsetting the sail COM using a countermass (attached via a rigid boom, see Figure~\ref{fig:geometric_stability_lit}a) changes the angle between the moment arm and forces on the cone surface, which can convert the destabilizing torque into a restoring torque. Offsetting the COM was shown to be a necessary requirement for conical-sail rotational stability unless other stabilization methods are introduced~\cite{Popova:2016aa}. This makes conical sails less viable lightsail candidates due to the difficulties in creating a sufficiently strong tether to the payload that stays within the mass budget.

Alternative geometric-sail and beam-profile combinations can generate restoring forces and torques for beam-riding stability. For example, spherical sails illuminated by beams with minimum intensity at their center (``hollow'' laser beams, depicted in Figure~\ref{fig:geometric_stability_lit}b) have been proposed that achieve the same beam-riding stability as conical sails~\cite{Popova:2016aa,Manchester:2017aa}. Spherical (or spherical sector) sails possess curvature that may reduce the bending stresses induced by the radiation pressure from the laser beam~\cite{Campbell:2022aa}. Additionally, for a full-sphere sail~\cite{Manchester:2017aa}, stability is much easier to achieve due to the decoupling between translational and rotational degrees of freedom. However, the optical-scattering-cross-section-to-mass ratio of a sphere is not ideal. Indeed the surface area of a sphere is 4 times larger than that of a disk with the same optical cross section, meaning the mass of the sphere is 4 times larger that that of a flat membrane (assuming the same materials and thickness). Furthermore,
spherical sails, thus far assumed rigid, may be susceptible to crumpling from the laser-beam-facing side. Indeed, using eq~\eqref{eq:rp_flat_reflection}, the radiation pressure from a $\SI{10}{\giga\watt\per\square\meter}$ laser corresponds to a photonic pressure of $\SI{67}{\pascal}$, which must be counteracted by inflating the sail, increasing the spacecraft mass. In reality, an even higher pressure (hence mass) is required to maintain the sail-surface tension against the intense laser beam. For instance, inflating a $\SI{10}{\meter^2}$-cross-section sphere with \ch{H2} at just $\SI{300}{\pascal}$ would add $\SI{6.3}{\gram}$ to the spacecraft mass at $\SI{273}{\kelvin}$. Finally, spherical sails must be made of a material that can withstand the inflation and radiation-induced pressure at only tens of nanometers in thickness.

Beyond spherical sails, configurations that rely on hollow beams, such as hyperboloids~\cite{Srinivasan:2016aa} or inverted cones~\cite{Rafat:2022aa}, are generally sub-optimal because non-centrally-concentrated beams require larger apertures for equivalent focal spot sizes, and most of the laser power is wasted at the edges of the sail to provide stability. Thus, configurations that involve centrally focused beams ({\em e.g.} Gaussians) are preferable because they focus most of the laser power on the sail to provide optimal longitudinal thrust, an important consideration given the immense costs associating with constructing and operating the laser~\cite{Parkin:2018aa}. Aside from the geometric shapes discussed here, more general shapes using surfaces of revolution can be conceived, which provide a moderately larger design space for self-stabilizing lightsails~\cite{Shirin:2021aa, Shirin:2022aa}.

\textbf{Geometric damping.}
In the literature discussed thus far, only marginal stability has been demonstrated. Here, we discuss implementations of damping processes that can contribute to the sail's necessary asymptotic stability. 

One proposed damping mechanism is parametric damping, whereby the laser is periodically turned off to create time-varying force coefficients, causing the amplitude of oscillation to decay. Parametric damping was observed experimentally by measuring such decay in oscillation amplitude of a sail attached to a torsional pendulum driven by radiation pressure~\cite{Chu:2021aa}. The ability to increase the oscillation amplitude was also demonstrated as part of the parametric oscillator technique. In a lightsail mission, parametric oscillation would be challenging to implement because it relies on active feedback by dynamically adjusting the laser output in response to oscillatory motion. Active feedback from Earth is difficult because the time scale of sail oscillations is likely far shorter than the travel time between laser source and sail. For instance, simulations of a spherical sail~\cite{Manchester:2017aa} (Figure~\ref{fig:geometric_stability_lit}b), assuming standard Starshot parameters, showed sail oscillations with a period of $\SI{0.1}{\second}$, an order of magnitude smaller than the light travel time between sail and laser at the midpoint of the acceleration phase. 

An alternative that does not require synchronization or feedback is to slowly vary the restoring-force coefficient on a time scale much slower than the oscillation period. Consider modeling the transverse oscillations as a classical oscillator with mass $m$ and spring constant $k$. If $k$ slowly decreases, time-invariance symmetry is broken, so energy in the transverse oscillation mode is not conserved, decreasing. Furthermore, the oscillation frequency $\omega=\sqrt{k/m}$ decreases, but in such a way that the {\em adiabatic invariant} $E/\omega$ remains constant~\cite{Landau:1976aa}. During such a slow relaxation of the stiffness, the spatial amplitude of oscillations \textit{increases} as $k^{-1/4}$, while the amplitude in velocity oscillations \textit{decreases} as $k^{1/4}$. Thus, this adiabatic method offers a way to decrease transverse-velocity oscillations, but at the cost of increasing the amplitude of spatial oscillations -- or vice versa. Based on  the simple-harmonic-oscillator model the method seems unsuitable for decreasing the amplitudes of both velocity and spatial oscillations simultaneously. Ultimately, decreasing the transverse-velocity oscillations is important to keep the sail on target, but minimizing spatial oscillations is required to stop the sail from exiting the stability region of the laser beam.

In the case of lightsails, slow changes to the restoring-force and torque coefficients can come from the sail dispersion~\cite{Taghavi:2022aa} (due to the overall Doppler effect in the COM reference frame), or from changes in laser-beam size. In principle, both methods could be designed to reduce the energy and amplitude of oscillations over the course of the acceleration. Evidence of such oscillation amplitude changes were generally found in works taking into account the Doppler effect for dispersive gratings~\cite{Salary:2020aa, Salary:2021aa, Taghavi:2022aa}, but so far have not shown simultaneous decrease in velocity and displacement amplitudes. 

Beyond time variable stiffness, a more direct way to dampen oscillations is to introduce explicitly velocity-dependent damping forces. 
A method to create a drag force that does not rely on active feedback and can be implemented on board the sail is an internal-damped degree of freedom~\cite{Rafat:2022aa}, shown in Figure~\ref{fig:geometric_stability_lit}c. The concept involves adding a damped mass-spring system into the sail, or by splitting the sail into multiple moving sections connected by damped hinges. In doing so, the internal damping couples to the external sail oscillations, reducing the transverse sail velocities. In particular, the damped-mass-spring system inserted into an inverted cone (or in the 2D case, a V-shaped mirror) showed full asymptotic stability with nearly complete velocity attenuation over roughly $\SI{2}{\second}$ of integration time. Although the damped internal degree of freedom is effective at reducing transverse velocities, it is unclear whether such a system could be included in the lightsail design within the low-mass budget. 

\begin{figure*}[htb]
    \centering
    \includegraphics[width=0.8\textwidth]{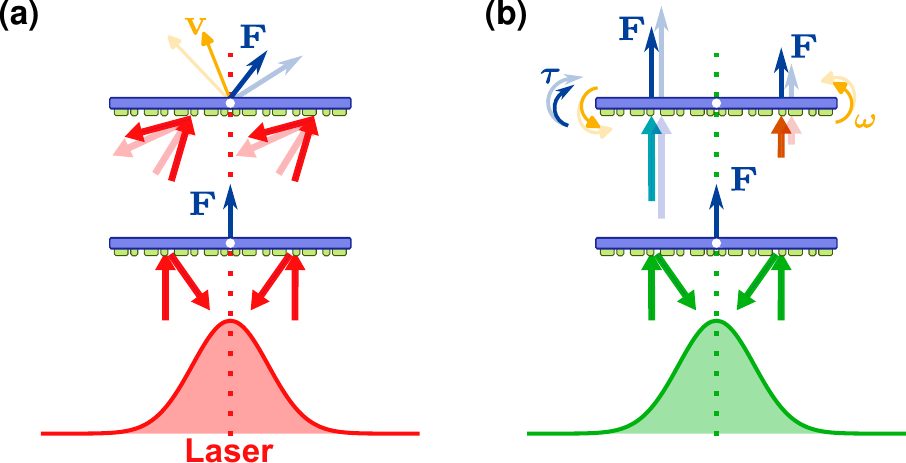}
    \caption{Relativistic effects on an appropriately designed lightsail (two connected gratings depicted here) can damp unwanted velocities. (a) Relativistic aberration changes the incident laser angle in a velocity-dependent manner, allowing for tailored reflections to generate a damping force. (b) The Doppler effect changes the intercepted laser power on sections of the sail rotating towards as opposed to rotating away from the laser source, resulting in a damping torque.}
    \label{fig:damping_explained}
\end{figure*}

In the same way that restoring forces and torques were generated by cleverly scattering laser light using judicious geometric or photonic designs, it is highly preferable that the same be achieved for damping forces and torques, essentially providing an ``all-in-one'' solution. Such an optical damping method has precedent in cavity optomechanics~\cite{Aspelmeyer:2014aa}. The solution for damping lightsails appeared in a somewhat fortuitous manner: in all the literature thus far, lightsail dynamics had been treated either in a completely Newtonian framework or by only modeling select special-relativistic effects (the relativistic Doppler shift). By incorporating special relativity rigorously for the entire dynamics, a damping method can be extracted which uses only the laser and a tailored sail design~\cite{Mackintosh:2024aa}. 
The mechanism is similar to the Poynting-Robertson effect~\cite{Poynting:1904aa,Robertson:1937aa,Klacka:2014aa,fuzfa:2020}, which is a consequence of the relativistic aberration of light experienced in the frame of reference of an object moving relative to a light source. A sail with an unwanted transverse velocity in the laser-source frame experiences relativistic aberration which, in the sail's own rest-frame, makes the incident laser light appear to originate with an incidence angle that depends on the transverse velocity (see Figure~\ref{fig:damping_explained}a). This change in incident angle results in asymmetric scattering and thus a velocity-dependent force when combined with appropriate sail designs such as V-shaped mirrors~\cite{Mackintosh:2024aa}. It was further shown that the Doppler effect can create a damping torque on the sail~\cite{Mackintosh:2024aa}, depicted in Figure~\ref{fig:damping_explained}d: the damping comes from the fact that, relative to the sail COM, the portion of the sail rotating towards (away from) the laser experiences a blueshift (redshift) of the laser wavelength and thus intercepts more (less) laser power. For simple geometries relying on mirrors, these effects are weak, but we will see that they can be enhanced considerably by innovative nanophotonic designs.

\subsection{Gratings\label{sec:stability_gratings}}
The versatility of nanostructured designs has created exciting possibilities for beam-riding stability. The main advantages of nanostructured designs for sail stability include the decoupling of the geometric sail shape from the optical scattering response, the ability to control the non-specular diffraction or anomalous scattering, and configurable angular dispersion and frequency dispersion. In addition, the opportunities for numerical optimization in advanced photonic structures is unmatched by geometric sails; in a sense, geometric reflectors can be considered a small subset of nanostructured designs.

\begin{figure*}[htb]
    \centering
    \includegraphics[width=0.65\linewidth]{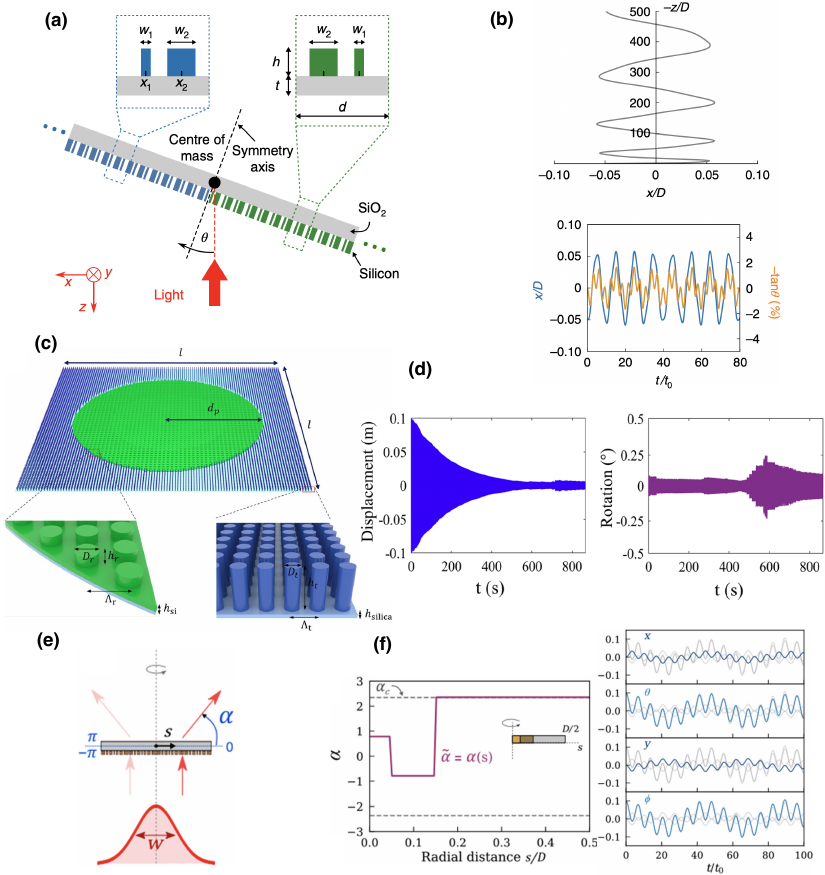}
    \caption{Nanostructured sail designs. (a-b) Bigrating design that shows bounded translation and rotation dynamics. Reproduced from Reference~\cite{Ilic:2019aa}. Copyright 2019 Springer Nature. (c-d) Metasurface design combining reflective and transmissive elements. Panel (d) shows the amplitude of displacement and rotation changing over time due to a Doppler-shift-induced adiabatically changing restoring force. Reproduced from Reference~\cite{Taghavi:2022aa} under a Creative Commons Attribution 4.0 International License. Copyright 2022 Springer Nature. (e-f) Radially-varying-reflection-profile metasurface. The reflection angle over radial distance and the sail dynamics are shown in panel (f). Reproduced from Reference~\cite{Kumar:2021aa}. Copyright 2021 American Physical Society.}
    \label{fig:nano_stability_lit}
\end{figure*}

Here, we focus on the diffraction-grating-based designs. Diffraction gratings were discussed in Section~\ref{sec:propulsion_designs} regarding propulsion, but there, the optimally propelled designs operated in the specular-reflection regime. In order for a grating to self-correct lateral displacements and rotations, it must operate in the diffractive regime, {\em i.e} $\lambda < \Lambda$. In this regime, light momentum is scattered parallel to the grating surface, generating transverse forces and allowing for stabilizing torques. The forces and torques can be calculated using a ray-optics approximation or by integrating the MST, with these approaches giving consistent results (see, for instance, Reference~\cite{Ilic:2019aa} Supplementary Information). With the ray approach, closed-form expressions can be readily obtained in terms of the reflection and transmission efficiencies (which are subsequently evaluated using, for example, RCWA solvers) and scattering angles (eq~\eqref{eq:grating_equation}).

The earliest candidates for self-stabilizing nanostructured lightsails were diffraction gratings~\cite{Ilic:2019aa, Srivastava:2019aa}. A single, uniform grating would not be stable in a Gaussian beam because the light-scattering angle and thus direction of radiation pressure is uniform across the entire grating surface. Instead, a ``bigrating'', composed of at least two gratings with mirror symmetry about their connection point (see Figure~\ref{fig:nano_stability_lit}a), can be made self stabilizing using the same physical principles as those involved in stabilizing the V-shaped mirror~\cite{Rafat:2022aa,Mackintosh:2024aa}. However, the bigrating is (ideally) flat, which is beneficial because flat designs require less material (hence mass) than a V-shaped mirror to achieve the same geometric cross section for intercepting the laser beam. Bigratings with appropriate design were demonstrated to exhibit marginally stable beam-riding behavior in 2D~\cite{Ilic:2019aa,Srivastava:2019aa,Srivastava:2024aa}, characterized by the translational and rotational sail displacements being bounded over the acceleration period (observe Figure~\ref{fig:nano_stability_lit}b for an example dynamics-simulation result). However, not all bigratings and Gaussian beams can combine to create stable-beam-riding dynamics, which is why linear stability analysis is typically used to guide and optimize grating designs. 

\textbf{Three-dimensional models.} 
The 2D models discussed thus far have been invaluable for illustrating the physical concepts behind the self-stabilizing mechanisms. However, for a realistic lightsail implementation that accounts for all dynamical effects, a 3D treatment is necessary. Extending the restoring forces from 1D bigratings to 2D surfaces was first accomplished with square-shaped sails composed of metagrating regions with non-trivial angles between regions, relying on laser-beam-axis restoring torques to align the sail with the incident polarization~\cite{Ilic:2019aa}. Alternative square-sail designs involve combining four triangular grating regions orthogonal to each other (similar in design to Figure~\ref{fig:flexible_lit}a), which operates under illumination by an unpolarized laser beam~\cite{Salary:2021aa}. Disk designs can also offer restoring motion: axicon-grating sails are created by rotating a bigrating about the longitudinal axis through the COM~\cite{Srivastava:2020aa, Chu:2021aa}. Stable axicon designs were identified using linear stability analysis, with the full stability subsequently verified by integrating the Newton-Euler equations of motion~\cite{Srivastava:2020aa}. All candidate 3D self-stabilizing designs have varying amounts of reflection and rotation symmetry about the sail COM, depending on the laser polarization they are designed to work with.

\textbf{Grating-enhanced damping.} 
Gratings, being intrinsically dispersive, are well paired with the acceleration-induced Doppler shift, enabling modulation of the restoring-force and torque coefficients that can lead to limited adiabatic attenuation of oscillations (as seen in Section~\ref{sec:geometric-stabilization}). However, the dispersive response of gratings makes them particularly promising for direct photonic damping forces: as discussed previously in the context of mirror-based sails, relativistic aberration and the Doppler effect contribute to Poynting-Robertson damping~\cite{Mackintosh:2024aa}. 
Because the effect depends on a changing incident laser angle, nanostructured sails are an excellent option for providing angularly resonant enhancement and thus, enhanced damping. The Poynting-Robertson damping that was demonstrated for a geometric-V-shaped sail had a crippling tradeoff: large damping was only possible if the longitudinal acceleration was minimized because the angle between mirrors needed to be small (more light is reflected into the transverse plane for damping, sacrificing the longitudinal components of reflection)~\cite{Mackintosh:2024aa}. Minimizing that tradeoff is possible with nanostructured designs. In particular, diffraction gratings optimized for a maximal damping coefficient showed enhanced Poynting-Robertson velocity attenuation, decreasing initial transverse sail velocities by a factor 8 compared to the V-shaped mirror~\cite{Lin:2024aa}. The improvement in  damping was a consequence of the substantial control over the reflected light directions (eq~\eqref{eq:grating_equation}) and efficiencies, courtesy of the finely tuned grating unit cell.

Beyond damping forces, the damping torque that comes from the rotation-induced differential Doppler effect~\cite{Mackintosh:2024aa} can, in principle, be implemented in nanostructured sail designs. In fact, because the damping torque arises from the sail's wavelength response to the Doppler shift, nanostructured designs are again more capable of providing resonant enhancement~\cite{Lin:2024aa}. However, it remains a current research question as to whether forces {\em and} torques (restoring and damping) can be sufficiently enhanced within a single sail architecture over the entire Doppler-broadened-laser spectrum.

\subsection{Metasurfaces\label{sec:stability_metasurfaces}}
Metasurfaces are a relatively new class of nanostructured materials that can be designed to exhibit unconventional optical phenomena by manipulating scattering strength and phase typically through subwavelength arrangements of nanoresonators (Figures~\ref{fig:rp_force}b and~\ref{fig:nano_stability_lit}c)~\cite{Yu:2011aa,Chen:2016aa,Khorasaninejad:2017aa}. Here we focus on the key physics that makes metasurfaces strong candidates for self-stabilizing lightsails. 

In recent literature, metasurfaces have often referred to any nanostructured surface that affects optical properties, which includes gratings and photonic crystals. However, having discussed gratings in the previous sections, we will concentrate on metasurfaces in which the spacing between resonators is subwavelength, so that no higher-order scattering occurs.
Such metasurfaces~\cite{Yu:2011aa} are then designed to change the phase of an incident plane wave across the metasurface interface such that the resultant refracted and reflected waves experience spatially varying phase shifts $\Phi(x)$ (see Figure~\ref{fig:rp_force}b). Due to this phase gradient, light incident on a metasurface is ``anomalously'' refracted and reflected, meaning the angles of refraction and reflection do not obey the traditional Snell's law and law of reflection, respectively. Instead, these traditional laws are respectively generalized to~\cite{Yu:2011aa}
\begin{align}
    n_t\sin(\theta_t) - n_i\sin(\theta) &= \frac{\lambda_0}{2\pi} \frac{d\Phi}{dx} \label{eq:generalized_refraction} \\
    \sin(\theta_r) - \sin(\theta) &= \frac{\lambda_0}{2\pi n_i} \frac{d\Phi}{dx} \,, \label{eq:generalized_reflection}
\end{align}
where $n_t$ and $n_i$ are the transmission- and incident-medium refractive indices, respectively, and $\theta_t$ and $\theta_r$ are the transmission and reflection angles, respectively. These anomalous refractions and reflections are a novel means of generating transverse scattering. 

In terms of radiation pressure, metasurfaces are different to gratings in two key ways. First, as long as the inter-resonator distance is subwavelength, there are no higher scattering orders. This seemingly simplifies the momentum-transfer analysis, if it weren't for a complication due to the second key difference, which is that metasurfaces are not periodic. Instead, they have an arrangement of resonators with a spatial gradient of resonance frequencies, making numerical analysis more challenging. In particular, ray-optics approaches for calculating radiation pressure, which work well for gratings due to their periodicity, are not necessarily appropriate for metasurfaces (unless the phase profile yields fixed refraction and reflection directions through eqs~\eqref{eq:generalized_refraction} and~\eqref{eq:generalized_reflection}). Therefore, metasurfaces often require a full MST analysis to calculate optical forces.

The most common approach for designing metasurfaces is through the simulation of a range of resonator dimensions to obtain a mapping between resonator design and resulting phase shift. This map is then used in reverse to design the metasurface that will provide the desired phase ramp. Metasurfaces could prove a powerful approach for providing stability in lightsails, however, their design is somewhat more complex than that of gratings due to their lack of periodicity and the addition of local-field effects leading to nanoresonator coupling.

Restoring forces and torques on metasurfaces can be generated by designing a symmetric phase profile $\Phi(x)$ (about the COM) to generate refracted and reflected light satisfying the conditions set out in Figure~\ref{fig:restoring_explained}. However, in order to obtain full control over the incident wavefront, the resonators of the metasurface must be capable of applying a full $2\pi$ phase shift with high-efficiency reflection, and do so across the entire Doppler-broadened spectrum. 

In the literature on lightsails, metasurfaces with an appropriate phase profile for linear stability were first demonstrated in 2D. The sail configuration was composed of a central region designed for high reflectivity and a perimeter designed to provide stabilizing forces and torques, with the laser being an annular beam (whose intensity is minimal at the laser-beam axis)~\cite{Siegel:2019aa}. The phase profile was designed to mimic reflections from a V-shaped mirror to provide stability in the hollow beam. The metasurface was simulated using Finite-Difference Time Domain simulations, integrating the MST (eq~\eqref{eq:MST}) in a box encompassing a small number (5-10) of resonators. The resultant numerical solution to Maxwell's equations and thus the local optical forces and torques (via eqs~\eqref{eq:MST_force} and~\eqref{eq:MST_torque}) were then extrapolated to the entire structure. Integrating over several resonators is an intermediate regime between integrating around a single resonator (computationally cheap, but imprecise) and integrating around the entire structure (computationally expensive but gives the full solution). The intermediate regime presents a tradeoff, allowing reasonable computation times with minor sacrifices in precision. Indeed, the effect of inter-resonator interactions are hardly captured with only a few resonators in the integration box, but the effect on the optical forces was shown to be minor~\cite{Siegel:2019aa}.

\textbf{Metasurfaces for 3D stability.} 
Conventional genetic algorithms have been applied to 3D metasurface lightsails~\cite{Salary:2020aa, Taghavi:2022aa} to find marginally stable designs (according to linear stability analysis and subsequently verified by integrating the equations of motion). For instance, a graded metasurface composed of Si nanodisks (with periodic spacing), depicted in Figure~\ref{fig:nano_stability_lit}c, can be constructed such that the nanodisk diameters vary across the sail surface to create conical or parabolic reflections~\cite{Taghavi:2022aa}. The design exploits a guided-mode resonance and magnetic-dipole resonance in the Doppler band to create near-unity reflectance over a broad spectral region. The problem of integrating the MST over the full sail is amplified in 3D compared to 2D, so instead of full integration, it was assumed that the local fields and scattering from the quasi-periodic metasurface at the small scale (containing a few unit cells) can be calculated using RCWA. RCWA thus provides the local scattering amplitudes (under the periodicity assumption), while the scattering directions are taken from the generalized Snell's law for metasurfaces (eqs~\eqref{eq:generalized_refraction} and~\eqref{eq:generalized_reflection}). 
In dynamics simulations, it was shown that oscillation amplitudes were smallest when the translation-rotation coupling was minimized~\cite{Taghavi:2022aa}. Weakening the coupling between translation and rotation can be done by adding a transmissive metasurface frame (blue region in Figure~\ref{fig:nano_stability_lit}c) around a reflective metasurface (green region in Figure~\ref{fig:nano_stability_lit}c). However, the transmissive elements come with diminished longitudinal thrust, another example of the tradeoff between longitudinal motion and transverse stability~\cite{Taghavi:2022aa}. The bounded dynamics for the reflective/transmissive sail are illustrated in Figure~\ref{fig:nano_stability_lit}d, which also showcases adiabatic reduction in oscillations as discussed in Section~\ref{sec:geometric-stabilization}.

The combination of transmissive and reflective elements was also shown to be beneficial in a study of radially symmetric metasurfaces~\cite{Kumar:2021aa}. It was shown that the sail's dynamical response could be completely characterized in terms of the radiation pressure on the structural building block (see Figure~\ref{fig:nano_stability_lit}e). The metasurface was assumed to reflect into spatially well-defined directions governed by the generalized Snell's law of eqs~\eqref{eq:generalized_reflection} and~\eqref{eq:generalized_refraction}. The validity of the resulting analytical forces and torques was backed by ray-tracing simulations on a conical reflector. The angle of generalized reflections as a function of radial distance from the sail center (shown in Figure~\ref{fig:nano_stability_lit}f) was optimized by running an MMA optimizer on the building-block structure with a longitudinal force figure of merit (eq~\eqref{eq:rp_flat_reflection} integrated over the surface) subject to the marginal stability constraints (eqs~\eqref{eq:c1}--\eqref{eq:c3}). The result was a marginally stable lightsail design, whose longitudinal force per unit incident-beam power was optimized by an order of magnitude compared to previous unoptimized marginally stable designs~\cite{Ilic:2019aa}.

It is interesting that two separate studies concluded that transmissive and reflective regions are needed for sail stability when using metasurfaces. This was not the case for gratings and photonic crystals because they can, in principle, provide independent control over transmission and reflection orders within the same unit cell. Ultimately, which nanophotonic design is more appropriate for sails is likely to depend on co-optimization with material, mass, thermal and payload constraints, as well as progress in large area nanofabrication technology.

\subsection{Flexible, curved sails\label{sec:flexible_sails}}
So far in this review, as in most of the published literature, sails were assumed to be rigid bodies, most often planar/flat structures due to the advantage of maximizing optical cross section for a fixed mass. However, planar structures are not compatible with the stresses imposed by the acceleration. 
Allowing some billowing of lightsails (when the sail radius of curvature is similar to or smaller than the sail diameter) lowers the mechanical stresses induced by the immense radiation pressure, allowing the sail to survive higher laser intensities that can lower the acceleration distance~\cite{Campbell:2022aa}.
Billowing can also help stability: initial investigations have shown that curved metasurfaces made of spherical c-Si scatterers operating at the magnetic-dipole resonance can be marginally stable when considering purely rotational dynamics~\cite{Gieseler:2021aa}. 

\begin{figure*}[htb]
    \centering
    \includegraphics[width=0.75\linewidth]{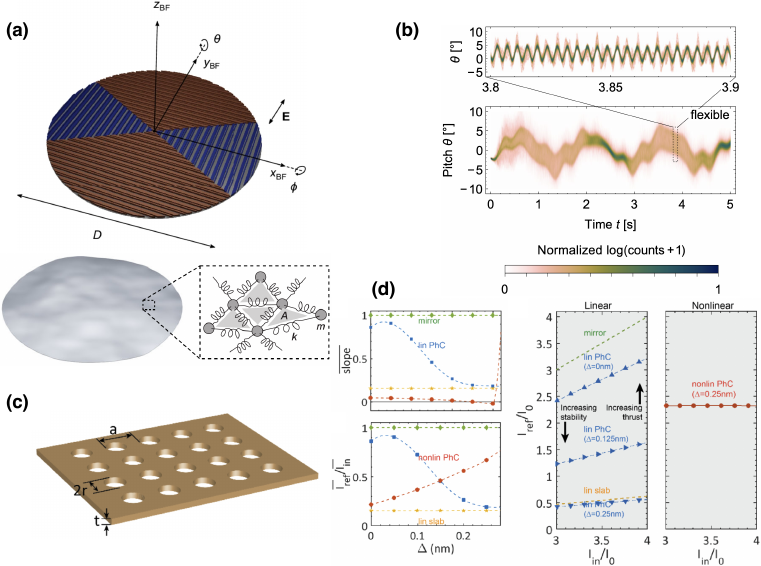}
    \caption{Novel sail designs. (a-b) Flexible, mesh-based grating sail. The angles of each mesh relative to the flat-sail configuration during the dynamics simulation are shown in the distribution of panel (b). Reproduced from Reference~\cite{Gao:2024aa} under a Creative Commons Attribution 4.0 International License. Copyright 2024 Springer Nature. (c-d) \ch{Si3N4} nonlinear photonic crystal. The relatively flat reflected intensity as a function of incident intensity in panel (d) demonstrates the agnostic response of the PhC to incident beam fluctuations. Reproduced from Reference~\cite{Myilswamy:2020aa}. Copyright 2020 Optica Publishing Group.}
    \label{fig:flexible_lit}
\end{figure*}

In reality, given how thin lightsails must be, they will naturally curve or billow because the laser intensity is nonuniform, thereby imparting nonuniform radiation pressure across the sail surface. 
Taking into account the mechanical flexibility of the sails adds considerable complexity to lightsail modeling in the form of coupling between the sail shape and optical response, with perturbations that are likely to be stochastic. The main approach to flexible sail modeling is to discretize the sail into small, rigid pieces, and calculate the optical forces and torques on each piece. For flexible sails, an additional stability criterion emerges in the form of structural stability. If perturbations in the sail's flexible surface exceed a threshold set by the material and structure, the sail risks being torn apart or destabilized further in its body motion. To prevent this~\cite{Savu:2022aa}, the sail can be made of a material with sufficiently large Young's modulus or thickness to increase bending stiffness. 

Beam-riding and structural stability can both be improved by spinning the lightsail about the laser-beam axis. A spinning sail experiences centrifugal tension, keeping it taut. This has the benefit of protecting against deformation instabilities and crumpling, all while maximizing the sail's optical cross section for increased longitudinal thrust. The benefits of spinning for stability were indeed shown in a first comprehensive study of flexible sail dynamics~\cite{Gao:2024aa}. As an example, parabolic sails in a Gaussian beam, which are unstable in the absence of sail spinning or an offset COM, were demonstrated to exhibit stable beam riding conditional on a sufficiently high frequency spin around the laser-beam axis. The model consisted of a flexible membrane sliced into a mesh of triangles with springs for edges and point masses for vertices~\cite{Gao:2024aa}. With high enough fidelity, the optical forces on the triangle are approximated as light rays, which allowed more complicated simulations including multiple reflections to be run. Multiple reflections within a parabolic reflector sail adds to instability because it acts against the intended restoring motion. However, a sufficiently large azimuthal spin frequency can counteract the destabilizing effects of multiple reflections, enabling self-stabilizing motion. Nanostructured designs were also investigated by assuming each triangle in the mesh was composed of a \ch{Si3N4} grating (see Figure~\ref{fig:flexible_lit}a), demonstrating stable, flexible 3D-grating dynamics. Indeed, as an example, the distribution of angles for all meshes relative to the ideal, flat-sail configuration are shown to be bounded in Figure~\ref{fig:flexible_lit}b. However, due to the computationally expensive nature of the simulation, stable dynamics were only displayed for up to $\SI{5}{\second}$ of laser-on time, less than 1\% of the required acceleration time. 
The flexible modeling also accounted for thermal effects, including: conduction in the mesh edges; absorption, radiative cooling and radiative heat transfer in the triangle surfaces and; the effect of ensuing mechanical strain due to differential thermal expansion. In $\SI{5}{\second}$ of laser acceleration, the sail heated up to a maximum of just $\SI{959}{\kelvin}$ (in the region closest to peak Gaussian intensity), which is below the \ch{Si3N4} melting point (Table~\ref{tab:materials}), but extended simulations are required to capture the complete thermal exchange between the sail and the environment. Interestingly, by setting optical absorption to zero and artificially simulating the sail with a constant temperature of $\SI{300}{\kelvin}$, it was discovered that the flexible-grating dynamics more closely matched the rigid-grating dynamics, suggesting that thermal effects play a substantial role in the lightsail shape stability. This is unsurprising, given that the authors' model incorporates, for example, linearized thermal expansion, which strongly influences the optical cross sections of each triangle. The results illustrate the complexity of the lightsail problem, requiring a full accounting of all propulsion, thermal and stability effects to gauge the success of a particular design. In future, all of these aspects must be optimized simultaneously to meet the mission objectives. This will be challenging given that the software developed for the multiphysics flexible-sail simulations~\cite{Gao:2024aa} already requires substantial computing power~\cite{kelzenberg:flexible_github}.

\textbf{Laser-beam instabilities.} In the lightsail mission, the laser beam presents a potentially significant source of instability. Spatial or temporal fluctuations in the beam intensity can disrupt the carefully balanced sail scattering, contributing to body-motion instability that is different than that of a simple displacement or rotation in the beam. 
One way to counteract local-beam fluctuations is to design a curved sail ({\em e.g.} hyperboloid) such that a potentially stable equilibrium exists despite the asymmetry~\cite{Srinivasan:2016aa}. Another method is to use a nonlinear sail material to harness the high laser-beam intensity inherent to the mission~\cite{Myilswamy:2020aa}. By designing a photonic crystal (Figure~\ref{fig:flexible_lit}c) to support a guided-resonance mode and exploit the Kerr nonlinearity, the sail's reflected intensity remains nearly constant for a range of input intensities (Figure~\ref{fig:flexible_lit}d), showing resilience against local beam fluctuations. However, such a scheme relies on photonic crystal resonances, which must be sufficiently broadband in the lightsail mission due to the Doppler shift. 

Some sails that are stable relative to displacement and rotation are also robust to small spatio-temporal perturbations in the beam profile: simulations showed adding white noise to the beam did not jeopardize stability for specific rigid, metasurface-based sails~\cite{Salary:2020aa}. However, this may be serendipitous and warrants more systematic study, in particular for flexible sails taking into account thermal effects. Any such study will also require a realistic model for sources of beam perturbations, including imperfect compensation of atmospheric fluctuations, active compensation for Earth's rotation and residual laser noise.

\section{Experiments\label{sec:experiments}}
Most of the work discussed so far has been theoretical and numerical. Experimental verification of many aspects of lightsails, such as deployment and stability, are difficult to achieve in laboratories subject to Earth's gravity, and may require extremely powerful lasers and extreme vacuum chambers. Many of the proposed structures are not yet able to be fabricated on the scales required, or rely on material properties that are insufficiently characterized~\cite{Atwater:2018aa}. Thus, before full sails can be made, let alone tested, it is imperative that experimental characterizations that can be achieved on Earth be conducted. To name a few, tests are needed for: more complete characterization of linear and nonlinear optical properties of candidate materials over the broad NIR-Doppler band and MIR band, and over the full range of temperatures a sail may encounter; initially small scale, then large scale nanostructure fabrication followed by complete optical characterization of their scattering; structural tests and; direct radiation-pressure measurements. In particular, low defect, high purity synthesis and nanostructuring of materials over square-meter scales will require substantial technological advancements, especially for the intricate designs found by inverse design that are often unintuitive. 

\begin{figure*}[htb]
    \centering
    \includegraphics[width=0.85\linewidth]{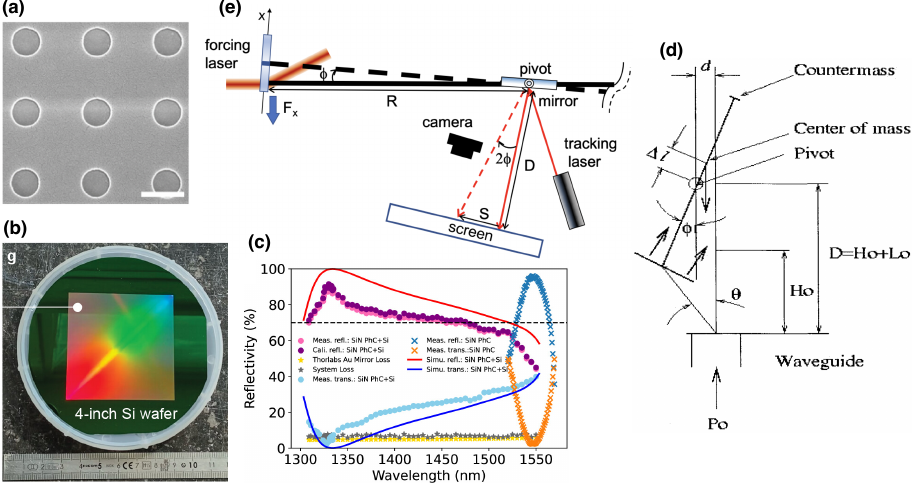}
    \caption{Lightsail experimental work. (a) Scanning-electron microscope image of fabricated \ch{Si3N4} photonic crystal. The white scale bar denotes 500~nm. Reproduced from Reference~\cite{Lien:2022aa}. Copyright 2022 Optica Publishing Group. (b-c) Image of fabricated \ch{Si3N4} circular-hole photonic crystal on a Si wafer, and measured reflectivity spectrum. Reproduced from Reference~\cite{Chang:2024aa}. Copyright 2024 Americal Chemical Society. (d) Balancing torque on a flat membrane connected to a countermass. Reproduced from Reference~\cite{Benford:2002aa}, with the permission of AIP Publishing. Copyright 2002 AIP Publishing. (e) Torsional pendulum setup. Reproduced from Reference~\cite{Chu:2019aa} with permission. \href{https://doi.org/10.1103/PhysRevLett.123.244302}{Copyright 2019 American Physical Society}.}
    \label{fig:experiments}
\end{figure*}

\subsection{Reflection and emissivity measurements}

The most immediate tests needed to characterize lightsail propulsion and thermal performance are specular reflection and emissivity measurements, respectively, to ensure theoretical lightsail designs can reproduce desired optical properties in practice. In one instance, a candidate \ch{Si3N4} photonic crystal sail design was fabricated using electron-beam lithography, off-normal electron-beam metal evaporation and reactive-ion etching~\cite{Lien:2022aa}. The final structure is shown in Figure~\ref{fig:experiments}a. By coating the \ch{Si3N4} photonic crystal with \ch{TiO2}, experimenters demonstrated the ability to spectrally shift the lightsail reflectance, which is important for tuning the maximal reflection towards the Doppler band for an as-yet-unknown laser-operating wavelength. However, the added mass from the \ch{TiO2} layer severely increases the acceleration distance (see Table~\ref{tab:acceleration_distance}). Other coatings like polydimethylsiloxane significantly enhance the multilayer-sail emissivity in the MIR wavelengths at the cost of adding substantial mass to the sail~\cite{Lien:2022aa}, resulting in a nearly five-fold increase of acceleration distance. The large acceleration distance can be attributed to the large membrane thickness compared to earlier numerically simulated designs, which can be seen across Table~\ref{tab:acceleration_distance}. The large thickness was chosen to accommodate \ch{Si3N4} Fabry-Perot reflections in the NIR-Doppler band, but the design ultimately suffered because \ch{Si3N4} has a lower refractive index and thus lower average reflectance than Si. Other investigations into \ch{Si3N4} photonic crystals optimized for maximal thrust and minimized cost are being performed, finding agreement between reflectivity simulations and experimental measurements~\cite{Norder:2024aa}. The increasing popularity of \ch{Si3N4} as a lightsail emissive-layer candidate has led to newer efforts for accurately measuring its absorption at proposed laser wavelengths~\cite{Feng:2024aa}.

One advantage of multilayers is the ability to overcome the inherently narrow-band and resonant nature of 2D photonic crystals' spectral reflectivity. For example, by combining a carefully designed \ch{Si3N4} photonic crystal with a Si substrate, more than 70\% reflection was measured experimentally over almost 200~nm bandwidth, corresponding to almost 70\% of the Doppler band initiated at 1300~nm~\cite{Chang:2024aa}. An image of the fabricated structure and measured reflectivity spectrum are shown in Figure~\ref{fig:experiments}b and~\ref{fig:experiments}c, respectively.

\subsection{Non-specular reflections}
For lightsail stability, experiments to measure angle-dependent reflections from nanostructured membranes are necessary to infer the optical forces and torques, and hence determine if restoring motion is possible. For a 1D \ch{Si3N4} grating, experimenters were able to find strong agreement between theory and experiment for both the diffracted angles (within $\pm\ang{1}$ of the angles predicted by eq~\eqref{eq:grating_equation}) and the functional form of the diffraction efficiencies over angle~\cite{Gao:2022aa}. Compared to finite-element-method-simulated diffraction efficiencies, the measured $r_{\pm1}$ efficiencies showed good agreement, but other orders showed deviations with magnitude up to 0.1. The discrepancy was attributed to fabrication imperfections and unwanted scattering from supporting anchor bars. This highlights one of the significant challenges to the lightsail mission: modern electromagnetic solvers can calculate the optical response of arbitrary structures to any desired accuracy, so the limiting factor for lightsail designs is the sophistication of fabrication techniques. Once such limitations are overcome, complete optical characterization of candidate sail designs are needed to ensure the correct scattering profiles for stability. In the case of Reference~\cite{Gao:2022aa}, the optical forces and torques inferred from the measured efficiencies and angles were integrated for approximately $\SI{5}{\second}$ (assuming $\SI{1}{\gram}$ sail mass and $\SI{10}{\giga\watt\per\square\meter}$ laser intensity) using the bigrating equations of motion~\cite{Ilic:2019aa}. For a region of initial conditions within $\pm\ang{10}$ sail tilt and $\pm25\%$ of the laser-beam width in lateral displacement, the simulated sail (combined with the experimental diffraction efficiencies) showed bounded dynamics, showcasing the promise of nanostructured designs for stable beam riding.

\subsection{Radiation pressure}
While radiation pressure and torques can be inferred by measuring scattering from grating-like structures, direct radiation-pressure measurements are also required to characterize the restoring motion. 

The earliest measurements of radiation pressure on sail-like membranes were conducted using carbon foils accelerated by a $\SI{150}{\kilo\watt}$ \ch{CO2} laser, where the foils were restricted to a guide wire in a large vacuum chamber~\cite{Myrabo:2002aa}. Sails were comprised of carbon truss disks ($\SI{5}{cm}$ diameter, order $\SI{10}{mg}$ mass) coated with molybdenum for high reflectivity at $\SI{10.6}{\micro\meter}$, with some experimental runs showing successful hovering and even acceleration against the Earth's gravity. In these experiments, a substantial difficulty was ensuring that propulsion was purely radiative, rather than from ablation of the sail: the sail samples reached temperatures in excess of $\SI{2500}{\kelvin}$ and a mass reduction was observed, suggesting propulsion was at least in part due to ablation. 

At around the same time, measurements of sail-stabilizing radiation pressure were carried out using microwave beams~\cite{Benford:2002aa,Benford:2003aa}. It was shown that a beam with spatial-intensity gradient could provide a balancing torque on a sail attached to a pivot and countermass (see Figure~\ref{fig:experiments}d). When the sail was rotated and displaced from its unstable equilibrium (where the countermass is directly over the sail and centered on the microwave-beam axis), there was a critical power at which the pendulum does not move, demonstrating the balancing torque.

Measuring radiation pressure of sails directly is notoriously difficult (as seen in, for instance, the carbon foil experiments~\cite{Myrabo:2002aa}), and typically requires ultra-high vacuum to avoid thermal or convective effects. The experimental setup of choice remains the torsional pendulum~\cite{Chu:2019aa,Chu:2021aa,Tang:2024aa} as pioneered by Lebedew and Nichols~\cite{Lebedew:1901,Nichols:1903_2}, an example of which is depicted in Figure~\ref{fig:experiments}e. 
To measure the photonic restoring force of a diffractive sail, Chu {\em et al.}~\cite{Chu:2019aa} attached a diffractive sheet (representing the lightsail) to one end of a torsional pendulum placed in a vacuum chamber. The other end of the pendulum is attached to a balancing mass, and the pendulum arms are aligned parallel to the floor. A mirror is attached at the pivot and a tracking laser pointed towards it, which, upon reflection, records the rotation of the pendulum and thus the (linear) displacement of the diffractive sheet. 
Once the pump laser illuminates the sail sample, the pendulum oscillates at a  frequency that differs from the natural oscillation frequency of the filament comprising the pendulum arm because the restoring force from the radiation pressure modifies the torsional spring constant and thus the oscillation frequency. The predicted increase in oscillation frequency showed close agreement with the measured value, serving as a demonstration of the restoring force created by diffractive sails.

Further experimental radiation-pressure measurements will be of fundamental importance for testing effects that appear almost uniquely in the lightsail mission. For instance, an important regime in the lightsail flight is when the laser beam partially misses the sail, resulting in the beam ``spilling'' over the edge of the reflector. Forces due to diffraction at edges are poorly understood: a theoretical study would require integration of the MST over large objects, which has been avoided by almost all theoretical and experimental investigations to date.  A recent experimental setup~\cite{Michaeli:2025aa} using a 50~nm \ch{Si3N4} membrane suspended by compliant springs allowed researchers to study the edge-diffraction regime for the first time. The setup enabled measurement of radiation pressures as small as \SI{70}{\femto\newton} using a collimated \SI{110}{\watt\per\square\centi\meter} beam. Using a modified apparatus, displacements due to {\em lateral} restoring forces on a metagrating were also measured~\cite{Michaeli:2025aa}, marking a key milestone in experimental tests for lightsails.

\section{Conclusions\label{sec:conclusion}}
Substantial scientific and technological advancements have put lightsails as the frontrunner for feasible interstellar spacecraft, making the lightsail endeavor one of the most promising and exciting scientific pursuits in the twenty-first century.  Looking forward, further progress must be achieved across numerous fields of science in order to bring lightsails to the next stage of viability -- and photonics lies at the core of the challenges, be it for the laser, the sail or communications. 

Regarding the sail, the most immediate technological barrier to the mission is the availability of candidate materials and nanofabrication methods. Even with a shortlist of sail materials, some of which were highlighted in this review, significant investigations into their complete mechanical, thermal and optical properties over a broad temperature range are of utmost importance, especially once they are worked into sub-micron membranes. 
Such data is currently unavailable, but serves as a fundamental first step towards identifying mission-ready materials. 

The next step in the endeavor is to finalize the design of the sail membrane. Simple stacked slabs or parachute-like geometric designs demonstrated modest potential for satisfying the mission's high-thrust and stable-beam-riding requirements, but are ultimately outclassed by nanostructured designs such as diffraction gratings and metasurfaces. The substantial advantage of nanophotonic architectures is the extensive design parameter space, in principle allowing multiple mission requirements such as thermal control, propulsion and stability to be addressed concurrently. However, it is only with recent advances in numerical optimization and inverse design that exploration of the full parameter space can begin. Designs optimized with such techniques have proven themselves as the leaders in propulsion and stability metrics, but there is clearly space for substantial improvement, especially for the thus-far unseen asymptotic stability. With the many constraints on sails, optimization must be performed over the entire problem space for lightsails simultaneously, including all aspects of propulsion, thermal management, stability, communication and more. Such optimization enables designs with adequate performance in all areas to be found, but poses the problem of defining figures of merit that can accurately capture the interplay between all aspects. 

Here, we stress the importance of the sail-stability problem to the overall success of the project. While substantial progress has been made in this front, there are several significant extensions that must be considered in future lightsail models. The most pressing extensions include the flexibility of the extremely thin sail membrane, and the rigorous incorporation of special relativity into sail dynamics. Both aspects have scarcely been addressed in lightsail literature thus far, yet are predicted to play a key role in the complex sail motion. From existing research on these topics, it has become apparent that photonic designs are the top contenders due to their robustness and versatility. Indeed, gratings and metasurfaces have proven their ability to readily exploit changes in optical conditions that are created by distorted sail shapes or generated by dispersive effects arising from near-relativistic velocities.

While the long-term goal of interstellar probes is the most inspiring application of laser-driven lightsails, there are smaller-scale missions in interplanetary exploration and precursor flights that can justify further research with tangible, short-term outcomes~\cite{Parkin:2024aa}. Nanophotonic {\em solar} sails, which do not require a ground-based laser, have also started attracting attention as a low-cost alternative for interplanetary exploration, and use many of the same principles discussed here~\cite{Chu:2024aa,Srivastava:2023aa,Srivastava:2023ab,Zhang:2022aa,Davoyan:2021aa,ullery:2018,swartzlander:2022}.

The immensity of the interstellar lightsail undertaking means that decades may pass before they come to fruition, depending on the development of several technologies. However, even without a single coordinated effort to create and launch a lightsail, advances in the requisite fields will naturally occur. For instance, progress in metalenses driven by mobile sensors, high-power laser miniaturization in industrial and defence industries, and continual innovation in fast numerical optimization techniques all have broad applications across disciplines. On the other hand, if lightsails gain significant interest in coming years as one of the most viable options for interstellar probes, then advancements in lightsails will symbiotically create breakthroughs in numerous fields. We see the lightsail project shifting the paradigm of space exploration in the twenty-first century, inspiring generations of scientists at the boundaries of scientific and technical capabilities.

\begin{acknowledgement}
This research was supported by an Australian Government Research Training Program (RTP) Scholarship.

\end{acknowledgement}




\bibliography{lightsails}

\providecommand{\latin}[1]{#1}
\makeatletter
\providecommand{\doi}
  {\begingroup\let\do\@makeother\dospecials
  \catcode`\{=1 \catcode`\}=2 \doi@aux}
\providecommand{\doi@aux}[1]{\endgroup\texttt{#1}}
\makeatother
\providecommand*\mcitethebibliography{\thebibliography}
\csname @ifundefined\endcsname{endmcitethebibliography}  {\let\endmcitethebibliography\endthebibliography}{}
\begin{mcitethebibliography}{139}
\providecommand*\natexlab[1]{#1}
\providecommand*\mciteSetBstSublistMode[1]{}
\providecommand*\mciteSetBstMaxWidthForm[2]{}
\providecommand*\mciteBstWouldAddEndPuncttrue
  {\def\EndOfBibitem{\unskip.}}
\providecommand*\mciteBstWouldAddEndPunctfalse
  {\let\EndOfBibitem\relax}
\providecommand*\mciteSetBstMidEndSepPunct[3]{}
\providecommand*\mciteSetBstSublistLabelBeginEnd[3]{}
\providecommand*\EndOfBibitem{}
\mciteSetBstSublistMode{f}
\mciteSetBstMaxWidthForm{subitem}{(\alph{mcitesubitemcount})}
\mciteSetBstSublistLabelBeginEnd
  {\mcitemaxwidthsubitemform\space}
  {\relax}
  {\relax}

\bibitem[Tsiolkovsky(1903)]{tsiolkovsky1903rocket}
Tsiolkovsky,~K.~E. Exploration of Outer Space by Means of Rocket Devices. \emph{Scientific Review} \textbf{1903}, Original in Russian\relax
\mciteBstWouldAddEndPuncttrue
\mciteSetBstMidEndSepPunct{\mcitedefaultmidpunct}
{\mcitedefaultendpunct}{\mcitedefaultseppunct}\relax
\EndOfBibitem
\bibitem[Lubin(2016)]{Lubin:2016aa}
Lubin,~P. A Roadmap to Interstellar Flight. \emph{Journal of the British Interplanetary Society} \textbf{2016}, \emph{69}, 40--72\relax
\mciteBstWouldAddEndPuncttrue
\mciteSetBstMidEndSepPunct{\mcitedefaultmidpunct}
{\mcitedefaultendpunct}{\mcitedefaultseppunct}\relax
\EndOfBibitem
\bibitem[Lubin \latin{et~al.}(2024)Lubin, Cohen, Meinhold, Srinivasan, Rupert, and Krogen]{Lubin:2024aa}
Lubin,~P.; Cohen,~A.~N.; Meinhold,~P.; Srinivasan,~P.; Rupert,~N.; Krogen,~P. In \emph{Laser Propulsion in Space}; Phipps,~C., Ed.; Elsevier, 2024; pp 205--225\relax
\mciteBstWouldAddEndPuncttrue
\mciteSetBstMidEndSepPunct{\mcitedefaultmidpunct}
{\mcitedefaultendpunct}{\mcitedefaultseppunct}\relax
\EndOfBibitem
\bibitem[Maxwell(1865)]{Maxwell:1865_dynamical}
Maxwell,~J.~C. VIII. A dynamical theory of the electromagnetic field. \emph{Philosophical Transactions of the Royal Society of London} \textbf{1865}, \emph{155}, 459--512\relax
\mciteBstWouldAddEndPuncttrue
\mciteSetBstMidEndSepPunct{\mcitedefaultmidpunct}
{\mcitedefaultendpunct}{\mcitedefaultseppunct}\relax
\EndOfBibitem
\bibitem[Lebedew(1901)]{Lebedew:1901}
Lebedew,~P. Untersuchungen über die Druckkräfte des Lichtes. \emph{Annalen der Physik} \textbf{1901}, \emph{311}, 433--458\relax
\mciteBstWouldAddEndPuncttrue
\mciteSetBstMidEndSepPunct{\mcitedefaultmidpunct}
{\mcitedefaultendpunct}{\mcitedefaultseppunct}\relax
\EndOfBibitem
\bibitem[Nichols and Hull(1903)Nichols, and Hull]{Nichols:1903_2}
Nichols,~E.~F.; Hull,~G.~F. The Pressure Due to Radiation. (Second Paper.). \emph{Physical Review (Series I)} \textbf{1903}, \emph{17}, 26--50, PRI\relax
\mciteBstWouldAddEndPuncttrue
\mciteSetBstMidEndSepPunct{\mcitedefaultmidpunct}
{\mcitedefaultendpunct}{\mcitedefaultseppunct}\relax
\EndOfBibitem
\bibitem[Ashkin(1970)]{ashkin:1970}
Ashkin,~A. Acceleration and Trapping of Particles by Radiation Pressure. \emph{Physical Review Letters} \textbf{1970}, \emph{24}, 156--159, PRL\relax
\mciteBstWouldAddEndPuncttrue
\mciteSetBstMidEndSepPunct{\mcitedefaultmidpunct}
{\mcitedefaultendpunct}{\mcitedefaultseppunct}\relax
\EndOfBibitem
\bibitem[Hänsch and Schawlow(1975)Hänsch, and Schawlow]{hansch:1975}
Hänsch,~T.~W.; Schawlow,~A.~L. Cooling of gases by laser radiation. \emph{Optics Communications} \textbf{1975}, \emph{13}, 68--69\relax
\mciteBstWouldAddEndPuncttrue
\mciteSetBstMidEndSepPunct{\mcitedefaultmidpunct}
{\mcitedefaultendpunct}{\mcitedefaultseppunct}\relax
\EndOfBibitem
\bibitem[Chu \latin{et~al.}(1986)Chu, Bjorkholm, Ashkin, and Cable]{chu:1986}
Chu,~S.; Bjorkholm,~J.~E.; Ashkin,~A.; Cable,~A. Experimental Observation of Optically Trapped Atoms. \emph{Physical Review Letters} \textbf{1986}, \emph{57}, 314--317, PRL\relax
\mciteBstWouldAddEndPuncttrue
\mciteSetBstMidEndSepPunct{\mcitedefaultmidpunct}
{\mcitedefaultendpunct}{\mcitedefaultseppunct}\relax
\EndOfBibitem
\bibitem[Aspelmeyer \latin{et~al.}(2014)Aspelmeyer, Kippenberg, and Marquardt]{Aspelmeyer:2014aa}
Aspelmeyer,~M.; Kippenberg,~T.~J.; Marquardt,~F. Cavity optomechanics. \emph{Reviews of Modern Physics} \textbf{2014}, \emph{86}, 1391--1452, RMP\relax
\mciteBstWouldAddEndPuncttrue
\mciteSetBstMidEndSepPunct{\mcitedefaultmidpunct}
{\mcitedefaultendpunct}{\mcitedefaultseppunct}\relax
\EndOfBibitem
\bibitem[Kippenberg and Vahala(2008)Kippenberg, and Vahala]{Kippenberg:2008aa}
Kippenberg,~T.~J.; Vahala,~K.~J. Cavity Optomechanics: Back-Action at the Mesoscale. \emph{Science} \textbf{2008}, \emph{321}, 1172--1176\relax
\mciteBstWouldAddEndPuncttrue
\mciteSetBstMidEndSepPunct{\mcitedefaultmidpunct}
{\mcitedefaultendpunct}{\mcitedefaultseppunct}\relax
\EndOfBibitem
\bibitem[Mori \latin{et~al.}(2010)Mori, Sawada, Funase, Morimoto, Endo, Yamamoto, Tsuda, Kawakatsu, Kawaguchi, Miyazaki, and Shirasawa]{Mori:2010aa}
Mori,~O.; Sawada,~H.; Funase,~R.; Morimoto,~M.; Endo,~T.; Yamamoto,~T.; Tsuda,~Y.; Kawakatsu,~Y.; Kawaguchi,~J.; Miyazaki,~Y.; Shirasawa,~Y. First Solar Power Sail Demonstration by IKAROS. \emph{Transactions of the Japan Society for Aeronautical and Space Sciences, Aerospace Technology Japan} \textbf{2010}, \emph{8}\relax
\mciteBstWouldAddEndPuncttrue
\mciteSetBstMidEndSepPunct{\mcitedefaultmidpunct}
{\mcitedefaultendpunct}{\mcitedefaultseppunct}\relax
\EndOfBibitem
\bibitem[Vulpetti(1996)]{Vulpetti:1996}
Vulpetti,~G. 3D high-speed escape heliocentric trajectories by all-metallic-sail low-mass sailcraft. \emph{Acta Astronautica} \textbf{1996}, \emph{39}, 161--170\relax
\mciteBstWouldAddEndPuncttrue
\mciteSetBstMidEndSepPunct{\mcitedefaultmidpunct}
{\mcitedefaultendpunct}{\mcitedefaultseppunct}\relax
\EndOfBibitem
\bibitem[Bailer-Jones(2021)]{bailer-jones:2021}
Bailer-Jones,~C. A.~L. The sun diver: Combining solar sails with the Oberth effect. \emph{American Journal of Physics} \textbf{2021}, \emph{89}, 235--243\relax
\mciteBstWouldAddEndPuncttrue
\mciteSetBstMidEndSepPunct{\mcitedefaultmidpunct}
{\mcitedefaultendpunct}{\mcitedefaultseppunct}\relax
\EndOfBibitem
\bibitem[Karlapp \latin{et~al.}(2024)Karlapp, Heller, and Tajmar]{Karlapp:2024}
Karlapp,~J.; Heller,~R.; Tajmar,~M. Ultrafast transfer of low-mass payloads to Mars and beyond using aerographite solar sails. \emph{Acta Astronautica} \textbf{2024}, \emph{219}, 889--895\relax
\mciteBstWouldAddEndPuncttrue
\mciteSetBstMidEndSepPunct{\mcitedefaultmidpunct}
{\mcitedefaultendpunct}{\mcitedefaultseppunct}\relax
\EndOfBibitem
\bibitem[Marx(1966)]{Marx:1966aa}
Marx,~G. Interstellar Vehicle Propelled By Terrestrial Laser Beam. \emph{Nature} \textbf{1966}, \emph{211}, 22--23\relax
\mciteBstWouldAddEndPuncttrue
\mciteSetBstMidEndSepPunct{\mcitedefaultmidpunct}
{\mcitedefaultendpunct}{\mcitedefaultseppunct}\relax
\EndOfBibitem
\bibitem[Forward(1984)]{Forward:1984aa}
Forward,~R.~L. Roundtrip interstellar travel using laser-pushed lightsails. \emph{Journal of Spacecraft and Rockets} \textbf{1984}, \emph{21}, 187--195\relax
\mciteBstWouldAddEndPuncttrue
\mciteSetBstMidEndSepPunct{\mcitedefaultmidpunct}
{\mcitedefaultendpunct}{\mcitedefaultseppunct}\relax
\EndOfBibitem
\bibitem[Redding(1967)]{Redding:1967aa}
Redding,~J.~L. Interstellar Vehicle propelled by Terrestrial Laser Beam. \emph{Nature} \textbf{1967}, \emph{213}, 588--589\relax
\mciteBstWouldAddEndPuncttrue
\mciteSetBstMidEndSepPunct{\mcitedefaultmidpunct}
{\mcitedefaultendpunct}{\mcitedefaultseppunct}\relax
\EndOfBibitem
\bibitem[{Breakthrough Foundation}(2023)]{Starshot:2024}
{Breakthrough Foundation} Starshot. 2023; \url{https://breakthroughinitiatives.org/initiative/3}\relax
\mciteBstWouldAddEndPuncttrue
\mciteSetBstMidEndSepPunct{\mcitedefaultmidpunct}
{\mcitedefaultendpunct}{\mcitedefaultseppunct}\relax
\EndOfBibitem
\bibitem[Parkin(2018)]{Parkin:2018aa}
Parkin,~K. L.~G. The Breakthrough Starshot system model. \emph{Acta Astronautica} \textbf{2018}, \emph{152}, 370--384\relax
\mciteBstWouldAddEndPuncttrue
\mciteSetBstMidEndSepPunct{\mcitedefaultmidpunct}
{\mcitedefaultendpunct}{\mcitedefaultseppunct}\relax
\EndOfBibitem
\bibitem[Parkin(2024)]{Parkin:2024aa}
Parkin,~K. L.~G. In \emph{Laser Propulsion in Space}; Phipps,~C., Ed.; Elsevier, 2024; pp 71--121\relax
\mciteBstWouldAddEndPuncttrue
\mciteSetBstMidEndSepPunct{\mcitedefaultmidpunct}
{\mcitedefaultendpunct}{\mcitedefaultseppunct}\relax
\EndOfBibitem
\bibitem[Tung and Davoyan(2022)Tung, and Davoyan]{Tung:2022aa}
Tung,~H.-T.; Davoyan,~A.~R. Low-Power Laser Sailing for Fast-Transit Space Flight. \emph{Nano Letters} \textbf{2022}, \emph{22}, 1108--1114, doi: 10.1021/acs.nanolett.1c04188\relax
\mciteBstWouldAddEndPuncttrue
\mciteSetBstMidEndSepPunct{\mcitedefaultmidpunct}
{\mcitedefaultendpunct}{\mcitedefaultseppunct}\relax
\EndOfBibitem
\bibitem[Kulkarni \latin{et~al.}(2018)Kulkarni, Lubin, and Zhang]{Kulkarni:2018aa}
Kulkarni,~N.; Lubin,~P.; Zhang,~Q. Relativistic Spacecraft Propelled by Directed Energy. \emph{The Astronomical Journal} \textbf{2018}, \emph{155}, 155--155\relax
\mciteBstWouldAddEndPuncttrue
\mciteSetBstMidEndSepPunct{\mcitedefaultmidpunct}
{\mcitedefaultendpunct}{\mcitedefaultseppunct}\relax
\EndOfBibitem
\bibitem[Worden \latin{et~al.}(2021)Worden, Green, Schalkwyk, Parkin, and Fugate]{Worden:2021aa}
Worden,~S.~P.; Green,~W.~A.; Schalkwyk,~J.; Parkin,~K.; Fugate,~R.~Q. Progress on the Starshot laser propulsion system. \emph{Applied Optics} \textbf{2021}, \emph{60}, H20--H23\relax
\mciteBstWouldAddEndPuncttrue
\mciteSetBstMidEndSepPunct{\mcitedefaultmidpunct}
{\mcitedefaultendpunct}{\mcitedefaultseppunct}\relax
\EndOfBibitem
\bibitem[Worden \latin{et~al.}(2024)Worden, Bandutunga, Sibley, Ireland, and Schalkwyk]{Worden:2024aa}
Worden,~S.~P.; Bandutunga,~C.; Sibley,~P.; Ireland,~M.; Schalkwyk,~J. In \emph{Laser Propulsion in Space}; Phipps,~C., Ed.; Elsevier, 2024; pp 39--70\relax
\mciteBstWouldAddEndPuncttrue
\mciteSetBstMidEndSepPunct{\mcitedefaultmidpunct}
{\mcitedefaultendpunct}{\mcitedefaultseppunct}\relax
\EndOfBibitem
\bibitem[Bandutunga \latin{et~al.}(2021)Bandutunga, Sibley, Ireland, and Ward]{Bandutunga:2021aa}
Bandutunga,~C.~P.; Sibley,~P.~G.; Ireland,~M.~J.; Ward,~R.~L. Photonic solution to phase sensing and control for light-based interstellar propulsion. \emph{Journal of the Optical Society of America B} \textbf{2021}, \emph{38}, 1477--1486\relax
\mciteBstWouldAddEndPuncttrue
\mciteSetBstMidEndSepPunct{\mcitedefaultmidpunct}
{\mcitedefaultendpunct}{\mcitedefaultseppunct}\relax
\EndOfBibitem
\bibitem[Sibley \latin{et~al.}(2020)Sibley, Ward, Roberts, Francis, and Shaddock]{Sibley:2020aa}
Sibley,~P.~G.; Ward,~R.~L.; Roberts,~L.~E.; Francis,~S.~P.; Shaddock,~D.~A. Crosstalk reduction for multi-channel optical phase metrology. \emph{Optics Express} \textbf{2020}, \emph{28}, 10400--10424\relax
\mciteBstWouldAddEndPuncttrue
\mciteSetBstMidEndSepPunct{\mcitedefaultmidpunct}
{\mcitedefaultendpunct}{\mcitedefaultseppunct}\relax
\EndOfBibitem
\bibitem[Peretz \latin{et~al.}(2022)Peretz, Mather, Hamilton, Pabarcius, Hall, Fugate, Green, and Klupar]{Peretz:2022aa}
Peretz,~E.; Mather,~J.; Hamilton,~C.; Pabarcius,~L.; Hall,~K.; Fugate,~R.; Green,~W.; Klupar,~P. Orbiting laser configuration and sky coverage: coherent reference for Breakthrough Starshot ground-based laser array. \emph{Journal of Astronomical Telescopes, Instruments, and Systems} \textbf{2022}, \emph{8}, 017004\relax
\mciteBstWouldAddEndPuncttrue
\mciteSetBstMidEndSepPunct{\mcitedefaultmidpunct}
{\mcitedefaultendpunct}{\mcitedefaultseppunct}\relax
\EndOfBibitem
\bibitem[Yoshida \latin{et~al.}(2023)Yoshida, Katsuno, Inoue, Gelleta, Izumi, De~Zoysa, Ishizaki, and Noda]{yoshida:PCSEL2023}
Yoshida,~M.; Katsuno,~S.; Inoue,~T.; Gelleta,~J.; Izumi,~K.; De~Zoysa,~M.; Ishizaki,~K.; Noda,~S. High-brightness scalable continuous-wave single-mode photonic-crystal laser. \emph{Nature} \textbf{2023}, \emph{618}, 727--732\relax
\mciteBstWouldAddEndPuncttrue
\mciteSetBstMidEndSepPunct{\mcitedefaultmidpunct}
{\mcitedefaultendpunct}{\mcitedefaultseppunct}\relax
\EndOfBibitem
\bibitem[Taghavi \latin{et~al.}(2022)Taghavi, Salary, and Mosallaei]{Taghavi:2022ab}
Taghavi,~M.; Salary,~M.~M.; Mosallaei,~H. Multifunctional metasails for self-stabilized beam-riding and optical communication. \emph{Nanoscale Advances} \textbf{2022}, \emph{4}, 1727--1740\relax
\mciteBstWouldAddEndPuncttrue
\mciteSetBstMidEndSepPunct{\mcitedefaultmidpunct}
{\mcitedefaultendpunct}{\mcitedefaultseppunct}\relax
\EndOfBibitem
\bibitem[Taghavi \latin{et~al.}(2024)Taghavi, Sabri, and Mosallaei]{Taghavi:2024aa}
Taghavi,~M.; Sabri,~R.; Mosallaei,~H. Active Metasurfaces for Non-Rigid Light Sail Interstellar Optical Communication. \emph{Advanced Theory and Simulations} \textbf{2024}, \emph{7}, 2300359--2300359\relax
\mciteBstWouldAddEndPuncttrue
\mciteSetBstMidEndSepPunct{\mcitedefaultmidpunct}
{\mcitedefaultendpunct}{\mcitedefaultseppunct}\relax
\EndOfBibitem
\bibitem[Mauskopf()]{Mauskopf:2023aa}
Mauskopf,~P. Communications receiver designs for interstellar probe missions. Proceedings of the International Astronautical Congress, IAC. Export Date: 15 December 2024; Cited By: 0\relax
\mciteBstWouldAddEndPuncttrue
\mciteSetBstMidEndSepPunct{\mcitedefaultmidpunct}
{\mcitedefaultendpunct}{\mcitedefaultseppunct}\relax
\EndOfBibitem
\bibitem[Maxwell and Library(1873)Maxwell, and Library]{Maxwell:1873aa}
Maxwell,~J.~C.; Library,~B. \emph{A Treatise on Electricity and Magnetism: by James Clerk Maxwell}; Clarendon Press: Oxford, 1873; Vol.~1\relax
\mciteBstWouldAddEndPuncttrue
\mciteSetBstMidEndSepPunct{\mcitedefaultmidpunct}
{\mcitedefaultendpunct}{\mcitedefaultseppunct}\relax
\EndOfBibitem
\bibitem[Einstein(1905)]{Einstein:1905aa}
Einstein,~A. Zur Elektrodynamik bewegter K{\"o}rper. \emph{Annalen der Physik} \textbf{1905}, \emph{322}, 891--921\relax
\mciteBstWouldAddEndPuncttrue
\mciteSetBstMidEndSepPunct{\mcitedefaultmidpunct}
{\mcitedefaultendpunct}{\mcitedefaultseppunct}\relax
\EndOfBibitem
\bibitem[Jackson(1962)]{Jackson:1962aa}
Jackson,~J.~D. \emph{Classical Electrodynamics}; John Wiley \& Sons, Inc.: United States of America, 1962\relax
\mciteBstWouldAddEndPuncttrue
\mciteSetBstMidEndSepPunct{\mcitedefaultmidpunct}
{\mcitedefaultendpunct}{\mcitedefaultseppunct}\relax
\EndOfBibitem
\bibitem[Yessenov and Abouraddy(2023)Yessenov, and Abouraddy]{Yessenov:2023aa}
Yessenov,~M.; Abouraddy,~A.~F. Relativistic transformations of quasi-monochromatic paraxial optical beams. \emph{Physical Review A} \textbf{2023}, \emph{107}, 042221--042221\relax
\mciteBstWouldAddEndPuncttrue
\mciteSetBstMidEndSepPunct{\mcitedefaultmidpunct}
{\mcitedefaultendpunct}{\mcitedefaultseppunct}\relax
\EndOfBibitem
\bibitem[Rezunkov(2021)]{Rezunkov:2021aa}
Rezunkov,~Y. \emph{High Power Laser Propulsion}; Atomic, Optical and Plasma Physics; Springer Nature Switzerland AG: Cham, 2021; Vol. 116\relax
\mciteBstWouldAddEndPuncttrue
\mciteSetBstMidEndSepPunct{\mcitedefaultmidpunct}
{\mcitedefaultendpunct}{\mcitedefaultseppunct}\relax
\EndOfBibitem
\bibitem[Atwater \latin{et~al.}(2018)Atwater, Davoyan, Ilic, Jariwala, Sherrott, Went, Whitney, and Wong]{Atwater:2018aa}
Atwater,~H.~A.; Davoyan,~A.~R.; Ilic,~O.; Jariwala,~D.; Sherrott,~M.~C.; Went,~C.~M.; Whitney,~W.~S.; Wong,~J. Materials challenges for the Starshot lightsail. \emph{Nature Materials} \textbf{2018}, \emph{17}, 861--867\relax
\mciteBstWouldAddEndPuncttrue
\mciteSetBstMidEndSepPunct{\mcitedefaultmidpunct}
{\mcitedefaultendpunct}{\mcitedefaultseppunct}\relax
\EndOfBibitem
\bibitem[Ilic \latin{et~al.}(2018)Ilic, Went, and Atwater]{Ilic:2018aa}
Ilic,~O.; Went,~C.~M.; Atwater,~H.~A. Nanophotonic Heterostructures for Efficient Propulsion and Radiative Cooling of Relativistic Light Sails. \emph{Nano Letters} \textbf{2018}, \emph{18}, 5583--5589\relax
\mciteBstWouldAddEndPuncttrue
\mciteSetBstMidEndSepPunct{\mcitedefaultmidpunct}
{\mcitedefaultendpunct}{\mcitedefaultseppunct}\relax
\EndOfBibitem
\bibitem[Jin \latin{et~al.}(2020)Jin, Li, Orenstein, and Fan]{Jin:2020aa}
Jin,~W.; Li,~W.; Orenstein,~M.; Fan,~S. Inverse Design of Lightweight Broadband Reflector for Relativistic Lightsail Propulsion. \emph{ACS Photonics} \textbf{2020}, \emph{7}, 2350--2355\relax
\mciteBstWouldAddEndPuncttrue
\mciteSetBstMidEndSepPunct{\mcitedefaultmidpunct}
{\mcitedefaultendpunct}{\mcitedefaultseppunct}\relax
\EndOfBibitem
\bibitem[Kudyshev \latin{et~al.}(2022)Kudyshev, Kildishev, Shalaev, and Boltasseva]{Kudyshev:2022aa}
Kudyshev,~Z.~A.; Kildishev,~A.~V.; Shalaev,~V.~M.; Boltasseva,~A. Optimizing Startshot Lightsail Design: A Generative Network-Based Approach. \emph{ACS Photonics} \textbf{2022}, \emph{9}, 190--196\relax
\mciteBstWouldAddEndPuncttrue
\mciteSetBstMidEndSepPunct{\mcitedefaultmidpunct}
{\mcitedefaultendpunct}{\mcitedefaultseppunct}\relax
\EndOfBibitem
\bibitem[Santi \latin{et~al.}(2022)Santi, Favaro, Corso, Lubin, Bazzan, Ragazzoni, Garoli, and Pelizzo]{Santi:2022aa}
Santi,~G.; Favaro,~G.; Corso,~A.~J.; Lubin,~P.; Bazzan,~M.; Ragazzoni,~R.; Garoli,~D.; Pelizzo,~M.~G. Multilayers for directed energy accelerated lightsails. \emph{Communications Materials} \textbf{2022}, \emph{3}, 16--16\relax
\mciteBstWouldAddEndPuncttrue
\mciteSetBstMidEndSepPunct{\mcitedefaultmidpunct}
{\mcitedefaultendpunct}{\mcitedefaultseppunct}\relax
\EndOfBibitem
\bibitem[Chang \latin{et~al.}(2024)Chang, Ji, Yao, van Run, and Gr{\"o}blacher]{Chang:2024aa}
Chang,~J.; Ji,~W.; Yao,~X.; van Run,~A.~J.; Gr{\"o}blacher,~S. Broadband, High-Reflectivity Dielectric Mirrors at Wafer Scale: Combining Photonic Crystal and Metasurface Architectures for Advanced Lightsails. \emph{Nano Letters} \textbf{2024}, \emph{24}, 6689--6695\relax
\mciteBstWouldAddEndPuncttrue
\mciteSetBstMidEndSepPunct{\mcitedefaultmidpunct}
{\mcitedefaultendpunct}{\mcitedefaultseppunct}\relax
\EndOfBibitem
\bibitem[Green and Keevers(1995)Green, and Keevers]{Green:1995aa}
Green,~M.~A.; Keevers,~M.~J. Optical properties of intrinsic silicon at 300 K. \emph{Progress in Photovoltaics: Research and Applications} \textbf{1995}, \emph{3}, 189--192\relax
\mciteBstWouldAddEndPuncttrue
\mciteSetBstMidEndSepPunct{\mcitedefaultmidpunct}
{\mcitedefaultendpunct}{\mcitedefaultseppunct}\relax
\EndOfBibitem
\bibitem[Palik(1998)]{Palik:1998aa}
Palik,~E.~D. \emph{Handbook of Optical Constants of Solids}; Academic Pres: San Diego, 1998; Vol. 1-2\relax
\mciteBstWouldAddEndPuncttrue
\mciteSetBstMidEndSepPunct{\mcitedefaultmidpunct}
{\mcitedefaultendpunct}{\mcitedefaultseppunct}\relax
\EndOfBibitem
\bibitem[Khashan and Nassif(2001)Khashan, and Nassif]{Kashan:2001aa}
Khashan,~M.~A.; Nassif,~A.~Y. {Dispersion of the optical constants of quartz and polymethyl methacrylate glasses in a wide spectral range: 0.2–3 $\mu$m}. \emph{Optics Communications} \textbf{2001}, \emph{188}, 129--139\relax
\mciteBstWouldAddEndPuncttrue
\mciteSetBstMidEndSepPunct{\mcitedefaultmidpunct}
{\mcitedefaultendpunct}{\mcitedefaultseppunct}\relax
\EndOfBibitem
\bibitem[Franta \latin{et~al.}(2016)Franta, Nečas, Ohlídal, and Giglia]{Franta:2016aa}
Franta,~D.; Nečas,~D.; Ohlídal,~I.; Giglia,~A. \emph{Optical characterization of SiO2 thin films using universal dispersion model over wide spectral range}; SPIE Photonics Europe; SPIE, 2016; Vol. 9890\relax
\mciteBstWouldAddEndPuncttrue
\mciteSetBstMidEndSepPunct{\mcitedefaultmidpunct}
{\mcitedefaultendpunct}{\mcitedefaultseppunct}\relax
\EndOfBibitem
\bibitem[Polyanskiy(2024)]{Polyanskiy:2024aa}
Polyanskiy,~M.~N. Refractiveindex.info database of optical constants. \emph{Scientific Data} \textbf{2024}, \emph{11}, 94\relax
\mciteBstWouldAddEndPuncttrue
\mciteSetBstMidEndSepPunct{\mcitedefaultmidpunct}
{\mcitedefaultendpunct}{\mcitedefaultseppunct}\relax
\EndOfBibitem
\bibitem[Brewer \latin{et~al.}(2022)Brewer, Campbell, Kumar, Kulkarni, Jariwala, Bargatin, and Raman]{Brewer:2022aa}
Brewer,~J.; Campbell,~M.~F.; Kumar,~P.; Kulkarni,~S.; Jariwala,~D.; Bargatin,~I.; Raman,~A.~P. Multiscale Photonic Emissivity Engineering for Relativistic Lightsail Thermal Regulation. \emph{Nano Letters} \textbf{2022}, \emph{22}, 594--601\relax
\mciteBstWouldAddEndPuncttrue
\mciteSetBstMidEndSepPunct{\mcitedefaultmidpunct}
{\mcitedefaultendpunct}{\mcitedefaultseppunct}\relax
\EndOfBibitem
\bibitem[Lien \latin{et~al.}(2022)Lien, Meng, Liu, Sakib, Tang, Wu, and Povinelli]{Lien:2022aa}
Lien,~M.~R.; Meng,~D.; Liu,~Z.; Sakib,~M.~A.; Tang,~Y.; Wu,~W.; Povinelli,~M.~L. Experimental characterization of a silicon nitride photonic crystal light sail. \emph{Optical Materials Express} \textbf{2022}, \emph{12}, 3032--3042\relax
\mciteBstWouldAddEndPuncttrue
\mciteSetBstMidEndSepPunct{\mcitedefaultmidpunct}
{\mcitedefaultendpunct}{\mcitedefaultseppunct}\relax
\EndOfBibitem
\bibitem[Philipp(1973)]{Philipp:1973aa}
Philipp,~H.~R. Optical Properties of Silicon Nitride. \emph{Journal of The Electrochemical Society} \textbf{1973}, \emph{120}, 295\relax
\mciteBstWouldAddEndPuncttrue
\mciteSetBstMidEndSepPunct{\mcitedefaultmidpunct}
{\mcitedefaultendpunct}{\mcitedefaultseppunct}\relax
\EndOfBibitem
\bibitem[Kumar \latin{et~al.}()Kumar, Feng, Yin, Lin, Mah, Fortman, Jaffe, Wan, Fang, Warzoha, Brar, Talghader, and Kats]{Kumar:2023aa}
Kumar,~T.; Feng,~D.; Yin,~S.; Lin,~P.; Mah,~M.; Fortman,~M.; Jaffe,~G.~R.; Wan,~C.; Fang,~C.; Warzoha,~R.; Brar,~V.~W.; Talghader,~J.~J.; Kats,~M.~A. Photothermal Commonpath Interferometry of Silicon Nitride Membranes for Laser Light Sails. 2023 Conference on Lasers and Electro-Optics (CLEO). pp 1--2\relax
\mciteBstWouldAddEndPuncttrue
\mciteSetBstMidEndSepPunct{\mcitedefaultmidpunct}
{\mcitedefaultendpunct}{\mcitedefaultseppunct}\relax
\EndOfBibitem
\bibitem[Luke \latin{et~al.}(2015)Luke, Okawachi, Lamont, Gaeta, and Lipson]{Luke:2015aa}
Luke,~K.; Okawachi,~Y.; Lamont,~M. R.~E.; Gaeta,~A.~L.; Lipson,~M. Broadband mid-infrared frequency comb generation in a ${\rm Si_3N_4}$ microresonator. \emph{Optics Letters} \textbf{2015}, \emph{40}, 4823--4826\relax
\mciteBstWouldAddEndPuncttrue
\mciteSetBstMidEndSepPunct{\mcitedefaultmidpunct}
{\mcitedefaultendpunct}{\mcitedefaultseppunct}\relax
\EndOfBibitem
\bibitem[Song \latin{et~al.}(2019)Song, Gu, Fang, Chen, Jiang, Wang, Zhai, Ho, and Liu]{Song:2019aa}
Song,~B.; Gu,~H.; Fang,~M.; Chen,~X.; Jiang,~H.; Wang,~R.; Zhai,~T.; Ho,~Y.-T.; Liu,~S. Layer-Dependent Dielectric Function of Wafer-Scale 2D MoS2. \emph{Advanced Optical Materials} \textbf{2019}, \emph{7}, 1801250\relax
\mciteBstWouldAddEndPuncttrue
\mciteSetBstMidEndSepPunct{\mcitedefaultmidpunct}
{\mcitedefaultendpunct}{\mcitedefaultseppunct}\relax
\EndOfBibitem
\bibitem[Ermolaev \latin{et~al.}(2020)Ermolaev, Stebunov, Vyshnevyy, Tatarkin, Yakubovsky, Novikov, Baranov, Shegai, Nikitin, Arsenin, and Volkov]{Ermolaev:2020aa}
Ermolaev,~G.~A.; Stebunov,~Y.~V.; Vyshnevyy,~A.~A.; Tatarkin,~D.~E.; Yakubovsky,~D.~I.; Novikov,~S.~M.; Baranov,~D.~G.; Shegai,~T.; Nikitin,~A.~Y.; Arsenin,~A.~V.; Volkov,~V.~S. Broadband optical properties of monolayer and bulk ${\rm MoS_2}$. \emph{npj 2D Materials and Applications} \textbf{2020}, \emph{4}, 21\relax
\mciteBstWouldAddEndPuncttrue
\mciteSetBstMidEndSepPunct{\mcitedefaultmidpunct}
{\mcitedefaultendpunct}{\mcitedefaultseppunct}\relax
\EndOfBibitem
\bibitem[Feldman(1976)]{Feldman:1976aa}
Feldman,~J.~L. Elastic constants of ${\rm 2H-MoS_2}$ and ${\rm 2H-NbSe_2}$ extracted from measured dispersion curves and linear compressibilities. \emph{Journal of Physics and Chemistry of Solids} \textbf{1976}, \emph{37}, 1141--1144\relax
\mciteBstWouldAddEndPuncttrue
\mciteSetBstMidEndSepPunct{\mcitedefaultmidpunct}
{\mcitedefaultendpunct}{\mcitedefaultseppunct}\relax
\EndOfBibitem
\bibitem[Hammond(2016)]{Hammond:2016aa}
Hammond,~C. In \emph{CRC Handbook of Chemistry and Physics}, 96th ed.; Haynes,~W.~M., Bruno,~T.~J., Lide,~D.~R., Eds.; CRC Press/Taylor \& Francis: Boca Raton, 2016\relax
\mciteBstWouldAddEndPuncttrue
\mciteSetBstMidEndSepPunct{\mcitedefaultmidpunct}
{\mcitedefaultendpunct}{\mcitedefaultseppunct}\relax
\EndOfBibitem
\bibitem[Gao \latin{et~al.}(2024)Gao, Kelzenberg, and Atwater]{Gao:2024aa}
Gao,~R.; Kelzenberg,~M.~D.; Atwater,~H.~A. Dynamically stable radiation pressure propulsion of flexible lightsails for interstellar exploration. \emph{Nature Communications} \textbf{2024}, \emph{15}, 4203--4203\relax
\mciteBstWouldAddEndPuncttrue
\mciteSetBstMidEndSepPunct{\mcitedefaultmidpunct}
{\mcitedefaultendpunct}{\mcitedefaultseppunct}\relax
\EndOfBibitem
\bibitem[Davoyan \latin{et~al.}(2021)Davoyan, Munday, Tabiryan, Swartzlander, and Johnson]{Davoyan:2021aa}
Davoyan,~A.~R.; Munday,~J.~N.; Tabiryan,~N.; Swartzlander,~G.~A.; Johnson,~L. Photonic materials for interstellar solar sailing. \emph{Optica} \textbf{2021}, \emph{8}, 722--734\relax
\mciteBstWouldAddEndPuncttrue
\mciteSetBstMidEndSepPunct{\mcitedefaultmidpunct}
{\mcitedefaultendpunct}{\mcitedefaultseppunct}\relax
\EndOfBibitem
\bibitem[Savu and Higgins(2022)Savu, and Higgins]{Savu:2022aa}
Savu,~D.-C.; Higgins,~A.~J. Structural stability of a lightsail for laser-driven interstellar flight. \emph{Acta Astronautica} \textbf{2022}, \emph{201}, 376--393\relax
\mciteBstWouldAddEndPuncttrue
\mciteSetBstMidEndSepPunct{\mcitedefaultmidpunct}
{\mcitedefaultendpunct}{\mcitedefaultseppunct}\relax
\EndOfBibitem
\bibitem[Guo(2007)]{Guo:2007aa}
Guo,~L. Nanoimprint Lithography: Methods and Material Requirements. \emph{Advanced Materials} \textbf{2007}, \emph{19}, 495--513\relax
\mciteBstWouldAddEndPuncttrue
\mciteSetBstMidEndSepPunct{\mcitedefaultmidpunct}
{\mcitedefaultendpunct}{\mcitedefaultseppunct}\relax
\EndOfBibitem
\bibitem[PlotDigitizer(2025)]{PlotDigitizer}
PlotDigitizer All-in-One Tool to Extract Data from Graphs, Plots \& Images. 2025; \url{https://plotdigitizer.com/}\relax
\mciteBstWouldAddEndPuncttrue
\mciteSetBstMidEndSepPunct{\mcitedefaultmidpunct}
{\mcitedefaultendpunct}{\mcitedefaultseppunct}\relax
\EndOfBibitem
\bibitem[Li(1980)]{Li:1980aa}
Li,~H.~H. Refractive index of silicon and germanium and its wavelength and temperature derivatives. \emph{Journal of Physical and Chemical Reference Data} \textbf{1980}, \emph{9}, 561--658\relax
\mciteBstWouldAddEndPuncttrue
\mciteSetBstMidEndSepPunct{\mcitedefaultmidpunct}
{\mcitedefaultendpunct}{\mcitedefaultseppunct}\relax
\EndOfBibitem
\bibitem[Li and Zhu(2015)Li, and Zhu]{Li:2015aa}
Li,~X.; Zhu,~H. Two-dimensional MoS2: Properties, preparation, and applications. \emph{Journal of Materiomics} \textbf{2015}, \emph{1}, 33--44\relax
\mciteBstWouldAddEndPuncttrue
\mciteSetBstMidEndSepPunct{\mcitedefaultmidpunct}
{\mcitedefaultendpunct}{\mcitedefaultseppunct}\relax
\EndOfBibitem
\bibitem[Jin \latin{et~al.}(2022)Jin, Li, Khandekar, Orenstein, and Fan]{Jin:2022aa}
Jin,~W.; Li,~W.; Khandekar,~C.; Orenstein,~M.; Fan,~S. Laser Cooling Assisted Thermal Management of Lightsails. \emph{ACS Photonics} \textbf{2022}, \emph{9}, 3384--3390\relax
\mciteBstWouldAddEndPuncttrue
\mciteSetBstMidEndSepPunct{\mcitedefaultmidpunct}
{\mcitedefaultendpunct}{\mcitedefaultseppunct}\relax
\EndOfBibitem
\bibitem[Lien \latin{et~al.}(2023)Lien, Meng, Liu, Sakib, Tang, Wu, and Povinelli]{Lien:2023aa}
Lien,~M.~R.; Meng,~D.; Liu,~Z.; Sakib,~M.~A.; Tang,~Y.; Wu,~W.; Povinelli,~M.~L. Experimental characterization of a silicon nitride photonic crystal light sail: erratum. \emph{Optical Materials Express} \textbf{2023}, \emph{13}, 2594--2594\relax
\mciteBstWouldAddEndPuncttrue
\mciteSetBstMidEndSepPunct{\mcitedefaultmidpunct}
{\mcitedefaultendpunct}{\mcitedefaultseppunct}\relax
\EndOfBibitem
\bibitem[Macleod(2018)]{Macleod:2018aa}
Macleod,~H. \emph{Thin-Film Optical Filters}, 5th ed.; Series in Optics and Optoelectronics; CRC Press: Boca Raton, 2018\relax
\mciteBstWouldAddEndPuncttrue
\mciteSetBstMidEndSepPunct{\mcitedefaultmidpunct}
{\mcitedefaultendpunct}{\mcitedefaultseppunct}\relax
\EndOfBibitem
\bibitem[Yeh(2005)]{yeh:optical_waves_in_layered_media}
Yeh,~P. \emph{Optical Waves in Layered Media}; Wiley, 2005\relax
\mciteBstWouldAddEndPuncttrue
\mciteSetBstMidEndSepPunct{\mcitedefaultmidpunct}
{\mcitedefaultendpunct}{\mcitedefaultseppunct}\relax
\EndOfBibitem
\bibitem[Qiu \latin{et~al.}(2021)Qiu, Zhang, Hu, and Kivshar]{QiuKivshar:2021_meta_review}
Qiu,~C.~W.; Zhang,~T.; Hu,~G.~W.; Kivshar,~Y. Quo Vadis, Metasurfaces? \emph{Nano Letters} \textbf{2021}, \emph{21}, 5461--5474\relax
\mciteBstWouldAddEndPuncttrue
\mciteSetBstMidEndSepPunct{\mcitedefaultmidpunct}
{\mcitedefaultendpunct}{\mcitedefaultseppunct}\relax
\EndOfBibitem
\bibitem[Zhou \latin{et~al.}(2024)Zhou, Wang, Yin, Wang, Manshaii, Xiao, Zhang, Bao, Jiang, and Chen]{Zhou:2024_meta_review}
Zhou,~Y.~L.; Wang,~S.~L.; Yin,~J.~Y.; Wang,~J.~J.; Manshaii,~F.; Xiao,~X.; Zhang,~T.~Q.; Bao,~H.; Jiang,~S.; Chen,~J. Flexible Metasurfaces for Multifunctional Interfaces. \emph{ACS Nano} \textbf{2024}, \emph{18}, 2685--2707\relax
\mciteBstWouldAddEndPuncttrue
\mciteSetBstMidEndSepPunct{\mcitedefaultmidpunct}
{\mcitedefaultendpunct}{\mcitedefaultseppunct}\relax
\EndOfBibitem
\bibitem[Petit \latin{et~al.}(1980)Petit, Botten, Cadilhac, Derrick, Maystre, McPhedran, Neviere, and Vincent]{Petit:1980aa}
Petit,~R.; Botten,~L.~C.; Cadilhac,~M.; Derrick,~G.~H.; Maystre,~D.; McPhedran,~R.~C.; Neviere,~M.; Vincent,~P. \emph{Electromagnetic Theory of Gratings}, 1st ed.; Springer Berlin Heidelberg: Berlin, Heidelberg, 1980; Vol.~22\relax
\mciteBstWouldAddEndPuncttrue
\mciteSetBstMidEndSepPunct{\mcitedefaultmidpunct}
{\mcitedefaultendpunct}{\mcitedefaultseppunct}\relax
\EndOfBibitem
\bibitem[Loewen and Popov(2017)Loewen, and Popov]{Loewen:2017aa}
Loewen,~E.~G.; Popov,~E. \emph{Diffraction Gratings and Applications}, 1st ed.; CRC Press: Boca Raton, 2017\relax
\mciteBstWouldAddEndPuncttrue
\mciteSetBstMidEndSepPunct{\mcitedefaultmidpunct}
{\mcitedefaultendpunct}{\mcitedefaultseppunct}\relax
\EndOfBibitem
\bibitem[Joannopoulos \latin{et~al.}(2008)Joannopoulos, Johnson, Winn, and Meade]{Joannopoulos:2008aa}
Joannopoulos,~J.~D.; Johnson,~S.~G.; Winn,~J.~N.; Meade,~R.~D. \emph{Photonic Crystals: Molding the Flow of Light}, 2nd ed.; Princeton University Press: Princeton, 2008\relax
\mciteBstWouldAddEndPuncttrue
\mciteSetBstMidEndSepPunct{\mcitedefaultmidpunct}
{\mcitedefaultendpunct}{\mcitedefaultseppunct}\relax
\EndOfBibitem
\bibitem[Ilic and Atwater(2019)Ilic, and Atwater]{Ilic:2019aa}
Ilic,~O.; Atwater,~H.~A. Self-stabilizing photonic levitation and propulsion of nanostructured macroscopic objects. \emph{Nature Photonics} \textbf{2019}, \emph{13}, 289--295\relax
\mciteBstWouldAddEndPuncttrue
\mciteSetBstMidEndSepPunct{\mcitedefaultmidpunct}
{\mcitedefaultendpunct}{\mcitedefaultseppunct}\relax
\EndOfBibitem
\bibitem[Song \latin{et~al.}(2018)Song, Catrysse, and Fan]{Song:2018aa}
Song,~A.~Y.; Catrysse,~P.~B.; Fan,~S. Broadband Control of Topological Nodes in Electromagnetic Fields. \emph{Physical Review Letters} \textbf{2018}, \emph{120}, 193903--193903\relax
\mciteBstWouldAddEndPuncttrue
\mciteSetBstMidEndSepPunct{\mcitedefaultmidpunct}
{\mcitedefaultendpunct}{\mcitedefaultseppunct}\relax
\EndOfBibitem
\bibitem[Kim and Lee(2023)Kim, and Lee]{kim:2023torwa}
Kim,~C.; Lee,~B. TORCWA: GPU-accelerated Fourier modal method and gradient-based optimization for metasurface design. \emph{Computer Physics Communications} \textbf{2023}, \emph{282}, 108552\relax
\mciteBstWouldAddEndPuncttrue
\mciteSetBstMidEndSepPunct{\mcitedefaultmidpunct}
{\mcitedefaultendpunct}{\mcitedefaultseppunct}\relax
\EndOfBibitem
\bibitem[Venter(2010)]{Venter:2010aa}
Venter,~G. \emph{Encyclopedia of Aerospace Engineering}; John Wiley \& Sons, Ltd.: Hoboken, 2010\relax
\mciteBstWouldAddEndPuncttrue
\mciteSetBstMidEndSepPunct{\mcitedefaultmidpunct}
{\mcitedefaultendpunct}{\mcitedefaultseppunct}\relax
\EndOfBibitem
\bibitem[Molesky \latin{et~al.}(2018)Molesky, Lin, Piggott, Jin, Vucković, and Rodriguez]{Molesky:2018aa}
Molesky,~S.; Lin,~Z.; Piggott,~A.~Y.; Jin,~W.; Vucković,~J.; Rodriguez,~A.~W. Inverse design in nanophotonics. \emph{Nature Photonics} \textbf{2018}, \emph{12}, 659--670\relax
\mciteBstWouldAddEndPuncttrue
\mciteSetBstMidEndSepPunct{\mcitedefaultmidpunct}
{\mcitedefaultendpunct}{\mcitedefaultseppunct}\relax
\EndOfBibitem
\bibitem[Bennet \latin{et~al.}(2024)Bennet, Langevin, Essoual, Khaireh-Walieh, Teytaud, Wiecha, and Moreau]{Bennet:2024optimization_tutorial}
Bennet,~P.; Langevin,~D.; Essoual,~C.; Khaireh-Walieh,~A.; Teytaud,~O.; Wiecha,~P.; Moreau,~A. Illustrated tutorial on global optimization in nanophotonics. \emph{Journal of the Optical Society of America B} \textbf{2024}, \emph{41}, A126--A145\relax
\mciteBstWouldAddEndPuncttrue
\mciteSetBstMidEndSepPunct{\mcitedefaultmidpunct}
{\mcitedefaultendpunct}{\mcitedefaultseppunct}\relax
\EndOfBibitem
\bibitem[Fan and Joannopoulos(2002)Fan, and Joannopoulos]{Fan:2002Fano}
Fan,~S.; Joannopoulos,~J.~D. Analysis of guided resonances in photonic crystal slabs. \emph{Physical Review B} \textbf{2002}, \emph{65}, 235112\relax
\mciteBstWouldAddEndPuncttrue
\mciteSetBstMidEndSepPunct{\mcitedefaultmidpunct}
{\mcitedefaultendpunct}{\mcitedefaultseppunct}\relax
\EndOfBibitem
\bibitem[Svanberg(1987)]{Svanberg:1987aa}
Svanberg,~K. The method of moving asymptotes---a new method for structural optimization. \emph{International Journal for Numerical Methods in Engineering} \textbf{1987}, \emph{24}, 359--373\relax
\mciteBstWouldAddEndPuncttrue
\mciteSetBstMidEndSepPunct{\mcitedefaultmidpunct}
{\mcitedefaultendpunct}{\mcitedefaultseppunct}\relax
\EndOfBibitem
\bibitem[Sparrow(2017)]{Sparrow:radiation_heat_trnsfer_book}
Sparrow,~E.~M. \emph{Radiation heat transfer}, first edition. ed.; Taylor and Francis: London, 2017\relax
\mciteBstWouldAddEndPuncttrue
\mciteSetBstMidEndSepPunct{\mcitedefaultmidpunct}
{\mcitedefaultendpunct}{\mcitedefaultseppunct}\relax
\EndOfBibitem
\bibitem[Cornelius and Dowling(1999)Cornelius, and Dowling]{Cornelius:1999aa}
Cornelius,~C.~M.; Dowling,~J.~P. Modification of Planck blackbody radiation by photonic band-gap structures. \emph{Physical Review A} \textbf{1999}, \emph{59}, 4736--4746, PRA\relax
\mciteBstWouldAddEndPuncttrue
\mciteSetBstMidEndSepPunct{\mcitedefaultmidpunct}
{\mcitedefaultendpunct}{\mcitedefaultseppunct}\relax
\EndOfBibitem
\bibitem[Baranov \latin{et~al.}(2019)Baranov, Xiao, Nechepurenko, Krasnok, Alù, and Kats]{baranov:2019}
Baranov,~D.~G.; Xiao,~Y.; Nechepurenko,~I.~A.; Krasnok,~A.; Alù,~A.; Kats,~M.~A. Nanophotonic engineering of far-field thermal emitters. \emph{Nature Materials} \textbf{2019}, \emph{18}, 920--930\relax
\mciteBstWouldAddEndPuncttrue
\mciteSetBstMidEndSepPunct{\mcitedefaultmidpunct}
{\mcitedefaultendpunct}{\mcitedefaultseppunct}\relax
\EndOfBibitem
\bibitem[Fan and Li(2022)Fan, and Li]{fan:2022review}
Fan,~S.; Li,~W. Photonics and thermodynamics concepts in radiative cooling. \emph{Nature Photonics} \textbf{2022}, \emph{16}, 182--190\relax
\mciteBstWouldAddEndPuncttrue
\mciteSetBstMidEndSepPunct{\mcitedefaultmidpunct}
{\mcitedefaultendpunct}{\mcitedefaultseppunct}\relax
\EndOfBibitem
\bibitem[Langevin \latin{et~al.}(2024)Langevin, Bennet, Khaireh-Walieh, Wiecha, Teytaud, and Moreau]{langevin:2024_pymoosh}
Langevin,~D.; Bennet,~P.; Khaireh-Walieh,~A.; Wiecha,~P.; Teytaud,~O.; Moreau,~A. PyMoosh: a comprehensive numerical toolkit for computing the optical properties of multilayered structures. \emph{Journal of the Optical Society of America B} \textbf{2024}, \emph{41}, A67--A78\relax
\mciteBstWouldAddEndPuncttrue
\mciteSetBstMidEndSepPunct{\mcitedefaultmidpunct}
{\mcitedefaultendpunct}{\mcitedefaultseppunct}\relax
\EndOfBibitem
\bibitem[Holdman \latin{et~al.}(2022)Holdman, Jaffe, Feng, Jang, Kats, and Brar]{holdman:2022aa}
Holdman,~G.~R.; Jaffe,~G.~R.; Feng,~D.; Jang,~M.~S.; Kats,~M.~A.; Brar,~V.~W. Thermal Runaway of Silicon-Based Laser Sails. \emph{Advanced Optical Materials} \textbf{2022}, \emph{10}, 2102835\relax
\mciteBstWouldAddEndPuncttrue
\mciteSetBstMidEndSepPunct{\mcitedefaultmidpunct}
{\mcitedefaultendpunct}{\mcitedefaultseppunct}\relax
\EndOfBibitem
\bibitem[Szidarovsky and Bahill(2017)Szidarovsky, and Bahill]{Szidarovsky:2017aa}
Szidarovsky,~F.; Bahill,~A.~T. \emph{Linear Systems Theory}, 2nd ed.; CRC Press: Boca Raton, 2017\relax
\mciteBstWouldAddEndPuncttrue
\mciteSetBstMidEndSepPunct{\mcitedefaultmidpunct}
{\mcitedefaultendpunct}{\mcitedefaultseppunct}\relax
\EndOfBibitem
\bibitem[Chu \latin{et~al.}(2021)Chu, Meem, Srivastava, Menon, and Swartzlander]{Chu:2021aa}
Chu,~Y.-J.~L.; Meem,~M.; Srivastava,~P.~R.; Menon,~R.; Swartzlander,~G.~A. Parametric control of a diffractive axicon beam rider. \emph{Optics Letters} \textbf{2021}, \emph{46}, 5141--5144\relax
\mciteBstWouldAddEndPuncttrue
\mciteSetBstMidEndSepPunct{\mcitedefaultmidpunct}
{\mcitedefaultendpunct}{\mcitedefaultseppunct}\relax
\EndOfBibitem
\bibitem[Shirin \latin{et~al.}(2021)Shirin, Schamiloglu, Sultan, Yang, Benford, and Fierro]{Shirin:2021aa}
Shirin,~A.; Schamiloglu,~E.; Sultan,~C.; Yang,~Y.; Benford,~J.; Fierro,~R. Modeling and Stability of a Laser Beam-Driven Sail. 2021 American Control Conference (ACC). 2021; pp 4269--4275\relax
\mciteBstWouldAddEndPuncttrue
\mciteSetBstMidEndSepPunct{\mcitedefaultmidpunct}
{\mcitedefaultendpunct}{\mcitedefaultseppunct}\relax
\EndOfBibitem
\bibitem[Singh(2000)]{Singh:2000aa}
Singh,~G. Characterization of Passive Dynamic Stability of a Microwave Sail. 2000\relax
\mciteBstWouldAddEndPuncttrue
\mciteSetBstMidEndSepPunct{\mcitedefaultmidpunct}
{\mcitedefaultendpunct}{\mcitedefaultseppunct}\relax
\EndOfBibitem
\bibitem[Schamiloglu \latin{et~al.}(2001)Schamiloglu, Abdallah, Miller, Georgiev, Benford, Benford, and Singh]{Schamiloglu:2001aa}
Schamiloglu,~E.; Abdallah,~C.~T.; Miller,~K.~A.; Georgiev,~D.; Benford,~J.; Benford,~G.; Singh,~G. 3-D simulations of rigid microwave-propelled sails including spin. \emph{AIP Conference Proceedings} \textbf{2001}, \emph{552}, 559--564\relax
\mciteBstWouldAddEndPuncttrue
\mciteSetBstMidEndSepPunct{\mcitedefaultmidpunct}
{\mcitedefaultendpunct}{\mcitedefaultseppunct}\relax
\EndOfBibitem
\bibitem[Popova \latin{et~al.}(2016)Popova, Efendiev, and Gabitov]{Popova:2016aa}
Popova,~E.; Efendiev,~M.; Gabitov,~I. On the stability of a space vehicle riding on an intense laser beam. \emph{Mathematical Methods in the Applied Sciences} \textbf{2016}, \emph{40}, 1346--1354\relax
\mciteBstWouldAddEndPuncttrue
\mciteSetBstMidEndSepPunct{\mcitedefaultmidpunct}
{\mcitedefaultendpunct}{\mcitedefaultseppunct}\relax
\EndOfBibitem
\bibitem[Manchester and Loeb(2017)Manchester, and Loeb]{Manchester:2017aa}
Manchester,~Z.; Loeb,~A. Stability of a Light Sail Riding on a Laser Beam. \emph{The Astrophysical Journal Letters} \textbf{2017}, \emph{837}, L20--L20\relax
\mciteBstWouldAddEndPuncttrue
\mciteSetBstMidEndSepPunct{\mcitedefaultmidpunct}
{\mcitedefaultendpunct}{\mcitedefaultseppunct}\relax
\EndOfBibitem
\bibitem[Goldstein \latin{et~al.}(2001)Goldstein, Poole, and Safko]{Goldstein:2001aa}
Goldstein,~H.; Poole,~C.; Safko,~J. \emph{Classical Mechanics}, 3rd ed.; Addison-Wesley: Boston, 2001\relax
\mciteBstWouldAddEndPuncttrue
\mciteSetBstMidEndSepPunct{\mcitedefaultmidpunct}
{\mcitedefaultendpunct}{\mcitedefaultseppunct}\relax
\EndOfBibitem
\bibitem[Siegel \latin{et~al.}(2019)Siegel, Wang, Menabde, Kats, Jang, and Brar]{Siegel:2019aa}
Siegel,~J.; Wang,~A.~Y.; Menabde,~S.~G.; Kats,~M.~A.; Jang,~M.~S.; Brar,~V.~W. Self-Stabilizing Laser Sails Based on Optical Metasurfaces. \emph{ACS Photonics} \textbf{2019}, \emph{6}, 2032--2040\relax
\mciteBstWouldAddEndPuncttrue
\mciteSetBstMidEndSepPunct{\mcitedefaultmidpunct}
{\mcitedefaultendpunct}{\mcitedefaultseppunct}\relax
\EndOfBibitem
\bibitem[Srivastava \latin{et~al.}(2019)Srivastava, Chu, and Swartzlander]{Srivastava:2019aa}
Srivastava,~P.~R.; Chu,~Y.-J.~L.; Swartzlander,~G.~A. Stable diffractive beam rider. \emph{Optics Letters} \textbf{2019}, \emph{44}, 3082--3085\relax
\mciteBstWouldAddEndPuncttrue
\mciteSetBstMidEndSepPunct{\mcitedefaultmidpunct}
{\mcitedefaultendpunct}{\mcitedefaultseppunct}\relax
\EndOfBibitem
\bibitem[Srivastava and Swartzlander(2020)Srivastava, and Swartzlander]{Srivastava:2020aa}
Srivastava,~P.~R.; Swartzlander,~G.~A. Optomechanics of a stable diffractive axicon light sail. \emph{The European Physical Journal Plus} \textbf{2020}, \emph{135}, 570--570\relax
\mciteBstWouldAddEndPuncttrue
\mciteSetBstMidEndSepPunct{\mcitedefaultmidpunct}
{\mcitedefaultendpunct}{\mcitedefaultseppunct}\relax
\EndOfBibitem
\bibitem[Gieseler \latin{et~al.}(2021)Gieseler, Rahimzadegan, and Rockstuhl]{Gieseler:2021aa}
Gieseler,~N.; Rahimzadegan,~A.; Rockstuhl,~C. Self-stabilizing curved metasurfaces as a sail for light-propelled spacecrafts. \emph{Optics Express} \textbf{2021}, \emph{29}, 21562--21575\relax
\mciteBstWouldAddEndPuncttrue
\mciteSetBstMidEndSepPunct{\mcitedefaultmidpunct}
{\mcitedefaultendpunct}{\mcitedefaultseppunct}\relax
\EndOfBibitem
\bibitem[Kumar \latin{et~al.}(2021)Kumar, Kindem, and Ilic]{Kumar:2021aa}
Kumar,~A.; Kindem,~D.; Ilic,~O. Optomechanical Self-Stability of Freestanding Photonic Metasurfaces. \emph{Physical Review Applied} \textbf{2021}, \emph{16}, 14053--14053\relax
\mciteBstWouldAddEndPuncttrue
\mciteSetBstMidEndSepPunct{\mcitedefaultmidpunct}
{\mcitedefaultendpunct}{\mcitedefaultseppunct}\relax
\EndOfBibitem
\bibitem[Gao \latin{et~al.}(2022)Gao, Kelzenberg, Kim, Ilic, and Atwater]{Gao:2022aa}
Gao,~R.; Kelzenberg,~M.~D.; Kim,~Y.; Ilic,~O.; Atwater,~H.~A. Optical Characterization of Silicon Nitride Metagrating-Based Lightsails for Self-Stabilization. \emph{ACS Photonics} \textbf{2022}, \emph{9}, 1965--1972\relax
\mciteBstWouldAddEndPuncttrue
\mciteSetBstMidEndSepPunct{\mcitedefaultmidpunct}
{\mcitedefaultendpunct}{\mcitedefaultseppunct}\relax
\EndOfBibitem
\bibitem[Rafat \latin{et~al.}(2022)Rafat, Dullin, Kuhlmey, Tuniz, Luo, Roy, Skinner, Alexander, Wheatland, and de~Sterke]{Rafat:2022aa}
Rafat,~M.~Z.; Dullin,~H.~R.; Kuhlmey,~B.~T.; Tuniz,~A.; Luo,~H.; Roy,~D.; Skinner,~S.; Alexander,~T.~J.; Wheatland,~M.~S.; de~Sterke,~C.~M. Self-Stabilization of Light Sails by Damped Internal Degrees of Freedom. \emph{Physical Review Applied} \textbf{2022}, \emph{17}, 024016--024016\relax
\mciteBstWouldAddEndPuncttrue
\mciteSetBstMidEndSepPunct{\mcitedefaultmidpunct}
{\mcitedefaultendpunct}{\mcitedefaultseppunct}\relax
\EndOfBibitem
\bibitem[Srivastava \latin{et~al.}(2024)Srivastava, Majumdar, Menon, and Swartzlander]{Srivastava:2024aa}
Srivastava,~P.~R.; Majumdar,~A.; Menon,~R.; Swartzlander,~G.~A. High forward thrust metasurface beam-riding sail. \emph{Optics Express} \textbf{2024}, \emph{32}, 1756--1763\relax
\mciteBstWouldAddEndPuncttrue
\mciteSetBstMidEndSepPunct{\mcitedefaultmidpunct}
{\mcitedefaultendpunct}{\mcitedefaultseppunct}\relax
\EndOfBibitem
\bibitem[Salary and Mosallaei(2020)Salary, and Mosallaei]{Salary:2020aa}
Salary,~M.~M.; Mosallaei,~H. Photonic Metasurfaces as Relativistic Light Sails for Doppler-Broadened Stable Beam-Riding and Radiative Cooling. \emph{Laser \& Photonics Reviews} \textbf{2020}, \emph{14}, 1900311--1900311\relax
\mciteBstWouldAddEndPuncttrue
\mciteSetBstMidEndSepPunct{\mcitedefaultmidpunct}
{\mcitedefaultendpunct}{\mcitedefaultseppunct}\relax
\EndOfBibitem
\bibitem[Salary and Mosallaei(2021)Salary, and Mosallaei]{Salary:2021aa}
Salary,~M.~M.; Mosallaei,~H. Inverse Design of Diffractive Relativistic Meta-Sails via Multi-Objective Optimization. \emph{Advanced Theory and Simulations} \textbf{2021}, \emph{4}, 2100047--2100047\relax
\mciteBstWouldAddEndPuncttrue
\mciteSetBstMidEndSepPunct{\mcitedefaultmidpunct}
{\mcitedefaultendpunct}{\mcitedefaultseppunct}\relax
\EndOfBibitem
\bibitem[Taghavi and Mosallaei(2022)Taghavi, and Mosallaei]{Taghavi:2022aa}
Taghavi,~M.; Mosallaei,~H. Increasing the stability margins using multi-pattern metasails and multi-modal laser beams. \emph{Scientific Reports} \textbf{2022}, \emph{12}, 20034--20034\relax
\mciteBstWouldAddEndPuncttrue
\mciteSetBstMidEndSepPunct{\mcitedefaultmidpunct}
{\mcitedefaultendpunct}{\mcitedefaultseppunct}\relax
\EndOfBibitem
\bibitem[Khalil(2002)]{Khalil:2002aa}
Khalil,~H. \emph{Nonlinear Systems}, 2nd ed.; Prentice-Hall, Inc.: Upper Saddle River, 2002\relax
\mciteBstWouldAddEndPuncttrue
\mciteSetBstMidEndSepPunct{\mcitedefaultmidpunct}
{\mcitedefaultendpunct}{\mcitedefaultseppunct}\relax
\EndOfBibitem
\bibitem[Garcia(2000)]{Garcia:2000aa}
Garcia,~A. \emph{Numerical Methods for Physics}; Prentice Hall: New Jersey, 2000\relax
\mciteBstWouldAddEndPuncttrue
\mciteSetBstMidEndSepPunct{\mcitedefaultmidpunct}
{\mcitedefaultendpunct}{\mcitedefaultseppunct}\relax
\EndOfBibitem
\bibitem[Benford \latin{et~al.}(2002)Benford, Benford, Gornostaeva, Garate, Anderson, Prichard, and Harris]{Benford:2002aa}
Benford,~J.; Benford,~G.; Gornostaeva,~O.; Garate,~E.; Anderson,~M.; Prichard,~A.; Harris,~H. Experimental tests of beam-riding sail dynamics. \emph{AIP Conference Proceedings} \textbf{2002}, \emph{608}, 457--461\relax
\mciteBstWouldAddEndPuncttrue
\mciteSetBstMidEndSepPunct{\mcitedefaultmidpunct}
{\mcitedefaultendpunct}{\mcitedefaultseppunct}\relax
\EndOfBibitem
\bibitem[Abdallah \latin{et~al.}(2003)Abdallah, Chahine, Georgiev, and Schamiloglu]{Abdallah:2003aa}
Abdallah,~C.~T.; Chahine,~E.; Georgiev,~D.; Schamiloglu,~E. Dynamics and Control of Microwave‐propelled Sails Using Delayed Measurements. \emph{AIP Conference Proceedings} \textbf{2003}, \emph{664}, 348--357\relax
\mciteBstWouldAddEndPuncttrue
\mciteSetBstMidEndSepPunct{\mcitedefaultmidpunct}
{\mcitedefaultendpunct}{\mcitedefaultseppunct}\relax
\EndOfBibitem
\bibitem[Benford \latin{et~al.}(2003)Benford, Gornostaeva, and Benford]{Benford:2003aa}
Benford,~G.; Gornostaeva,~O.; Benford,~J. Experimental Tests Of Beam‐Riding Sail Dynamics. \emph{AIP Conference Proceedings} \textbf{2003}, \emph{664}, 325--335\relax
\mciteBstWouldAddEndPuncttrue
\mciteSetBstMidEndSepPunct{\mcitedefaultmidpunct}
{\mcitedefaultendpunct}{\mcitedefaultseppunct}\relax
\EndOfBibitem
\bibitem[Campbell \latin{et~al.}(2022)Campbell, Brewer, Jariwala, Raman, and Bargatin]{Campbell:2022aa}
Campbell,~M.~F.; Brewer,~J.; Jariwala,~D.; Raman,~A.~P.; Bargatin,~I. Relativistic Light Sails Need to Billow. \emph{Nano Letters} \textbf{2022}, \emph{22}, 90--96\relax
\mciteBstWouldAddEndPuncttrue
\mciteSetBstMidEndSepPunct{\mcitedefaultmidpunct}
{\mcitedefaultendpunct}{\mcitedefaultseppunct}\relax
\EndOfBibitem
\bibitem[Srinivasan \latin{et~al.}(2016)Srinivasan, Hughes, Lubin, Zhang, Madajian, Brashears, Kulkarni, Cohen, and Griswold]{Srinivasan:2016aa}
Srinivasan,~P.; Hughes,~G.; Lubin,~P.; Zhang,~Q.; Madajian,~J.; Brashears,~T.; Kulkarni,~N.; Cohen,~A.; Griswold,~J. \emph{Stability of laser-propelled wafer satellites}; SPIE Optical Engineering + Applications; SPIE, 2016; Vol. 9981\relax
\mciteBstWouldAddEndPuncttrue
\mciteSetBstMidEndSepPunct{\mcitedefaultmidpunct}
{\mcitedefaultendpunct}{\mcitedefaultseppunct}\relax
\EndOfBibitem
\bibitem[Shirin \latin{et~al.}(2022)Shirin, Schamiloglu, and Fierro]{Shirin:2022aa}
Shirin,~A.; Schamiloglu,~E.; Fierro,~R. Towards Stable Interstellar Flight: Levitation of a Laser-Propelled Sailcraft. 2022 American Control Conference (ACC). 2022; pp 1828--1834\relax
\mciteBstWouldAddEndPuncttrue
\mciteSetBstMidEndSepPunct{\mcitedefaultmidpunct}
{\mcitedefaultendpunct}{\mcitedefaultseppunct}\relax
\EndOfBibitem
\bibitem[Landau and Lifshitz(1976)Landau, and Lifshitz]{Landau:1976aa}
Landau,~L.~D.; Lifshitz,~E.~M. In \emph{Mechanics (Third Edition)}; Landau,~L.~D., Lifshitz,~E.~M., Eds.; Course of Theoretical Physics; Butterworth-Heinemann: Oxford, 1976; Chapter 7, pp 131--167\relax
\mciteBstWouldAddEndPuncttrue
\mciteSetBstMidEndSepPunct{\mcitedefaultmidpunct}
{\mcitedefaultendpunct}{\mcitedefaultseppunct}\relax
\EndOfBibitem
\bibitem[Mackintosh \latin{et~al.}(2024)Mackintosh, Lin, Wheatland, and Kuhlmey]{Mackintosh:2024aa}
Mackintosh,~R.; Lin,~J.~Y.; Wheatland,~M.~S.; Kuhlmey,~B.~T. Relativistic damping of laser-beam-driven light sails. \emph{Physical Review Applied} \textbf{2024}, \emph{21}, 64032--64032\relax
\mciteBstWouldAddEndPuncttrue
\mciteSetBstMidEndSepPunct{\mcitedefaultmidpunct}
{\mcitedefaultendpunct}{\mcitedefaultseppunct}\relax
\EndOfBibitem
\bibitem[Poynting(1904)]{Poynting:1904aa}
Poynting,~J.~H. XII. Radiation in the solar system: its effect on temperature and its pressure on small bodies. \emph{Philosophical Transactions of the Royal Society of London. Series A, Containing Papers of a Mathematical or Physical Character} \textbf{1904}, \emph{202}, 525--552\relax
\mciteBstWouldAddEndPuncttrue
\mciteSetBstMidEndSepPunct{\mcitedefaultmidpunct}
{\mcitedefaultendpunct}{\mcitedefaultseppunct}\relax
\EndOfBibitem
\bibitem[Robertson and Russell(1937)Robertson, and Russell]{Robertson:1937aa}
Robertson,~H.~P.; Russell,~H.~N. Dynamical Effects of Radiation in the Solar System. \emph{Monthly Notices of the Royal Astronomical Society} \textbf{1937}, \emph{97}, 423--437\relax
\mciteBstWouldAddEndPuncttrue
\mciteSetBstMidEndSepPunct{\mcitedefaultmidpunct}
{\mcitedefaultendpunct}{\mcitedefaultseppunct}\relax
\EndOfBibitem
\bibitem[Kla{\v c}ka \latin{et~al.}(2014)Kla{\v c}ka, Petr{\v z}ala, P{\'a}stor, and K{\'o}mar]{Klacka:2014aa}
Kla{\v c}ka,~J.; Petr{\v z}ala,~J.; P{\'a}stor,~P.; K{\'o}mar,~L. The Poynting--Robertson effect: A critical perspective. \emph{Icarus} \textbf{2014}, \emph{232}, 249--262\relax
\mciteBstWouldAddEndPuncttrue
\mciteSetBstMidEndSepPunct{\mcitedefaultmidpunct}
{\mcitedefaultendpunct}{\mcitedefaultseppunct}\relax
\EndOfBibitem
\bibitem[Fűzfa \latin{et~al.}(2020)Fűzfa, Dhelonga-Biarufu, and Welcomme]{fuzfa:2020}
Fűzfa,~A.; Dhelonga-Biarufu,~W.; Welcomme,~O. Sailing towards the stars close to the speed of light. \emph{Physical Review Research} \textbf{2020}, \emph{2}, 043186, PRRESEARCH\relax
\mciteBstWouldAddEndPuncttrue
\mciteSetBstMidEndSepPunct{\mcitedefaultmidpunct}
{\mcitedefaultendpunct}{\mcitedefaultseppunct}\relax
\EndOfBibitem
\bibitem[Lin \latin{et~al.}(2024)Lin, de~Sterke, Wheatland, Song, and Kuhlmey]{Lin:2024aa}
Lin,~J.~Y.; de~Sterke,~C.~M.; Wheatland,~M.~S.; Song,~A.~Y.; Kuhlmey,~B.~T. All-optical damping forces enhanced by metasurfaces for stable relativistic lightsail propulsion. \emph{Physical Review Applied} \textbf{2024}, \emph{22}, 064028\relax
\mciteBstWouldAddEndPuncttrue
\mciteSetBstMidEndSepPunct{\mcitedefaultmidpunct}
{\mcitedefaultendpunct}{\mcitedefaultseppunct}\relax
\EndOfBibitem
\bibitem[Yu \latin{et~al.}(2011)Yu, Genevet, Kats, Aieta, Tetienne, Capasso, and Gaburro]{Yu:2011aa}
Yu,~N.; Genevet,~P.; Kats,~M.~A.; Aieta,~F.; Tetienne,~J.-P.; Capasso,~F.; Gaburro,~Z. Light Propagation with Phase Discontinuities: Generalized Laws of Reflection and Refraction. \emph{Science} \textbf{2011}, \emph{334}, 333--337\relax
\mciteBstWouldAddEndPuncttrue
\mciteSetBstMidEndSepPunct{\mcitedefaultmidpunct}
{\mcitedefaultendpunct}{\mcitedefaultseppunct}\relax
\EndOfBibitem
\bibitem[Chen \latin{et~al.}(2016)Chen, Taylor, and Yu]{Chen:2016aa}
Chen,~H.-T.; Taylor,~A.~J.; Yu,~N. A review of metasurfaces: physics and applications. \emph{Reports on Progress in Physics} \textbf{2016}, \emph{79}, 076401\relax
\mciteBstWouldAddEndPuncttrue
\mciteSetBstMidEndSepPunct{\mcitedefaultmidpunct}
{\mcitedefaultendpunct}{\mcitedefaultseppunct}\relax
\EndOfBibitem
\bibitem[Khorasaninejad and Capasso(2017)Khorasaninejad, and Capasso]{Khorasaninejad:2017aa}
Khorasaninejad,~M.; Capasso,~F. Metalenses: Versatile multifunctional photonic components. \emph{Science} \textbf{2017}, \emph{358}, eaam8100\relax
\mciteBstWouldAddEndPuncttrue
\mciteSetBstMidEndSepPunct{\mcitedefaultmidpunct}
{\mcitedefaultendpunct}{\mcitedefaultseppunct}\relax
\EndOfBibitem
\bibitem[Myilswamy \latin{et~al.}(2020)Myilswamy, Krishnan, and Povinelli]{Myilswamy:2020aa}
Myilswamy,~K.~V.; Krishnan,~A.; Povinelli,~M.~L. Photonic crystal lightsail with nonlinear reflectivity for increased stability. \emph{Optics Express} \textbf{2020}, \emph{28}, 8223--8232\relax
\mciteBstWouldAddEndPuncttrue
\mciteSetBstMidEndSepPunct{\mcitedefaultmidpunct}
{\mcitedefaultendpunct}{\mcitedefaultseppunct}\relax
\EndOfBibitem
\bibitem[Kelzenberg and Gao(2024)Kelzenberg, and Gao]{kelzenberg:flexible_github}
Kelzenberg,~M.~D.; Gao,~R. Flexible Lightsail Simulator (2024). 2024; \url{https://github.com/Starshot-Lightsail/FlexSailSim}\relax
\mciteBstWouldAddEndPuncttrue
\mciteSetBstMidEndSepPunct{\mcitedefaultmidpunct}
{\mcitedefaultendpunct}{\mcitedefaultseppunct}\relax
\EndOfBibitem
\bibitem[Chu \latin{et~al.}(2019)Chu, Tabiryan, and Swartzlander]{Chu:2019aa}
Chu,~Y.-J.~L.; Tabiryan,~N.~V.; Swartzlander,~G.~A. Experimental Verification of a Bigrating Beam Rider. \emph{Physical Review Letters} \textbf{2019}, \emph{123}, 244302--244302\relax
\mciteBstWouldAddEndPuncttrue
\mciteSetBstMidEndSepPunct{\mcitedefaultmidpunct}
{\mcitedefaultendpunct}{\mcitedefaultseppunct}\relax
\EndOfBibitem
\bibitem[Norder \latin{et~al.}(2024)Norder, Yin, De~Jong, Stallone, Aydogmus, Sberna, Bessa, and Norte]{Norder:2024aa}
Norder,~L.; Yin,~S.; De~Jong,~M.~J.; Stallone,~F.; Aydogmus,~H.; Sberna,~P.~M.; Bessa,~M.~A.; Norte,~R.~A. Pentagonal Photonic Crystal Mirrors: Scalable Lightsails with Enhanced Acceleration via Neural Topology Optimization. \emph{arXiv} \textbf{2024}, \relax
\mciteBstWouldAddEndPunctfalse
\mciteSetBstMidEndSepPunct{\mcitedefaultmidpunct}
{}{\mcitedefaultseppunct}\relax
\EndOfBibitem
\bibitem[Feng \latin{et~al.}(2024)Feng, Kumar, Yin, Mah, Lin, Fortman, Jaffe, Wan, Mei, Xiao, Synowicki, Warzoha, Brar, Talghader, and Kats]{Feng:2024aa}
Feng,~D.; Kumar,~T.; Yin,~S.; Mah,~M.; Lin,~P.; Fortman,~M.; Jaffe,~G.~R.; Wan,~C.; Mei,~H.; Xiao,~Y.; Synowicki,~R.; Warzoha,~R.~J.; Brar,~V.~W.; Talghader,~J.~J.; Kats,~M.~A. Self-referencing photothermal common-path interferometry to measure absorption of ${\rm Si_3N_4}$ membranes for laser-light sails. \emph{arXiv preprint arXiv:2404.04449} \textbf{2024}, \relax
\mciteBstWouldAddEndPunctfalse
\mciteSetBstMidEndSepPunct{\mcitedefaultmidpunct}
{}{\mcitedefaultseppunct}\relax
\EndOfBibitem
\bibitem[Myrabo \latin{et~al.}()Myrabo, Knowles, Bagford, I, and Harris]{Myrabo:2002aa}
Myrabo,~L.~N.; Knowles,~T.~R.; Bagford,~J.~O.; I,~D. B. S.~I.; Harris,~H.~M. Laser-boosted light sail experiments with the 150-kW LHMEL II ${\rm CO_2}$ laser. Proc. SPIE. pp 774--798\relax
\mciteBstWouldAddEndPuncttrue
\mciteSetBstMidEndSepPunct{\mcitedefaultmidpunct}
{\mcitedefaultendpunct}{\mcitedefaultseppunct}\relax
\EndOfBibitem
\bibitem[Tang \latin{et~al.}(2024)Tang, Li, Du, Zhang, Gong, Guo, Pu, and Luo]{Tang:2024aa}
Tang,~T.; Li,~L.; Du,~A.; Zhang,~F.; Gong,~P.; Guo,~Y.; Pu,~M.; Luo,~X. Precision measurement of optical force based on torsion balance. Proc. SPIE. 2024; p 131045S\relax
\mciteBstWouldAddEndPuncttrue
\mciteSetBstMidEndSepPunct{\mcitedefaultmidpunct}
{\mcitedefaultendpunct}{\mcitedefaultseppunct}\relax
\EndOfBibitem
\bibitem[Michaeli \latin{et~al.}(2025)Michaeli, Gao, Kelzenberg, Hail, Merkt, Sader, and Atwater]{Michaeli:2025aa}
Michaeli,~L.; Gao,~R.; Kelzenberg,~M.~D.; Hail,~C.~U.; Merkt,~A.; Sader,~J.~E.; Atwater,~H.~A. Direct radiation pressure measurements for lightsail membranes. \emph{Nature Photonics} \textbf{2025}, \relax
\mciteBstWouldAddEndPunctfalse
\mciteSetBstMidEndSepPunct{\mcitedefaultmidpunct}
{}{\mcitedefaultseppunct}\relax
\EndOfBibitem
\bibitem[Chu and Gong(2024)Chu, and Gong]{Chu:2024aa}
Chu,~Y.; Gong,~S. Minimum-time rendezvous for Sun-facing diffractive solar sails with diverse deflection angles. \emph{Astrodynamics} \textbf{2024}, \relax
\mciteBstWouldAddEndPunctfalse
\mciteSetBstMidEndSepPunct{\mcitedefaultmidpunct}
{}{\mcitedefaultseppunct}\relax
\EndOfBibitem
\bibitem[Srivastava \latin{et~al.}(2023)Srivastava, Crum, and Swartzlander]{Srivastava:2023aa}
Srivastava,~P.~R.; Crum,~R.~M.; Swartzlander,~G.~A. Broadband Diffractive Solar Sail. \textbf{2023}, \relax
\mciteBstWouldAddEndPunctfalse
\mciteSetBstMidEndSepPunct{\mcitedefaultmidpunct}
{}{\mcitedefaultseppunct}\relax
\EndOfBibitem
\bibitem[Srivastava and Swartzlander~Jr(2023)Srivastava, and Swartzlander~Jr]{Srivastava:2023ab}
Srivastava,~P.~R.; Swartzlander~Jr,~G.~A. Diffractive Light and Solar Sails. PhD Thesis, 2023\relax
\mciteBstWouldAddEndPuncttrue
\mciteSetBstMidEndSepPunct{\mcitedefaultmidpunct}
{\mcitedefaultendpunct}{\mcitedefaultseppunct}\relax
\EndOfBibitem
\bibitem[Zhang \latin{et~al.}(2022)Zhang, Firuzi, Yuan, Gong, and Gong]{Zhang:2022aa}
Zhang,~P.; Firuzi,~S.; Yuan,~C.; Gong,~X.; Gong,~S. General passive stability criteria for a Sun-pointing attitude using the metasurface sail. \emph{Aerospace Science and Technology} \textbf{2022}, \emph{122}, 107380--107380\relax
\mciteBstWouldAddEndPuncttrue
\mciteSetBstMidEndSepPunct{\mcitedefaultmidpunct}
{\mcitedefaultendpunct}{\mcitedefaultseppunct}\relax
\EndOfBibitem
\bibitem[Ullery \latin{et~al.}(2018)Ullery, Soleymani, Heaton, Orphee, Johnson, Sood, Kung, and Kim]{ullery:2018}
Ullery,~D.~C.; Soleymani,~S.; Heaton,~A.; Orphee,~J.; Johnson,~L.; Sood,~R.; Kung,~P.; Kim,~S.~M. Strong Solar Radiation Forces from Anomalously Reflecting Metasurfaces for Solar Sail Attitude Control. \emph{Scientific Reports} \textbf{2018}, \emph{8}, 10026\relax
\mciteBstWouldAddEndPuncttrue
\mciteSetBstMidEndSepPunct{\mcitedefaultmidpunct}
{\mcitedefaultendpunct}{\mcitedefaultseppunct}\relax
\EndOfBibitem
\bibitem[Swartzlander(2022)]{swartzlander:2022}
Swartzlander,~G.~A. Theory of radiation pressure on a diffractive solar sail. \emph{Journal of the Optical Society of America B} \textbf{2022}, \emph{39}, 2556\relax
\mciteBstWouldAddEndPuncttrue
\mciteSetBstMidEndSepPunct{\mcitedefaultmidpunct}
{\mcitedefaultendpunct}{\mcitedefaultseppunct}\relax
\EndOfBibitem
\end{mcitethebibliography}

\end{document}